\documentclass[aps,prd,nofootinbib,superscriptaddress,floatfix]{revtex4}
\usepackage{graphicx}
\usepackage{bm}
\usepackage{braket}
\usepackage{bbold}
\usepackage{amssymb}
\usepackage{amsmath,latexsym}
\usepackage{subfigure}
\usepackage{slashed}
\usepackage[T1]{fontenc}
\usepackage{multirow,array}
\usepackage{mathtools}
\usepackage{mathrsfs}
\usepackage{dsfont}
\usepackage[colorlinks=false,linktocpage=true]{hyperref}
\usepackage{hyperref}
\usepackage[utf8]{inputenc}
\usepackage[export]{adjustbox}
\usepackage{float}
\usepackage[makeroom]{cancel}
\usepackage{wrapfig}
\usepackage{ulem}
\usepackage{color}

\newcommand{\red}[1]{{\color{red} #1}}

\newcommand{\spacer}{ {$\displaystyle{\phantom{\int}}$} }
\newcommand{\be}{\begin{equation}}
\newcommand{\ee}{\end{equation}}
\newcommand{\ba}{\begin{eqnarray}}
\newcommand{\ea}{\end{eqnarray}}
\newcommand{\sub}[1]{   \begin{subequations}
                        #1
                        \end{subequations} }

\newcommand{\uncomment}[1]{{ \red{\bf[text removed]} }}  
\newcommand{\di}{\!{\rm d}}
\newcommand{\la}{\langle}
\newcommand{\ra}{\rangle}
\newcommand{\uvec}[1]{\boldsymbol{#1}}

\newcommand{\fslash}[1] {{\not\! #1\,}}

\newcommand{\mx}{{m_{\rm max}}}

\begin{document}

\newcommand*{\UConn}{Department of Physics, University of Connecticut,
  Storrs, CT 06269, U.S.A.}
\newcommand*{\Poly}{CPHT, CNRS, Ecole Polytechnique, Institut 
  Polytechnique de Paris, Route de Saclay, 91128 Palaiseau, France}  
\newcommand*{\BNL}{Department of Physics, Brookhaven National Laboratory, Upton, NY 11973, U.S.A.}

\title{\boldmath
2D energy-momentum tensor distributions of nucleon in a 
large-$N_c$ quark model \\
\vspace{2mm} from ultra-relativistic to non-relativistic limit}
\author{C\'edric Lorc\'e}\affiliation{\Poly}
\author{Peter Schweitzer}\affiliation{\UConn}
\author{Kemal Tezgin}\affiliation{\BNL}
\date{02/02/2022}
\begin{abstract}
\noindent
Form factors of the energy-momentum tensor (EMT) can be 
interpreted in certain frames in terms of spatial distributions
of energy, stress, linear and angular momentum, based on 2D or 3D Fourier transforms. This interpretation is in general subject to ``relativistic recoil corrections'', except when the nucleon moves at the speed of light like e.g. in the infinite-momentum frame. We show that it is possible to formulate a large-$N_c$ limit in 
which the probabilistic interpretation of the nucleon EMT distributions holds also in other frames. We use the bag model formulated in the
large-$N_c$ limit as an internally consistent quark model framework to
visualize the information content associated with the 2D EMT distributions. In order to provide more 
intuition, we present results in the physical situation and in three different limits: by considering a 
heavy-quark limit, a large system-size limit and 
a constituent-quark limit.
The visualizations of the distributions in these  
extreme limits will help to interpret the results from experiments,
lattice QCD, and other models or effective theories.
\end{abstract}
\maketitle


\maketitle

\section{Introduction}
\label{sec-1:Introduction}

In the recent years, the energy-momentum tensor (EMT) 
has been recognized as a key object by the hadronic physics community and attracted accordingly a lot of attention. It is directly related to the questions of the nucleon mass and spin decompositions which constitute two of the three pillars of the Electron-Ion Collider project in the U.S.A.~\cite{Accardi:2012qut,Aschenauer:2017jsk,AbdulKhalek:2021gbh}. High-energy scattering experiments and calculations in lattice QCD and models can be used to constrain matrix elements of the EMT, allowing us to study the mass~\cite{Ji:1994av,Ji:1995sv,Lorce:2017xzd,Hatta:2018sqd,Rodini:2020pis,Metz:2020vxd,Ji:2021mtz,Liu:2021gco,Lorce:2021xku}, spin~\cite{Ji:1996ek,Bakker:2004ib,Leader:2013jra,Wakamatsu:2014zza,Ji:2020ena}, and spatial distributions of energy, momentum and stress inside the nucleon~\cite{Polyakov:2002yz,Perevalova:2016dln,Polyakov:2018zvc,Lorce:2018egm,Cosyn:2019aio,Freese:2021czn,Panteleeva:2021iip,Freese:2021qtb,Freese:2021mzg,Metz:2021lqv,Ji:2021mfb}. This offers an unprecedented picture of the nucleon structure and even a glimpse into the question of its stability.

While both experimental 
~\cite{Airapetian:2001yk,Stepanyan:2001sm,
Ellinghaus:2002bq,Chekanov:2003ya,Aktas:2005ty,Airapetian:2006zr,
Camacho:2006qlk,Mazouz:2007aa,Airapetian:2007aa,Girod:2007aa,
Airapetian:2008aa,Airapetian:2009ac,Airapetian:2009bm,Airapetian:2009cga,
Airapetian:2009aa,Airapetian:2010ab,Airapetian:2010aa,Airapetian:2011uq,
Airapetian:2012mq,Airapetian:2012pg,Jo:2015ema,CLAS:2017udk,Burkert:2018bqq,CLAS:2018ddh,Kumericki:2019ddg,Benali:2020vma,CLAS:2021ovm,Dutrieux:2021nlz,Burkert:2021ith,JeffersonLabHallA:2022pnx} 
and lattice QCD data~\cite{Mathur:1999uf,Hagler:2003jd,Gockeler:2003jfa,Hagler:2007xi,Deka:2013zha,Yang:2018nqn,Shanahan:2018nnv,Shanahan:2018pib,Alexandrou:2020sml}
are accumulating,
numerous fundamental questions are addressed and studied from the theory side, ranging from the proper definition of the renormalized EMT in QCD, the various possibilities for decomposing the mass and the spin of a composite system, the understanding of relativistic effects and frame dependence, and many more aspects (see~\cite{Lorce:2021xku} for the most recent account), to the identification and suggestion of new processes and experimental observables. At the present stage of our knowledge, model calculations inspired by QCD are particularly useful since they provide valuable predictions guiding experimental studies. They also allow one to test explicitly general relations derived from formal considerations. A large number of models
and approaches have been developed over the years and used to study particular parton distributions or observables~\cite{Ji:1997gm,Petrov:1998kf,Schweitzer:2002nm,Ossmann:2004bp,
Goeke:2007fp,Goeke:2007fq,Wakamatsu:2007uc,
Cebulla:2007ei,Jung:2013bya,Kim:2012ts,Jung:2014jja,
Mai:2012yc,Mai:2012cx,Cantara:2015sna,Gulamov:2015fya,Nugaev:2019vru,
Donoghue:1991qv,Kubis:1999db,Belitsky:2002jp,Ando:2006sk,Diehl:2006ya,
Alharazin:2020yjv,Pasquini:2014vua,
Grigoryan:2007vg,Pasquini:2007xz,Hwang:2007tb,Abidin:2008ku,Abidin:2008hn,Abidin:2009hr,Broniowski:2008hx,
Brodsky:2008pf,Chakrabarti:2015lba,
Kumar:2017dbf,Mondal:2017lph,Hudson:2017xug,Hudson:2017oul,Kumano:2017lhr,Anikin:2019kwi,Granados:2019zjw,Freese:2019bhb,Freese:2019eww,Neubelt:2019sou,Azizi:2019ytx,Ozdem:2019pkg,Ozdem:2020ieh,Varma:2020crx,Kim:2021jjf,Kim:2022syw,Owa:2021hnj}.

In this work we push further the study of the EMT using the bag model in the large-$N_c$ limit studied in Ref.~\cite{Neubelt:2019sou}. We focus here on the 2D spatial distributions which are defined for arbitrary values of the nucleon average three-momentum 
$\vec P$~\cite{Lorce:2017wkb,Lorce:2018egm,Lorce:2020onh,Lorce:2021gxs,Lorce:2022jyi,Kim:2021jjf,Kim:2022bia,Kim:2022syw}. 
Besides obtaining a 2D picture of the nucleon
in the physical situation, we will also discuss in detail three insightful limits, namely a heavy-quark
limit, a large system-size limit, and a constituent
quark limit. While representing very different physical
situations, the limits have in common that the quarks
become effectively non-relativistic and the 
quark Compton wavelength becomes much smaller 
than the system size. We will study the behavior 
of the EMT distributions in these situations.
This will show how, within a quark model framework, 
the internal nucleon structure changes as one goes 
away from the real-world situation with tightly bound
ultra-relativistic quarks forming a compact nucleon, 
and approaches the different limits.

The paper is organized as follows. In Section~\ref{Sec-2:EMT-in-general} we present in a nutshell how 3D and 2D spatial distribution associated with the EMT are constructed, along with various general properties and the large-$N_c$ limit. Then we remind in Section~\ref{Sec-3:bag-model} the analytical results of Ref.~\cite{Neubelt:2019sou} for the 3D distributions in the bag model, and introduce in Section~\ref{sec:limits-overview} the various limits we will consider later (namely heavy quarks, large system size, and constituent quarks). After showing in Section~\ref{Sec:EMT-physical-situation} the 2D distributions in the physical situation, we discuss in detail three different limits in Sections~\ref{Sec:heavy-mass-limit-B-fix}-\ref{Sec:constituent-limit}. Finally we study the mass structure of the nucleon within the bag model picture in Section~\ref{Sec:M-decompose}, and summarize our findings in Section~\ref{Sec:conclusions}. Additional discussions can be found in Appendices.

\newpage

\section{EMT form factors and spatial distributions}
\label{Sec-2:EMT-in-general}

In this section we introduce the EMT form factors, define the
2D and 3D distributions in different reference frames, review their relations, 
and discuss the description of these EMT form factors and distributions in the
large-$N_c$ limit.

\subsection{Energy-momentum tensor and  form factors}

In QCD, the local gauge-invariant quark and gluon contributions to the EMT are defined as\footnote{See Refs.~\cite{Hatta:2018sqd,Tanaka:2018nae,Rodini:2020pis,Metz:2020vxd} for the case of a symmetric EMT renormalized in $\overline{\text{MS}}$ scheme up to three loops.}
\sub{\ba
    T^{\mu\nu}_q&=&\overline\psi_q\gamma^\mu\,\tfrac{i}{2}\overset{\leftrightarrow}{D}\!\!\!\!\!\phantom{D}^\nu\psi_q,\\
    T^{\mu\nu}_g&=&-F^{\mu\lambda}F^{\nu}_{\phantom{\nu}\lambda}+\tfrac{1}{4}\,g^{\mu\nu} F^2,
\ea}
where $\overset{\leftrightarrow}{D}_\mu=\overset{\rightarrow}{\partial}_\mu-\overset{\leftarrow}{\partial}_\mu-2ig A_\mu$ is the symmetric covariant derivative in the fundamental representation, $F^{\mu\nu}$ is the gluon field-strength tensor in the adjoint representation, and $g_{\mu\nu}=\text{diag}(+1,-1,-1,-1)$ is the Minkowski metric. The EMT is a key object since many current fundamental questions about the hadronic stucture are related to its components. Namely, the $00$ component addresses the question of the origin of the hadron mass~\cite{Ji:1994av,Ji:1995sv,Lorce:2017xzd,Hatta:2018sqd,Metz:2020vxd,Ji:2021mtz,Lorce:2021xku}, the $0i$ and $i0$ components address the question of the origin of the hadron spin~\cite{Ji:1996ek,Leader:2013jra,Wakamatsu:2014zza}, and the $ij$ components contain information about pressure forces inside the nucleon~\cite{Polyakov:2002yz,Polyakov:2018zvc,Lorce:2018egm,Freese:2021czn}.

The corresponding generalized angular momentum (AM) tensor is given by \cite{Leader:2013jra,Wakamatsu:2014zza,Lorce:2017wkb}
\be\label{genkindec1}
    J^{\mu\alpha\beta}= \sum_q L_{q}^{\mu\alpha\beta} + \sum_q  S_{q}^{\mu\alpha\beta} +  J_{g}^{\mu\alpha\beta},
\ee 
where ($\epsilon_{0123}=+1$)
\sub{\ba
     L_{q}^{\mu\alpha\beta} &=& x^\alpha\, T^{\mu\beta}_{q}\,-\,x^\beta\, T^{\mu\alpha}_{q}\, , \\
     S_{q}^{\mu\alpha\beta}\label{quark-spin-operator} &=& \tfrac{1}{2}\,\epsilon^{\mu\alpha\beta\lambda}\,
        \overline{\psi}_q\gamma_\lambda\gamma_5\psi_q\, , \\
     J_{g}^{\mu\alpha\beta} &=& x^\alpha\, T^{\mu\beta}_{g}\,-\,x^\beta\, T^{\mu\alpha}_{g}
\ea}
represent the quark orbital, quark spin, and gluon total AM contributions. The tensors $ L_{q}^{\mu\alpha\beta}$ and $J_{g}^{\mu\alpha\beta}$ are covariant forms of $\vec r\times\vec p$ and will accordingly be qualified as orbital-like. Lorentz symmetry implies that the generalized AM tensor is conserved $\partial_\mu J^{\mu\alpha\beta}=0$, and in turn relates the antisymmetric part of the quark EMT to the quark spin contribution
\be\label{antisymmid}
T^{[\alpha\beta]}_q\equiv \tfrac{1}{2}\left(T^{\alpha\beta}_q-T^{\beta\alpha}_q\right) = -\tfrac{1}{2}\,\partial_\mu S_{q}^{\mu\alpha\beta}.
\ee
In the literature, one often uses a symmetric EMT, known as the Belinfante EMT, 
which in QCD is related to the general asymmetric EMT as follows~\cite{Leader:2013jra}
\ba\label{Beldef}
T^{\mu\nu}_{\text{Bel,}a}=T^{\{\mu\nu\}}_{a}\equiv\tfrac{1}{2}\left(T^{\mu\nu}_{a}+T^{\nu\mu}_{a}\right).
\ea
The Belinfante generalized AM tensor reads
\be\label{genkindec2}
    J^{\mu\alpha\beta}_\text{Bel}= \sum_q J_{\text{Bel},q}^{\mu\alpha\beta} +  J_{\text{Bel},g}^{\mu\alpha\beta},
\ee
with
\be
J_{\text{Bel},a}^{\mu\alpha\beta} = x^\alpha\, T^{\mu\beta}_{\text{Bel},a}\,-\,x^\beta\, T^{\mu\alpha}_{\text{Bel},a},\qquad a=q,g.
\ee
Contrary to the kinetic generalized AM tensor $J^{\mu\alpha\beta}$, the Belinfante version $J^{\mu\alpha\beta}_\text{Bel}$ is purely orbital-like.

For a spin-$1/2$ target with mass $M_N$, the matrix elements of the general asymmetric EMT evaluated at the space-time origin $x=0$ can be parametrized in the following way \cite{Bakker:2004ib,Leader:2013jra,Cotogno:2019vjb}
\begin{equation}
\begin{aligned}
\label{Eq:def-EMT-kinetic}
    \la p^\prime,\vec s^{\,\prime}|  T_a^{\mu\nu}(0) |p,\vec s\rangle
    = \bar u(p^\prime,\vec s^{\,\prime})\biggl[A_a(t)\,\frac{P^\mu P^\nu}{M_N} &+
    J_a(t)\ \frac{P^{\{\mu}i\sigma^{\nu\}\lambda}
    \Delta_\lambda}{M_N}
    + D_a(t)\,
    \frac{\Delta^\mu\Delta^\nu-g^{\mu\nu}\Delta^2}{4M_N}\\
    &-S_a(t)\ \frac{P^{[\mu}i\sigma^{\nu]\lambda}
    \Delta_\lambda}{M_N} + \bar{C}_a(t)\,M_N\,g^{\mu\nu}\biggr]u(p,\vec s),
\end{aligned}
\end{equation}
where the kinematic variables are defined as
\be
        P=\tfrac{1}{2}(p'+p),        \quad
        \Delta=p'-p,          \quad
        t=\Delta^2.
        \label{Eq:kin-variables}
\ee
The unit vector $\vec s$ ($\vec s^{\,\prime}$) indicates the direction along which the initial (final) rest-frame spin is aligned. 
The form factors for different parton species depend on the renormalization 
scale $\mu$, e.g.\  $A_a(t)\equiv A_a(t,\mu^2)$, which is usually omitted for
brevity. The total EMT form factors $A(t)\equiv \sum_a A_a(t,\mu^2)$ and analogs 
for $J(t)$, $D(t)$ are renormalization scale invariant. 
The form factors $\bar{C}_a(t,\mu^2)$ account for the non-conservation 
of the separate quark and gluon EMTs. The total EMT being conserved, it 
follows that $\sum_a\bar{C}_a(t,\mu^2)=0$. Moreover, Poincar\'e symmetry implies that $A(0)=1$ and $J(0)=1/2$~\cite{Ji:1996ek,Lowdon:2017idv,Cotogno:2019xcl,Lorce:2019sbq}. Unlike the gluon spin, the quark spin operator can be expressed in way that is both local and gauge-invariant. As a result, the quark contribution to the EMT receives in general an antisymmetric contribution described by the form factor $S_q(t)$. For the Belinfante EMT, the latter drops out owing to Eq.~\eqref{Beldef}.

\subsection{3D spatial distributions in the Breit frame}
\label{Subsec:3D-distributions-in-BF}

For a nucleon state with rest-frame polarization in the $\vec s$-direction, a 3D spatial distribution of the EMT can be defined in the Breit frame (BF) where $P^\mu=(P^0,\vec{0})$ and $\Delta^\mu=(0,\vec{\Delta})$ as follows~\cite{Donoghue:2001qc,Polyakov:2002yz,Lorce:2018egm,Lorce:2021gxs}
\be\label{BFdef}
    \langle T^{\mu\nu}_{a}\rangle_\text{BF}(\vec r)=\int\frac{d^3\Delta}{(2\pi)^3}\,
    e^{-i\vec\Delta\cdot\vec r}\left[\frac{\la p^\prime,\vec s| 
    T_a^{\mu\nu}(0) |p,\vec s\rangle}{2P^0}\right]_{\vec P=\vec 0},
\ee
and can be expressed in terms of 3D Fourier transforms of the EMT form factors. Its components give access to a wealth of physical information. 


The $00$ component corresponds to the quark and gluon energy distributions  
\be
\langle T^{00}_{a}\rangle_\text{BF}(\vec r)=\epsilon_a(r),
\ee
which are related to the nucleon mass by
\be
    \sum_{a=q,g}\int d^3r\,\epsilon_a(r)\,=\,M_N.
\ee


The $0i$ and $i0$ components are related to the AM distributions inside the nucleon
\sub{\label{OAMdef}\ba
\epsilon^{ijk}r^j\langle T^{0k}_{q}\rangle_\text{BF}(\vec{r})&=&L^i_q(\vec r),\\
\epsilon^{ijk}r^j\langle T^{[0k]}_{q}\rangle_\text{BF}(\vec{r})&=&-\tfrac{1}{2}\!\left[\vec r\times(\vec\nabla\times\vec S_q(\vec r))\right]^i, \\
\epsilon^{ijk}r^j\langle T^{\{0k\}}_{q}\rangle_\text{BF}(\vec{r})&=&J^i_{\text{Bel},q}(\vec r),\\
\epsilon^{ijk}r^j\langle T^{0k}_{g}\rangle_\text{BF}(\vec{r})&=&J^i_g(\vec r)=J^i_{\text{Bel},g}(\vec r),
\ea}
which satisfy the AM sum rule~\cite{Ji:1996ek,Lorce:2021gxs}
\be
\sum_{a=q,g}\int d^3r\,\vec J_{\text{Bel},a}(\vec r)=\frac{\vec s}{2}.
\ee
A similar sum rule holds for the asymmetric EMT
\be
\int d^3r\left[\sum_q \vec L_q(\vec r)+\sum_q \vec S_q(\vec r)+\vec J_{g}(\vec r)\right]=\frac{\vec s}{2},
\ee
and involves the 3D distribution of quark spin in the BF
\be\label{Eq:spin-density-def}
    S^i_q(\vec r)=\frac{1}{2}\int\frac{d^3\Delta}{(2\pi)^3}\,
    e^{-i\vec\Delta\cdot\vec r}\left[\frac{\la p^\prime,\vec s| 
    \overline\psi(0)\gamma^i\gamma_5\psi(0) |p,\vec s\rangle}{2P^0}\right]_{\vec P=\vec 0}.
\ee
Note that $J^i_{\text{Bel},q}(\vec r)$ and $L^i_q(\vec r)+S^i_q(\vec r)$ differ by a total derivative~\cite{Lorce:2017wkb}
\be\label{Eq:relation-AM-distributions}
L^i_q(\vec r)+S^i_q(\vec r)-J^i_{\text{Bel},q}(\vec r)=\tfrac{1}{2}\nabla^j\!\left[r^j S^i_q(\vec r)-\delta^{ji}\,\vec r\cdot\vec S_q(\vec r)\right]
\ee
which vanishes under spatial integration.

For a nucleon target polarized along $\vec s$, the spatial dependence of any AM distribution (generically denoted by $J^i_a$) can be decomposed into monopole \cite{Goeke:2007fp} and quadrupole \cite{Lorce:2017wkb,Schweitzer:2019kkd} contributions 
\be
        J_a^i(\vec{r}) =
       \left[\delta^{ij}J^\text{mono}_a(r)
        +\left(\frac{r^ir^j}{r^2}-\frac13\,\delta^{ij}\right)J^\text{quad}_a(r)
        \right] s^j.
\ee
The monopole and quadrupole contributions are 
related to each other as~\cite{Schweitzer:2019kkd}
\be\label{Eq:ang-mom-dens}
        J^\text{mono}_a(r)
        =-\tfrac{2}{3}\,J^\text{quad}_a(r) \equiv J_a(r)
\ee
for the orbital-like contributions $J_a\in\{L^i_q,J^i_{\text{Bel},q},J^i_{g}\}$.
However, for the quark spin contribution $S^i_q$ the monopole and quadrupole
contributions are independent. 

The symmetric stress tensor can similarly be decomposed into monopole and quadrupole contributions~\cite{Polyakov:2002yz} 
\begin{equation}
    \langle T^{\{ij\}}_{a}\rangle_\text{BF}(\vec{r}) = \delta^{ij}p_a(r)+\left( \frac{r^i r^j}{r^2} - \frac13\,\delta^{ij} \right)s_a(r),
\end{equation}
which are interpreted as the (spin-independent) distributions of isotropic pressure and pressure anisotropy (or shear forces), respectively. The so-called radial and tangential pressures are then given by the combinations~\cite{Polyakov:2018zvc,Lorce:2018egm}
\be
p_{r,a}(r)=p_a(r)+\tfrac{2}{3}\,s_a(r),\qquad
p_{t,a}(r)=p_a(r)-\tfrac{1}{3}\,s_a(r).
\ee
The conservation of total EMT $\partial_\mu T^{\mu\nu}=0$ relates total pressure anisotropy $s(r)=\sum_a s_a(r)$ and total isotropic pressure $p(r)=\sum_a p_a(r)$ through a differential
equation
\be\label{Eq:p(r)+s(r)}
    \frac23\,\frac{d s(r)}{d r\;}+
    \frac{2s(r)}{r} + \frac{d p(r)}{d r\;} = 0\;.
\ee
It indicates in particular that the variation of the radial pressure is caused by shear forces\footnote{For macroscopic fluids in hydrostatic equilibrium and subjected to an external gravitational field, the bulk pressure is isotropic and decreases with height because of the external anisotropic gravitational force. Isotropic pressure also suddenly changes at the gas-liquid interface where anisotropic forces are modeled in terms of a surface tension.}.
Other consequences of EMT conservation are the following conditions:
\sub{\label{Eq:von-Laue}
\ba
    \int_0^\infty dr\,r^2 p(r) &=& 0\;,   
    \label{Eq:von-Laue-3}\\
    \int_0^\infty dr\,r 
    \biggl[-\frac13\,s(r)+p(r)\biggr] &=& 0\;, \label{Eq:von-Laue-2}\\
    \int_0^\infty dr\,
    \biggl[-\frac43\,s(r)+p(r)\biggr] &=& 0\;, \label{Eq:von-Laue-1}
\ea}
where Eq.~(\ref{Eq:von-Laue-3}) is called the von Laue
condition (or sometimes, more loosely, the equilibrium condition), while the
Eqs.~(\ref{Eq:von-Laue-2},~\ref{Eq:von-Laue-1}) are sometimes 
called the respective lower-dimensional von Laue conditions
(though they are pertinent to the 3D pressure distribution,
and should not be confused with the 2D conditions discussed
in App.~\ref{App-2D-constraints}).
The relations~\eqref{Eq:von-Laue} are necessary conditions
for the mechanical stability of an extended particle.

\subsection{2D spatial distributions with arbitrary momentum}

3D spatial distributions are restricted to the BF, where the target has vanishing average momentum $\vec P=\vec 0$. The concept of relativistic spatial distribution can however be extended to the more general case $\vec P\neq \vec 0$, at the price of losing one spatial dimension. Choosing for convenience the $z$-direction along $\vec P$, 2D spatial distributions of the EMT can be defined in the class of elastic frames (EF), where the energy transfer vanishes $\Delta^0=0$, as follows~\cite{Lorce:2017wkb,Lorce:2018egm,Lorce:2020onh,Lorce:2021gxs}
\begin{equation}\label{Eq:EMT-2D-elastic-frame}
\langle T^{\mu\nu}_{a}\rangle_\text{EF}(\uvec b;P_z)=\int\frac{d^2\Delta_\perp}{(2\pi)^2}\,e^{-i\uvec\Delta_\perp\cdot\uvec b}\left[\frac{\la p^\prime,\vec s|  T_a^{\mu\nu}(0) |p,\vec s\rangle}{2P^0}\right]_{|\uvec P_\perp|=\Delta_z=0} \,.
\end{equation}
The BF corresponds to the special EF where $P_z\to0$. In that case, the 2D distributions simply reduce to the projection of 3D distributions onto the transverse plane
\begin{equation}\label{Eq:EMT-2D-Breit-frame}
    \langle T^{\mu\nu}_{a}\rangle_\text{EF}(\uvec b;0)=\int dz\,
    \langle T^{\mu\nu}_{a}\rangle_\text{BF}(\vec r)
\end{equation}
with $\vec r=(\uvec b, z)$. We can then easily relate the 2D distributions in the BF to the 3D ones
~\cite{Lorce:2017wkb,Lorce:2018egm} 
\sub{\label{Eqs:2D-3D-coonetction}
\begin{eqnarray}
  \epsilon_a(b) & = & \int dz \, \epsilon_a(r) \, , \label{Eq:2D-energy} \\
  J^{i}_a (\uvec{b}) & = & \int dz \,J^{i} _a (\vec{r}) \, , \label{Eq:2D-am} \\
  p_a(b) & =&  \int dz  \left[p_a(r) + \frac{b^2 - 2 z^2}{6 r^2}\,s_a(r) \right] 
  \, , \label{Eq:2D-pressure} \\
  s_a(b) & = &\int dz \, \frac{b^2}{r^2}\,s_a(r)\, , \label{Eq:2D-shear} \\
  {p_{r,a}(b)} & =&  \int dz \, \frac{b^2\,p_{r,a}(r) + z^2 \, p_{t,a}(r)}{r^2} 
  \, , \\
  {p_{t,a}(b)} & =&  \int dz \, p_{t,a}(r)
  \, , 
\end{eqnarray}}
where $J^{i}_a$ denotes either $L_{q}^{i}$, $S_{q}^{i}$, $J_{\text{Bel},q}^{i}$ or $J_{g}^{i}=J_{\text{Bel},g}^{i}$.
The transformation from the 3D to 2D distributions with spherical symmetry
is invertible and known as Abel transformation~\cite{Panteleeva:2021iip}. 
The pressure distributions $p(b)$ and $s(b)$ correspond to the 2D monopole 
and quadrupole contributions to the transverse part ($i,j=1,2$) of the 
symmetric stress tensor
\be
 \langle T^{\{ij\}}_{a}\rangle_\text{EF}(\uvec{b};0) = \delta^{ij}_\perp p_a(b)+\left( \frac{b^i b^j}{b^2} - \frac12\,\delta^{ij}_\perp \right)s_a(b).
\ee
Like in the 3D case, it follows from the conservation of the total EMT that
\be\label{Eq:diff-eq-p-s-2D}
\frac12\,\frac{d s(b)}{d b\;}+
    \frac{s(b)}{b} + \frac{d p(b)}{db\;} = 0\;.
\ee

For a longitudinally polarized nucleon, these 2D distributions satisfy 
the relations 
\sub{
\ba
      \sum\limits_a\int d^2 b\,\epsilon_a(b) &=& M_N, \\
      \sum\limits_a\int d^2 b\,J^z_{\text{Bel},a}(b) &=& \frac12 \, , \\
      \sum\limits_a\int d^2 b\,p_a(b) &=& 0 \, , \label{Eq:von-Laue-2D}\\
      2M_N \sum\limits_a\int d^2 b\;b^2 p_a(b) &=& D(0) \, , \\
      -\frac{M_N}{2} \sum\limits_a\int d^2 b\;b^2 s_a(b) &=& D(0)\,,
\ea
}
where $D(0)$ is the $D$-term \cite{Polyakov:1999gs}.
Since relativistic boosts do not commute 
with each other, 2D distributions get more and more distorted as we 
increase $P_z$. In the infinite-momentum frame (IMF), they coincide 
(up to a trivial Jacobian factor) with the light-front (LF) spatial
distributions~\cite{Lorce:2018egm,Freese:2021czn} in the symmetric
Drell-Yan frame defined by $\Delta^+=0$ and 
$P^\mu=(P^0,\uvec 0_\perp,P_z)$,
\begin{equation}\label{Eq:lim-LC}
\lim_{P_z\to \infty}\langle T^{\mu\nu}_{a}\rangle_\text{EF}(\uvec b;P_z)/\sqrt{2}=\lim_{P^+\to\infty}\langle T^{\mu\nu}_{a}\rangle_\text{LF}(\uvec b),
\end{equation}
where
\begin{equation}\label{Eq:EMT-LC}
\langle T^{\mu\nu}_{a}\rangle_\text{LF}(\uvec b)=\int\frac{d^2\Delta_\perp}{(2\pi)^2}\,e^{-i\uvec\Delta_\perp\cdot\uvec b}\left[\frac{\langle p^\prime,\lambda| T_a^{\mu\nu}(0) |p,\lambda\rangle}{2P^+}\right]_{|\uvec P_\perp|=\Delta^+=0}.
\end{equation}
Here the LF components are defined as $a^\pm=(a^0\pm a^3)/\sqrt{2}$, and the LF momentum eigenstates with LF helicity $\lambda$ are normalized as $\langle p',\lambda'|p,\lambda\rangle=2P^+(2\pi)^3\delta(p'^+-p^+)\,\delta^{(2)}(\uvec p^{\,\prime}_\perp-\uvec p_\perp)\,\delta_{\lambda'\lambda}$. 

\subsection{Stability requirements for 2D BF distributions}
\label{Sec-IID}

The 3D EMT distributions satisfy certain criteria which are necessary 
(but not sufficient) requirements for mechanical stability. Namely, in a 3D stable system, it is expected (at least classically)~\cite{Lorce:2018egm} that at $r=0$ 
one has 
$\epsilon(0) < \infty$, $p(0) < \infty$, 
$s(0) = 0$, while at $r>0$ the following inequalities
hold
\ba
&&     \epsilon(r) > 0, 
\quad  p_r(r) > 0 ,
\quad  \frac{d\epsilon(r)}{dr} < 0, 
\quad  \frac{dp_r(r)}{dr} < 0, \nonumber\\
&&     
       \epsilon(r) + p_i(r) \geq 0,
\quad  \epsilon(r) + 3\,p(r) \geq 0, 
\quad   \epsilon(r) \geq |p_i(r)|,
\phantom{\frac11}
\ea 
where $i=r,\,t$. (We remind that throughout this work we use natural units with
$c=1$ and $\hbar = 1$.)

These constraints on the 3D distributions can be translated into 
2D stability conditions. At $b=0$ we expect the following to hold:
$\epsilon(0) < \infty$, 
$p(0) < \infty$ and $s(0) = 0$.
For $b>0$ the other constraints are
\ba
&&    \epsilon(b) \geq 0,
\quad p_r(b) \geq 0,
\quad \frac{d\epsilon(b)}{db} \leq 0,
\quad \frac{dp_r(b)}{db} \leq 0, \nonumber\\
&& \epsilon(b) + p_i(b) \geq 0,
\quad \epsilon(b) + 2\,p(b) \geq 0,
\quad \epsilon(b) \geq |p_i(b)|.
\phantom{\frac11}\ea
While alluded to in Ref.~\cite{Lorce:2018egm}, 
to the best our knowledge these constraints on the
2D BF distributions have not been discussed explicitly 
before in literature, except the positivity of radial 
pressure expressed as $p(b)+\frac{1}{2}\,s(b)\geq 0$ 
\cite{Freese:2021czn}.
The proofs of these relations, relying on the validity of the corresponding 3D counterparts, are given in Appendix~\ref{App-2D-constraints}.

\subsection{Large-$N_c$ limit}
\label{Sec-III-E:large-Nc}

In the large-$N_c$ limit the nucleon mass behaves as $M_N\sim N_c$,
while the nucleon 3-momenta are assumed to scale like $N_c^0$. 
This implies the hierarchy 
$P^0\sim N_c\gg |\vec P|\sim|\vec\Delta|\sim N_c^0\gg \Delta^0\sim N_c^{-1}$.
The initial four-momentum is given by $p^\mu\approx M_N(1,\vec v)$ with the initial nucleon 
velocity $\vec{v}\approx \vec p/M_N \sim N_c^{-1}$, and similarly for the final state. Thus, the  motion of the nucleon is slow and non-relativistic. Independently of the
nucleon being non-relativistic as a whole, the motion of its constituents 
may however range from non-relativistic (e.g. heavy quarks in non-relativistic
quark models) to ultra-relativistic (e.g. light quarks in relativistic 
models or QCD) as we shall discuss below.

The leading terms in the large-$N_c$ expansions of the nucleon matrix elements polarized along $\vec s$ for the different quark EMT components are given by
\begin{subequations}\label{Eqs:EMT-expansion-CORR}
\begin{align}
    \la p^\prime,\vec s|  T_Q^{00}(0) |p,\vec s\rangle
    & = 
    2M_N^2\left[A_Q(t)+\bar C_Q(t)
    +\frac{\vec\Delta^2}{4M_N^2}\,D_Q(t)\right]
     +\, {\cal O}(N_c^0)\,,
    \\
    \la p^\prime,\vec s| T_Q^{\{0i\}}(0) |p,\vec s\rangle
    & = 
    2M_N\left[P^iA_Q(t)
    +\frac{i(\vec s\times\vec\Delta)^i}{2}\,J_Q(t)\right]
     +\, {\cal O}(N_c^0)\,,
    \label{Eq:trouble-some-term}\\
    \la p^\prime,\vec s|  T_Q^{[0i]}(0) |p,\vec s\rangle
    & = -M_N\,i(\vec s\times\vec\Delta)^iS_Q(t)
     +\, {\cal O}(N_c^0)\, ,  
    \\
    \la p^\prime,\vec s|  T_Q^{\{ij\}}(0) |p,\vec s\rangle
    & = 
    2M_N^2\left[-\delta^{ij}\bar C_Q(t)+
    \frac{\Delta^i\Delta^j-\delta^{ij}\vec\Delta^2}{4M_N^2}\,D_Q(t)\right]
     +\, {\cal O}(N_c^0)\,, \label{symmstress}
    \\
    \la p^\prime,\vec s|  T_Q^{[ij]}(0) |p,\vec s\rangle
    & = {\cal O}(N_c^0)\,. \label{shear}
\end{align}
\end{subequations}
The large-$N_c$ behavior of the EMT form factors for the 
different $u\pm d$ flavor combinations of the light quarks is
as follows:
$A^{u+d}(t)\sim N_c^0$,
$J^{u-d}(t)\sim N_c$, 
$S^{u-d}(t)\sim N_c$, 
$D^{u+d}(t)\sim N_c^2$,
$\bar{C}^{u+d}(t)\sim N_c^0$
are leading, while 
$A^{u-d}(t)\sim N_c^{-1}$,
$J^{u+d}(t)\sim N_c^0$, 
$S^{u+d}(t)\sim N_c^0$, 
$D^{u-d}(t)\sim N_c$, 
$\bar{C}^{u-d}(t)\sim N_c^{-1}$ are respectively subleading 
in the $1/N_c$ expansion~\cite{Polyakov:2018zvc}. 
The total quark contribution denoted by the index $Q$ in
Eqs.~(\ref{Eqs:EMT-expansion-CORR}) is already exhausted
by the $u+d$ flavor combination when working in a model 
in the SU(2) flavor sector which we shall do in the following.

In the large-$N_c$ limit, $P_z$ remains always much smaller than $M_N$. Distortions of spatial distributions induced by the motion of the target are therefore subleading in the $1/N_c$ expansion. The fundamental reason for this is that the Lorentz group becomes the Galilean group in the limit $P^0\propto N_c\to\infty$. 
An exception are the AM distributions
for a transversely polarized nucleon due to the appearance of the term $P^i A_Q(t)$ 
in Eq.~\eqref{Eq:trouble-some-term}. This term is expected because it is associated with the center-of-mass  motion of the system. Indeed, let us consider a rigid block
of matter moving at some constant velocity without rotation, and hence with vanishing internal AM. The spatial distribution of momentum is nonzero inside the body, and the AM distribution does not vanish. Integrating over space, one finds that total AM is given by $\vec J_\text{CM}=\vec R\times\vec P$, where $\vec R$ is the position of the center of mass relative to the origin of the coordinate system. Choosing the origin along the trajectory of the center of mass eliminates this external contribution to the total AM, but does not set the corresponding spatial distribution to zero.
Notice that this contribution drops out when considering
a longitudinally polarized nucleon (which we shall do throughout
in the following).
Therefore, in the large $N_c$ limit, the Breit frame and elastic frame 
2D distributions coincide for a longitudinally polarized nucleon, and in 
the case of a transversely polarized nucleon they differ for the AM distribution by a trivial expected effect due to the center-of-mass motion. 

Note that we may also consider the infinite-momentum limit 
$P_z\to\infty$, but since the large-$N_c$ limit was taken first, 
the nucleon will never move with relativistic velocities, and hence 
will never coincide with the corresponding LF spatial distributions. 
In the following we will discuss a set of 2D distributions in the bag model 
in the large-$N_c$ limit with the understanding that for them no distinction needs to be made between BF, EF and IMF distributions.

\section{The bag model, and a recap of the associated 3D EMT distributions}
\label{Sec-3:bag-model}

In the bag model quarks are confined inside a spherical cavity (``bag'') 
of radius $R$ by appropriate boundary conditions on its surface $S$. Baryons (mesons) are described by placing $N_c=3$ non-interacting quarks (a $\bar qq$ pair) 
in a color-singlet state inside the cavity \cite{Chodos:1974je,Chodos:1974pn}. 
The Lagrangian of the bag model can be written as \cite{Thomas:2001kw}
\be\label{Eq:Lagrangian-bag}
        {\cal L} = \sum\limits_q\biggl[\overline\psi_q\biggl(
        \tfrac{i}{2}\overset{\leftrightarrow}{\fslash{\partial}}
        -m\biggl)\psi_q\biggr]\Theta_V 
        + 
        \frac12\,\sum\limits_q
        \overline\psi_q\,\psi_q\,\eta^\mu\partial_\mu\Theta_V
        -
        B\,\Theta_V\,,
\ee
where $\overset{\leftrightarrow}{\partial}_\mu=\overset{\rightarrow}{\partial}_\mu-\overset{\leftarrow}{\partial}_\mu$ and $B>0$ is the energy density inside the bag.  It is convenient
to define (in the rest frame of the bag)
\be
        \Theta_V=\Theta(R-r),   \quad
        \eta^\mu=(0,\vec{e}_r), \quad
        \vec{e}_r = \vec{r}/r,  \quad
        r=|\vec{r}\,|.
\ee
From the Lagrangian~\eqref{Eq:Lagrangian-bag} one obtains the equations 
of motion for the (free) quarks $(i\fslash{\partial}-m)\psi_q=0$ for 
$r<R$ inside the bag, as well as the linear boundary condition 
$i\fslash{\eta}\,\psi_q = \psi_q$ for $\vec{r}\in S$ and the 
non-linear boundary condition 
$-\,\frac12\,\sum_q\eta_\mu\partial^\mu(\overline\psi_q\psi_q)= B$. The boundary conditions are such that there is no energy-momentum flowing out of the bag, i.e.\ $\eta_\mu T^{\mu\nu}(t,\vec{r})=0$ for
$\vec{r}\in S$ \cite{Chodos:1974je}. The ground state has 
positive parity and is described by the wave function
\be\label{Eq:bag-wave-function}
    \psi_{s}(t,\vec{r}) = e^{-i\varepsilon_it}
    \,\phi_s(\vec{r})
        \, , \;\;\;
    \phi_s(\vec{r})   = \frac{A}{\sqrt{4\pi}}\,
    \left(\begin{array}{l}
        \alpha_+j_0(\omega_ir/R)\,\chi_s \phantom{\displaystyle\frac11}\\
        \alpha_-j_1(\omega_ir/R)\,i\vec{\sigma}\cdot\vec{e}_r\chi_s
    \end{array}\right)
        \, , \;\;\;
        A = \biggl(\frac{\Omega_i(\Omega_i- m R)}
        {R^3j_0^2(\omega_i)(2\Omega_i(\Omega_i-1)+mR)}\biggr)^{\!1/2},
\ee
where $\alpha_\pm=\sqrt{1\pm mR/\Omega_i}$ with
$\Omega_i=\sqrt{\omega_i^2+m^2R^2}$, $\sigma^i$ are $2\times2$
Pauli matrices, $\chi_s$ are two-component Pauli spinors.
The single-quark energies are given by $\varepsilon_i=\Omega_i/R$.
The $\omega_i$ denote solutions of the transcendental equation
\be\label{Eq:omega-transcendental-eq}
        \omega_i = (1-mR-\Omega_i)\,\tan\omega_i \,.
\ee
The ground-state solution for massless quarks is $\omega_0\approx 2.04$, 
and swipes the interval $2.04 \lesssim \omega_0(mR) \le \pi$ when the 
product $mR$ is varied from 0 to infinity. 
The constant $A$ in Eq.~(\ref{Eq:bag-wave-function}) is such that 
$\int\di^3r\,\phi^\dag_{s^\prime}(\vec{r}^{\,})\,\phi^{ }_{s^{ }}(\vec{r}^{\,})
=\delta_{s^\prime\!s}$.

The nucleon mass is due to contributions from quarks and the bag, and is given by 
\be\label{virial-massive-M}
       M_N = N_c\,\frac{\Omega_0}{R}+\frac{4\pi}{3}\,R^3B.
\ee
The condition 
$M_N'(R)=0$ is sometimes referred to as the virial theorem and yields 
the relation
\be\label{virial-massive-MR}
    4\pi R^4 B =N_c\,\frac{2(\Omega_0-1)\omega_0^2}
                            {2\Omega_0(\Omega_0-1)+m R}.
\ee
Assuming SU(4) spin-flavor symmetry, the nucleon matrix elements of
quark operators are related to those of the single quark by spin-flavor 
factors: $N_q$ for nucleon spin-independent matrix elements and $P_q$ for 
spin-dependent matrix elements. For the proton we have $N_u = \frac{N_c+1}{2}$, 
$N_d = \frac{\;N_c-1}{2}$, $P_u = \frac{N_c+5}{6}$, $P_d = \frac{-N_c+1}{6}$ 
where $N_c=3$ is the number of colors. For the neutron the labels $u$ and $d$ 
are interchanged~\cite{Karl:1984cz}.

The bag model belongs to the class of so-called ``independent-particle models''
in which one encounters technical difficulties when evaluating one-body operators 
such as the EMT~\cite{Ji:1997gm}. The large-$N_c$ limit allows one to avoid these
problems and to consistently evaluate EMT form factors~\cite{Neubelt:2019sou}.
In the following we shall therefore assume that we work in the large-$N_c$ 
limit (when presenting numerical results we of course set $N_c=3$).
One important advantage of working in the large-$N_c$ limit is that
the system as a whole moves with non-relativistic velocities, so that the 2D and 3D distributions can be thought of as actual densities, and not only as quasidensities~\cite{Lorce:2020onh}.

In the bag model the kinetic quark EMT operator is given by
\be\label{Eq:EMT-kinetic}
        T^{\mu\nu}_{q} = \overline{\psi}_q\gamma^\mu\tfrac{i}{2}\overset{\leftrightarrow}{\partial}\!\!\!\!\phantom{\partial}^\nu\psi_q\,.
\ee
The expressions for EMT form factors associated with the symmetric part were derived for 
$N_c=3$ in~\cite{Ji:1997gm} and in the large-$N_c$ limit in~\cite{Neubelt:2019sou}. 
When calculating matrix elements of local operators in the 
large-$N_c$ limit, one naturally obtains expressions for the
form factors which are given by Fourier transforms of 3D 
distributions~\cite{Goeke:2007fp}. 
(We do not repeat here the expressions for the EMT form 
factors derived in the bag model in large-$N_c$ limit in~\cite{Neubelt:2019sou}
but 
other examples can be found in the Appendices \ref{Appendix:Sz} and \ref{Appendix:charge-distribution}, namely the electric and axial form factors included for comparison.) 

The 3D quark and ``gluon'' EMT distributions 
are given by  
\begin{subequations}
\label{Eq:EMT-kin}
\ba
    T^{00}_{q}(r) &=&
    \frac{N_q\,A^2}{4\pi}\;\frac{\Omega_0}{R}\,
    \biggl(\alpha_+^2j_0^2+\alpha_-^2j_1^2\biggr)\,\Theta_V\,,
    \label{Eq:EMT-kin-T00q}\\
    T^{0k}_{q}(\vec{r}) &=&
        -\,\frac{P_q\,A^2}{4\pi}\,
        \biggl(\alpha_-^2\frac{j_1^2}{r}\biggr)
        \epsilon^{klm}\,e_r^l\,S^m \,\Theta_V \,,
    \label{Eq:EMT-kin-T0kq}\\
    T^{k0}_{q}(\vec{r}) &=&
        -\,\frac{P_q\,A^2}{4\pi}\,
        \biggl(2\alpha_{+}\alpha_{-}\frac{\Omega_0}{R}\,j_0\,j_1\biggr)
        \epsilon^{klm}\,e_r^l\,S^m \,\Theta_V \,,
    \label{Eq:EMT-kin-Tk0q}\\
    T^{ik}_{q}(\vec{r}) &=&
    \frac{N_q\,A^2}{4\pi}\;\alpha_+\alpha_-\left[
    \biggl(j_0\,j_1^\prime-j_0^\prime\,j_1-\frac{j_0\,j_1}{r}\;\biggr)
    e_r^ie_r^k \; + \; \frac{j_0\,j_1}{r}\;\delta^{ik} \right]\Theta_V \,,
    \label{Eq:EMT-kin-Tijq}\\
    T^{\mu\nu}_{g}(r) &=& g^{\mu\nu}B\,\Theta_V \,.
    \phantom{\frac11}
    \label{Eq:EMT-kin-G}
\ea
\end{subequations}
The arguments of the spherical Bessel functions are
$j_i=j_i(\omega_0\, r/R)$, primes denote differentiation with 
respect to $r$. The contribution $T^{\mu\nu}_g(r)=g^{\mu\nu}B\,\Theta_V $ 
is due to the bag, i.e.\ due to non-fermionic degrees of freedom. It 
is essential to bind the quarks, and in this sense it can be associated 
with ``gluonic'' effects in QCD  \cite{Ji:1997gm,Neubelt:2019sou}.
The derivation of the results in (\ref{Eq:EMT-kin}) is described 
in detail in Ref.~\cite{Neubelt:2019sou},
except that the antisymmetric contribution related to the spin distribution~\eqref{antisymmid}
was not computed. These are new results obtained
in this work. The Eqs.~(\ref{Eq:EMT-kin}) are
the starting point for the developments in this
work. 

For completeness let us summarize in the following the explicit results
for the EMT distributions. 
The total energy distribution $\epsilon(r)$ inside the nucleon is the sum of the contributions to the $T^{00}$ component of the EMT. Hence, both quarks and the bag contribute to the energy distribution. Their overall contribution is given by
\sub{\label{Eqs:3D-review}
\begin{equation}
  \epsilon(r) = 
  \Biggl[\frac{N_c\,A^2}{4\pi}\;\frac{\Omega_0}{R}\,
  \biggl(\alpha_+^2j_0^2+\alpha_-^2j_1^2\biggr)+B\Biggr]\,\Theta_V\,.
  \label{Eq:3D-energy} 
\end{equation}

The AM distribution is determined from the $T^{0k}$ components 
of the asymmetric EMT. It receives no contribution from the bag and consists 
only of spin and orbital angular momentum (OAM) contributions due to quarks.
Choosing the nucleon polarization along the $z$-direction the total AM, OAM and
spin distributions are given by
\ba
    J^z(\vec{r}) &=&  
        \sum_q\biggl[L_{q}^z (\vec{r}) + S_{q}^z (\vec{r})\biggr], \\
    \label{Eq:OAM-z-distribution} L_{q}^z  (\vec{r}) &=& \frac{P_q\,A^2}{4\pi}\,
        \biggl[\alpha_{-}^2\,j_1^2\,\bigl( 1-\cos^2\theta\bigr)\biggr]\,\Theta_V, \\
    \label{Eq:Spin-z-distribution} S_{q}^z (\vec{r}) &=& \frac{P_q\,A^2}{8\pi}\,
        \biggl[\alpha_{+}^2\,j_0^2\,+\,\alpha_{-}^2\,j_1^2\,
        \bigl(2\,\cos^2\theta\,-\,1\,\bigr)\biggr]\,\Theta_V,\\
    \label{Eq:Belinfante-z-distribution} J^z_{\rm Bel,q}(\vec{r}) &=& 
    \frac{P_q\,A^2}{8\pi}\,\biggl[\frac{2\Omega_0}{R}\,
    \alpha_+\alpha_-\,r\,j_0\,j_1+\alpha_-^2\,j_1^2\biggr]
    (1-\cos^2\theta)\,\Theta_V\,, 
\ea
where the angle $\theta$ is defined by the projection of $\vec{r}$
on the $z$-axis (with the unit vector $\vec{e}_z$)
as $\vec{e}_z\cdot\vec{r}=r\,\cos\theta$.

The isotropic pressure and pressure anisotropy distributions are related 
to the symmetric part of $T^{ij}$ (the antisymmetric contribution
to $T^{ij}$ is zero in the leading order of the large-$N_c$ expansion).
Both the bag and quark degrees of freedom contribute to the isotropic
pressure, which is related to the trace of $T^{ij}$. The pressure 
anisotropy $s(r)$, being related to the symmetric traceless part 
of $T^{ij}$, is due to quarks only. The model expressions are given by
\begin{eqnarray}
  p(r) & = &
  \Biggl[\frac{N_{c}\,A^{2}}{12\pi}\;\alpha_{+}\alpha_{-}
  \biggl(j_{0}j_{1}^{\prime}-j_{0}^{\prime}j_{1}+\frac{2}{r}\,j_{0}j_{1}\biggr)-B\Biggr]
  \,\Theta_V 
\, =\, p_q(r) - B \,\Theta_V,
  \label{Eq:3D-pressure} \\
  s(r) & = &
  \Biggl[\frac{N_{c}\,A^{2}}{4\pi}\;\alpha_{+}\alpha_{-}
  \biggl(j_{0}j_{1}^{\prime}-j_{0}^{\prime}j_{1}-\frac{1}{r}\,j_{0}j_{1}\biggr)\Biggr]
  \,\Theta_V \label{Eq:3D-shear}
\end{eqnarray}}
which satisfy the differential relation~\eqref{Eq:p(r)+s(r)},
and $p(r)$ satisfies the conditions~\eqref{Eq:von-Laue}. 
In Eq.~(\ref{Eq:3D-pressure}) we defined 
the quark contribution $p_q(r)$ to the total pressure for later convenience.

\section{Limits within the bag model}
\label{sec:limits-overview}

It will be instructive to study 2D EMT distributions 
not only in the physical situation (which we shall do in
Sec.~\ref{Sec:EMT-physical-situation}), but also in various limiting situations within the bag model
(in Secs.~\ref{Sec:heavy-mass-limit-B-fix}, 
\ref{Sec:large-R-limit-m-fix}, \ref{Sec:constituent-limit}).
For that we will explore three limits corresponding to three 
different physical situations as explained in this section.

The bag model is uniquely defined by specifying two out 
of the following three parameters: the bag constant $B$ 
representing QCD properties in the vacuum sector, 
the quark mass $m$ reflecting QCD properties in the 
quark sector, and the bag radius $R$ which 
is a key property characterizing hadronic properties.
The nucleon mass plays a special role because the bag
solution is determined by minimizing the nucleon mass
as a function of the bag radius, $M_N(R)$. Moreover, 
in the physical situation one can choose the parameters to
reproduce the experimental value of $M_N$ (this can and will be relaxed in some of the limits).
All the other hadronic properties are then automatically determined.

The limits are therefore uniquely defined by specifying 
one parameter which will be taken to infinity, and one
quantity which will be kept fixed.
The three limits considered in this work will be 
referred to as L1, L2, L3.
In the limit L1, the quark mass $m$ will be taken to 
infinity keeping the bag constant $B$ fixed. 
In the limit L2, the bag radius will be taken to 
infinity while the quark mass $m$ is fixed.
In the limit L3, we finally will take the quark mass to
approach $1/N_c$ of the  nucleon mass with the latter 
kept fixed at its physical value (in the limits
discussed here, $N_c$ is always a constant). The limits 
are summarized in Table~\ref{Tab1} which features the 
quantities $B,\,R,\,m,\,M_N$ showing which is varied,
which is kept fixed, and the behavior (``response'') of 
the respectively other quantities in these limits.
Some comments are in order.

In a general situation, the exact relation between the
parameters
is complicated and governed by two equations, namely the 
transcendental equation (\ref{Eq:omega-transcendental-eq}) 
determining the frequency $\omega_0$ of the ground state 
bag solution for given $m$ and $R$, and 
the virial theorem (\ref{virial-massive-MR}) which 
determines the minimum of the nucleon mass $M_N$ understood
as a function of $R$ for specified\footnote{In
this system of equations, the four quantities $B$, $R$, $m$,
and $\omega_0$ are connected by two equations, 
Eqs.~(\ref{Eq:omega-transcendental-eq})
and (\ref{virial-massive-MR}), meaning that two of 
these four quantities can be eliminated. This leaves 
two free parameters which must be specified or fixed
in some way, as described in the text. Notice that
in the text $\omega_0$ is not considered to be
a model parameter and is always implicitly assumed to 
be eliminated.}
$m$ and $B$.  
Therefore, in the general case, no analytic relations 
exist between the parameters. However, in each of the 
three limits, the dimensionless variable $mR\to\infty$ 
goes to infinity. 

Physically, this means that the quark Compton
wavelength becomes much smaller than the system size.
In the three limits the dynamics becomes effectively
non-relativistic. This may not be intuitive at first
glance, especially in the limit L2 where we can choose
the quarks to have any (non-zero) mass, and light quarks
are always associated with relativistic effects. 
However, a clear criterion revealing that a system is non-relativistic is that the quark mass $m$ makes a
dominant contribution to the quark energy $\Omega_0/R$.
This condition is met in all three limits, i.e.\ we have
\be
        \frac{\Omega_0/R-m}{m} \ll 1.
\ee
Notice, that $\omega_0=\omega_0(mR)$ is a function of $mR$.
The situation simplifies considerably in the limit 
$mR\to \infty$ because 
the transcendental bag
equation~\eqref{Eq:omega-transcendental-eq} 
can then be solved analytically with $\omega_0(mR)=\pi-
\pi/(2mR) + {\cal O}(1/(mR)^3)$~\cite{Neubelt:2019sou},
and the virial theorem (\ref{virial-massive-MR}) 
assumes the form
\be\label{eq:virial-limit-mR}
    4mBR^5 = N_c\pi + \dots
\ee
where the dots indicate subleading terms suppressed 
by powers of $1/(mR)$ for large $mR$ (notice that 
power corrections in Eq.~(\ref{eq:virial-limit-mR}) 
can be determined analytically if
needed~\cite{Neubelt:2019sou}). 

From Eq.~(\ref{eq:virial-limit-mR}) we see that in the 
heavy quark limit L1, $m\to\infty$ with $B$ fixed,
the bag radius decreases like $R \propto m^{-1/5}$,
while the nucleon mass in Eq.~(\ref{virial-massive-M})
approaches the limit $M_N\to N_cm\to\infty$, 
cf.\ the  ``response column'' in Table~\ref{Tab1}. Notice 
that in this limit the inertia of the quarks increases, and 
the dynamics of the system  becomes non-relativistic.
We will comment more on this limit in 
Sec.~\ref{Sec:heavy-mass-limit-B-fix}.

In the large system size limit L2, 
$R\to\infty$ with $m$ fixed, we read off from 
Eq.~(\ref{eq:virial-limit-mR}) that $B$ decreases like
$R^{-5}$. The bag contribution to the 
nucleon mass $\frac43\,\pi\,R^3B\sim R^{-2}$ decreases 
in the large-$R$ limit. The nucleon mass becomes smaller
and approaches $M_N\to N_c m$ similarly to the limit L1,
albeit now $m$ is fixed and (if we choose to work 
with light quarks) $M_N$ can be small. 
Interestingly, even though $m>0$ can be chosen to be 
small, one deals with a non-relativistic dynamics also
in this case. This can be understood by considering that 
as the system size increases, the uncertainty on the 
quark positions $\Delta x\sim R$ grows while the momenta
$\sim 1/R$ decrease according to Heisenberg's uncertainty
principle. We will discuss further features of this limit
in Sec.~\ref{Sec:large-R-limit-m-fix}.

In the constituent quark mass limit L3, we will
keep the nucleon mass fixed (at its physical value)
and make $m$ approach one third of the nucleon mass.
Hence, in this limit the system has the mass of the 
physical nucleon, but its mass is asymptotically given 
by the masses of the ``constituent quarks'' added up.
This in turn means that the system size must grow
$R\to\infty$ which must be accompanied by a 
decreasing strength of the interaction with
$B\sim R^{-5}$ per Eq.~(\ref{eq:virial-limit-mR}).
We will come back to this limit
in Sec.~\ref{Sec:constituent-limit}.

\begin{table}[t!]
\begin{tabular}{clrllll}
\hline
\hline
    \hspace{5mm} Acronym \hspace{5mm}
    & Limit, varied parameter
    & \hspace{5mm} Fixed quantity 
    & \hspace{6mm} 
    & \multicolumn{2}{l}{
    Response of other quantities \spacer}\\
\hline
L1  & heavy quark limit, {\boldmath $m\to \infty$ }
    & $B=\rm fixed$ \hspace{2mm}
    & & $R\sim m^{-1/5}$  \hspace{3mm}
    & $M_N \to N_c m \to\infty$  
    \spacer \\
L2  & large system size limit, {\boldmath $R\to \infty$} 
    & $m \; = \rm fixed$ \hspace{2mm}       
    & & $B \sim R^{-5}$  \hspace{3mm}
    & $M_N \to N_cm=\rm fixed$ \\
L3  & constituent quark limit, {\boldmath $m\to M_N/N_c$ } 
    \hspace{5mm}
    & $M_N = \rm  fixed$ \hspace{2mm}
    & & $R\to\infty$ \hspace{2mm}
    & $B\sim R^{-5}$
    \spacer \\
\hline
\hline
\end{tabular}
\caption{\label{Tab1} 
Limits within the bag model
considered in this work.
L1: heavy quark limit with 
    the bag constant $B$ kept fixed.
L2: large system size limit with 
    quark mass $m$ kept fixed.
L3: constituent quark limit with the
    nucleon mass $M_N$ kept fixed. 
The varied parameters are stressed in bold
in column 2. The behavior of unconstrained
quantities is shown in columns 4 and~5.}
\end{table}

In the limit L1 the strength of the bag interactions 
remains constant. The limits L2 and L3 have in common that 
in both cases the strength of the interactions decreases,
which makes the system size large. The general
connection between system size and strength of interaction 
is nicely illustrated in Bohr's semi-classical 
H-atom model, where the electron moves with ``velocity'' 
$v_n=\alpha\,c/n$ in the $n^{\rm th}$ ``orbit''
with the ``radius'' $r_n=\lambda_e\,n^2/\alpha$, 
where $\lambda_e =\hbar/(m_ec)$ denotes the electron
Compton wavelength and $m_e$ the (reduced) mass. 
Thus, atoms have large sizes of ${\cal O}(1\,\mbox{\AA})$ 
and can be described to a good approximation in terms 
of a non-relativistic Schr\"odinger equation, 
because the electromagnetic coupling constant 
$\alpha\simeq 1/137$ is small. 

In the bag model, the strength of the interaction 
is encoded in the bag constant $B$. This can be
intuitively understood in various ways. 
For instance, taking $B\to 0$ at the Lagrangian 
level in Eq.~(\ref{Eq:Lagrangian-bag}) 
one recovers the free Dirac theory.
Another way to convince oneself that $B$ is
responsible for producing a finite-size bound
state is to notice that setting $B\to0$ in 
Eq.~(\ref{virial-massive-M}) yields $M_N(R)\propto R^{-1}$ (using massless quarks for sake of simplicity in this
argument), and the nucleon mass as function of $R$ assumes
its minimum at $R\to\infty$ which means that the quarks 
are unbound. Yet another way to see that no bound state
exists when $B$ is absent is provided by the 
von Laue condition (\ref{Eq:von-Laue-3}): when 
$B=0$ the 3D pressure has no node, and one finds
$\int_0^\infty dr\,r^2 p(r)>0$ meaning that 
the nucleon explodes~\cite{Neubelt:2019sou}.
This corresponds to the situation in the Bogoliubov 
model~\cite{Bogo:1967} which can be viewed historically 
as a predecessor of the bag model~\cite{Thomas:2001kw}.

These three limits represent very different physical 
situations, but as already mentioned they have in common that the product 
$mR\to\infty$, even though $m$ and $R$ behave differently 
in each case. As a consequence the EMT distributions have
common leading expressions in these three limits which can 
be expressed as~\cite{Neubelt:2019sou}
\sub{\label{Eqs:limits-3D}
\ba
      \epsilon(r) &=& N_c\,m\;c_0\,j_0(\kappa r)^2\,\Theta_V 
      + \dots \;,\label{Eq:3D-energy-mR} \phantom{\frac11}\\
      J_{{\rm Bel}}^z(\vec{r}) &=& \frac12\;{c_0}\;\kappa r\;j_0(\kappa r)\;j_1(\kappa r)\;(1-\cos^2\theta)\,\Theta_V + \dots \;, \label{Eq:3D-JzBEL-mR}\\
      S^z(\vec{r}) &=& \frac12\;{c_0}\,j_0(\kappa r)^2 \,\Theta_V + \dots \;, \label{spin-mR}\\
      L^z  (\vec{r}) &=& \frac{\pi^2}{4\,(mR)^2}\,c_0\,j_1(\kappa r)^2\;\bigl( 1-\cos^2\theta\bigr)\,\Theta_V + \dots \;,
      \label{Eq:3D-angularmomentum-mR} \\
      s(r) &=& \frac{N_c\,\pi}{2\,mR}\;c_0\biggl(
      -j_0^\prime(\kappa r)j_1(\kappa r)
      -\frac1r\,j_0(\kappa r)j_1(\kappa r)
      +j_0(\kappa r)j_1^\prime(\kappa r)\biggr)\Theta_V 
      + \dots \;,\label{Eq:3D-shear-mR} \\
p(r) &=& \frac{N_c\,\pi}{6\,mR} \;c_0\biggl(
      -j_0^\prime(\kappa r)j_1(\kappa r)
      +\frac2r\,j_0(\kappa r)j_1(\kappa r)
      +j_0(\kappa r)j_1^\prime(\kappa r)\biggr)\Theta_V
      - 
      B\,\Theta_V+ \ldots 
      \;,\label{Eq:3D-pressure-mR}
\ea}
where $c_0 = \pi/(2R^3)$, $\kappa = \pi/R$, and the 
normalization is such that $\int d^3r\,c_0\,j_0(\kappa r)^2\,\Theta_V = 1$. 
The dots indicate in each case subleading terms that are suppressed by 
$1/mR$ with respect to the corresponding leading contributions. 
The leading expression for the energy distribution in Eq.~\eqref{Eq:3D-energy-mR}
satisfies $\int d^3r\,\epsilon(r)=N_c m$ which is the mass of the nucleon in each of the three limits.
The leading expression for the Belinfante AM in Eq.~\eqref{Eq:3D-JzBEL-mR} satisfies 
$\int d^3r\,J_{{\rm Bel}}^z(\vec{r})=\frac12$.
In the limit of $mR\to\infty$, the leading term of the total kinetic AM $\int d^3 r\left[L^z(\vec{r})+S^z(\vec r)\right]=\frac12$ is dominated by the spin contribution in Eq.~\eqref{spin-mR}  
with the OAM being suppressed by two orders of
the small parameter $1/(mR)$. 
The kinetic AM $J^z$ and intrinsic spin distribution
$S^z$ become equal and isotropic. In contradistinction to that,
the Belinfante AM retains its monopole and quadrupole
decompositions for $mR\to\infty$.

For the following discussions it is of importance to 
note that 
in the expression for the 3D pressure the bag constant enters as 
$p(r)=\dots -B\,\Theta_V$, see 
Eqs.~(\ref{Eq:3D-pressure},~\ref{Eq:3D-pressure-mR}).
The practical implication of this is that $p(r)$ has the same behavior 
as $B$ in the limits in Table~\ref{Tab1}. Being tightly connected to the 
pressure by the Eqs.~(\ref{Eq:p(r)+s(r)},~\ref{Eq:von-Laue}), $s(r)$ must 
also scale like $B$ in the different limits.

For completeness, let us remark that one could
formulate further limits in the bag model. 
For instance, in Ref.~\cite{Neubelt:2019sou} the limit
$m\to\infty$ with $R$ fixed was considered, which is 
different from the L1 limit discussed here. 
(However, the limits L2 and L3 were defined in
\cite{Neubelt:2019sou} exactly as in this work,
and used to study 3D EMT distributions and the
$D$-term.)

After discussing the physical situation in the next section, we shall 
investigate the behavior of 2D EMT distributions in the limits introduced here.

\section{2D EMT distributions in the bag model in the physical situation}
\label{Sec:EMT-physical-situation}

\begin{figure}[t]
\begin{centering}
\includegraphics[height=4.0cm]{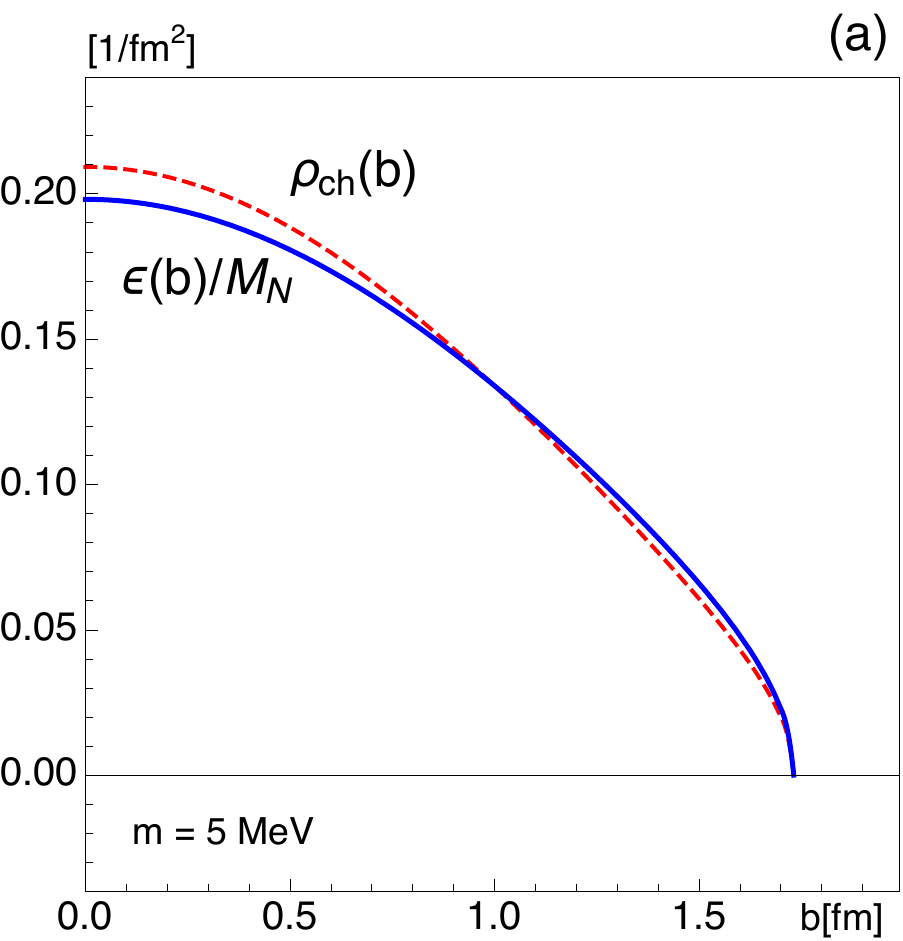} \hspace{4mm} \
\includegraphics[height=4.0cm]{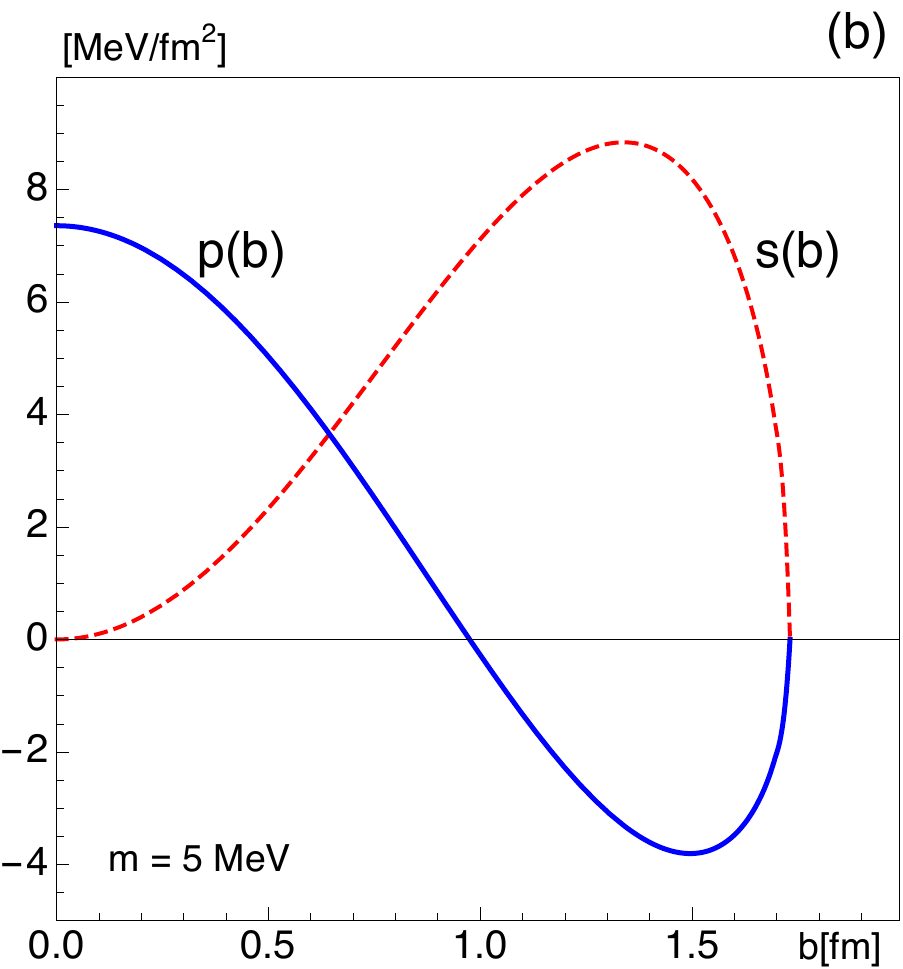} \hspace{4mm} \
\includegraphics[height=4.0cm]{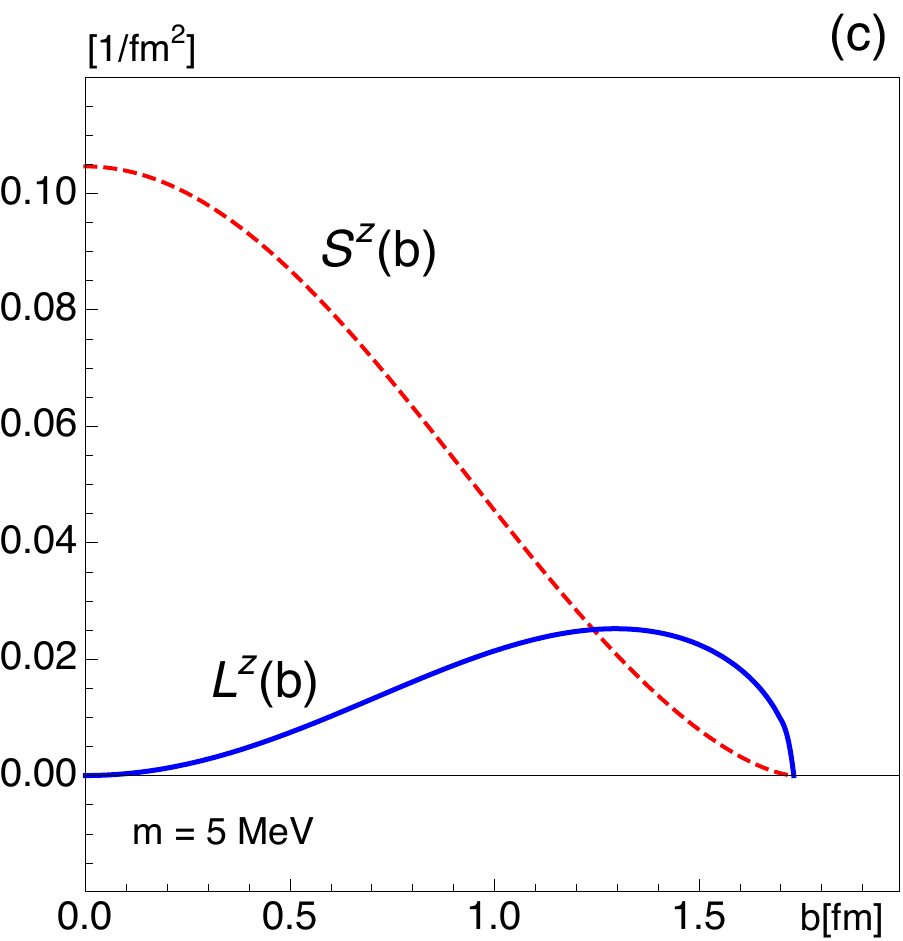} \hspace{4mm} \
\includegraphics[height=4.0cm]{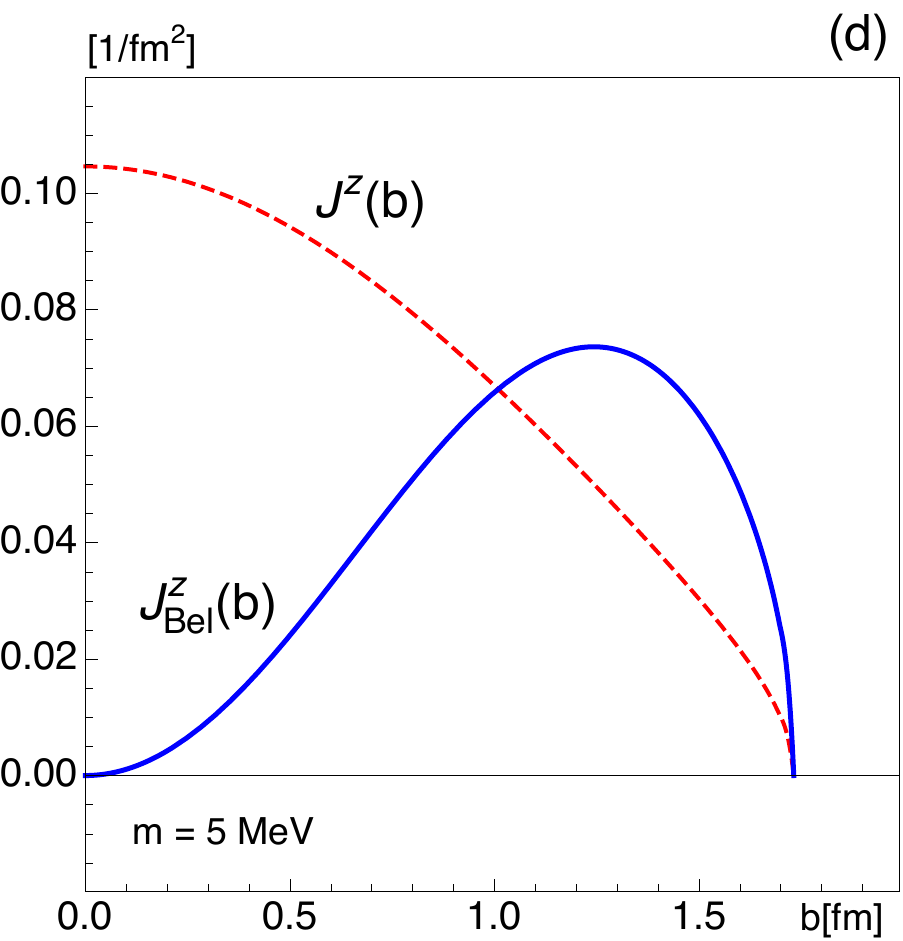} \hspace{-1mm} \
\par\end{centering}
\caption{\label{Fig-1:physical-situation} 
2D distributions in the bag model in the physical 
situation with $m=5\,\rm MeV$, $M_N=938\,{\rm MeV}$:
(a) electric charge distribution and energy distribution 
normalized to unity, 
(b) pressure and shear forces,
(c) spin and kinetic OAM distributions,
(d) total kinetic and Belinfante form of AM. 
All distributions vanish 
at the bag radius $R=1.7\,\rm fm$. }
\end{figure}

In the physical situation the proton is made of light quarks. 
For definiteness we choose $m = 5\,\rm MeV$ and neglect 
isospin breaking effects. The physical nucleon mass is 
reproduced for the bag radius $R=1.7\,\rm fm$.
The Fig.~\ref{Fig-1:physical-situation} shows the results for 
the 2D distribution of energy, pressure, shear force, kinetic and Belinfante form 
of AM.
The 2D energy distribution has the physical dimension of energy per 
unity area, the 2D pressure and shear force have the dimensions of 
force per unit length, and all three distributions can be
expressed in units of $\rm MeV/fm^2$.
The AM distributions have the physical
dimension (area)$^{-1}$ and can be
expressed in units of $\rm 1/fm^2$
(we use $\hbar=c=1$).

In the bag model all (2D or 3D) spatial distributions are
non-zero only inside the bag, which is expected in this model. 
A first and generic observation regarding the 2D distributions 
is that they go to zero at the bag radius $R$. This is in
contradistinction to 3D distributions which in general do not vanish 
at the bag boundary. In fact, there is no reason why 3D spatial
distributions should drop to zero at ``the edge of a system.'' 
The bag model 3D distributions exhibit characteristic
discontinuities due to the $\Theta(R-r)$ functions in 
(\ref{Eqs:3D-review}) at $r=R$. Such discontinuities may
seem ``unphysical'' at first glance, but this is a 
consistent description of 3D spatial distributions in this model
\cite{Neubelt:2019sou}. 

One notable exception is the normal 
force where $\frac23s(r)+p(r)>0$ must hold for all values of $r$
within a system, and the point where the normal force becomes 
zero defines the ``edge of the system.'' This necessary
condition for mechanical stability~\cite{Perevalova:2016dln} 
is the only physical constraint for 3D EMT distributions
for $r\to R$ we are aware of, and the bag model complies
with it \cite{Neubelt:2019sou}.
In other cases the 3D EMT distributions are not constrained to 
vanish at $r=R$ and do not do so. 
This is different in the case of 2D distributions. From their
relations to 3D distributions (\ref{Eqs:2D-3D-coonetction}) it
follows that 2D distributions must vanish when
$b\to R$ as we will see in the following.

The energy distribution $\epsilon(b)$ is largest in the center 
($b=0$) and decreases monotonously until it becomes zero 
at $b=R$, see Fig.~\ref{Fig-1:physical-situation}a. At small
$b$ we find the behavior
$\epsilon(b) = \epsilon(0)- a_\epsilon b^2 + {\cal O}(b^4)$. 
The coefficients $a_i$ (here $i=\epsilon$) are defined as positive 
quantities here and in the following. The short distance physics 
is, however, beyond what non-perturbative approaches 
like the bag model can meaningfully describe. As $b\to R$ 
the behavior of 2D distributions is determined by the integral
relations (\ref{Eqs:2D-3D-coonetction}). 
For instance, if we denote by $T^{00}(R)\neq 0$ the value of the 
3D energy distribution at $r=R$, then the behavior of the 2D EMT
distribution is given by 
$\epsilon(b) = T^{00}(R)\sqrt{R^2-b^2}$ modulo 
subleading terms when approaching the bag boundary 
from the inside. In particular the slope of $\epsilon(b)$ diverges for
$b\to R$. 

It is instructive to compare the energy distribution 
to the electric charge distribution 
of the proton whose expression is derived in 
App.~\ref{Appendix:charge-distribution}. 
For that we plot in Fig.~\ref{Fig-1:physical-situation}a the 
energy distribution $\epsilon(b)$ normalized with 
respect to the nucleon mass, such that the integrals 
$\int d^2b\,\dots$ yield unity in both cases. The bag 
model predicts that the 2D distributions of electric
charge and energy in the nucleon are similar. 
It will be interesting to test this prediction in 
other models and lattice QCD. 

The pressure and shear force are shown in
Fig.~\ref{Fig-1:physical-situation}b. They behave like 
$p(b) = p(0)-a_p b^2 + {\cal O}(b^4)$ and 
$s(b) = a_s b^2 + {\cal O}(b^4)$ close to the
center. The behavior when $b$ approaches the bag boundary is
analogous to that of the energy distribution discussed above.
The shear force is positive for $0< b <R$. The pressure is 
positive in the inner region and is negative in the outer region
with a node at $b = 1.1\,\rm fm$. 
The 2D pressure obeys the von Laue condition
(\ref{Eq:von-Laue-2D}), and the 2D shear forces and pressure 
satisfy the differential relation  (\ref{Eq:diff-eq-p-s-2D}). 

In Fig.~\ref{Fig-1:physical-situation}c the spin $S^z(b)$ 
and kinetic OAM $L^z(b)$ distributions are shown. 
The former is larger and finite at $b=0$, while the latter 
is smaller and vanishes for $b\to0$. This is 
to be expected from the 
factor of $r^j$ appearing in the definition of OAM 
distribution~\eqref{OAMdef}.
The magnitudes of these distributions reflect the fact that 
$65.8\,\%$ of nucleon AM is due to quark spin, and $34.2\,\%$ due to 
OAM.
These are typical values in relativistic quark models. 
The spin distribution does not exhibit the characteristic vertical 
slope as $b\to R$ like the other distributions in Fig.~\ref{Fig-1:physical-situation}, 
because the corresponding 3D distribution $S^z_q(\vec{r})$ vanishes 
for $|\vec{r}|=R$ (for any value of the quark mass $m$).

The total (kinetic) AM distribution 
$J^z(b)=L^z(b)+S^z(b)$ is depicted in
Fig.~\ref{Fig-1:physical-situation}d. For comparison the
Belinfante AM distribution $J^z_{\rm Bel}(b)$
is shown. Both distributions have the same normalization  
$\int d^2b\,J^z(b) = 
 \int d^2b\,J^z_{\rm Bel}(b) = \frac12$
but have much different shapes,
see the discussion in App.~\ref{App:Jkin-vs-JBel}.
This has been observed also in other 
models~\cite{Adhikari:2016dir,Lorce:2017wkb}. 
The key difference is that the Belinfante OAM distribution has 
by definition a pure orbital form~\eqref{OAMdef}, whereas the kinetic AM distribution receives both spin and orbital contributions.

\section{2D EMT distributions in the heavy quark limit}
\label{Sec:heavy-mass-limit-B-fix}

In this section, we discuss 2D EMT distributions in the 
limit L1 in which $m\to \infty$ with the bag constant $B$ fixed. 
From Eq.~(\ref{eq:virial-limit-mR}) we conclude
that the bag radius decreases as 
$R \propto m^{-1/5}$ for $m\to \infty$, cf.\ Table~\ref{Tab1}.
Consequently, the
size of heavy hadrons decreases\footnote{Notice 
that the proton size can be characterized e.g.\ in terms of 
the mean square charge radius and does not coincide with the 
bag radius. But the latter effectively sets the length scale 
in the bag model. Thus, if $R$ decreases as $m\to\infty$, 
so does the hadron size.}
with increasing $m$.
This feature is intuitively expected, although in QCD the 
hadron size goes like $1/m$ in the heavy quark limit.
It is important to keep in mind that here we deal with a
simplistic implementation of a heavy quark limit within
a quark model. 

The masses of the hadrons, however, scale correctly in 
this limit: the nucleon mass is given by $M_N = N_c m$ 
up to subleading terms suppressed by powers of $1/mR$
\cite{Neubelt:2019sou}.
(This general result holds also for mesons where the number 
of colors $N_c$ is replaced by the number of constituents 
$N_{\rm const}=2$.) In principle, one could implement a 
``more correct'' heavy quark limit, where hadron masses grow
linearly with $m$ and hadron radii decrease as $1/m$, 
by keeping $BR^4$ fixed which implies via
Eq.~(\ref{eq:virial-limit-mR}) 
that the system size would decrease like $1/m$.
While this might be an interesting exercise 
in itself, it is not obvious whether such an approach 
would yield a more realistic heavy quark limit in 
the bag model.
We therefore content ourselves with the $m\to\infty$ limit
with $B=\rm fixed$. This is sufficient for our purposes to   
study the behavior of the EMT properties in a system where 
the constituents become massive.

Dimensional analysis tells us that $\epsilon(r)\sim M_N/R^3$, $J^z(\vec r)\sim R^{-3}$.
As shown in Sec.~\ref{sec:limits-overview},
the 3D distributions $p(r)$ and $s(r)$ have the same behavior 
as the bag constant $B$ which is kept fixed in the limit L1. 
It then follows that the 3D distributions scale like 
$\epsilon(r) \sim m^{8/5}$, 
$J^{z}(\vec{r}) \sim m^{3/5}$,
$s(r)$ and $p(r) \sim m^0$
when $m\to\infty$.
This is consistent with Eq.~(\ref{eq:virial-limit-mR}) 
and the scaling relations (\ref{Eqs:limits-3D}). 
Hence, the 3D energy and AM distributions 
increase, while the mechanical 3D forces do not scale
when $m\to\infty$.
A similar analysis can be applied to 
2D distributions. As one spatial dimension is integrated 
out, the large-$m$ scaling of 2D distributions
differs from that of the respective 3D distributions 
by one power of $R\propto m^{-1/5}$. In particular, 
one obtains 
$\epsilon(b) \sim m^{7/5}$,
$J^z(b) \sim m^{2/5}$,
$s(b)$ and $p(b) \sim m^{-1/5}$. 
We see that as $m\to\infty$, the 2D energy and 
AM distributions increase, but 
the mechanical 2D forces inside the nucleon decrease. 
It should be stressed that these are 
``geometric effects'' due to looking at EMT properties 
through ``3D-glasses'' or ``2D-glasses.''

\begin{figure}[t!]
\begin{centering}
\includegraphics[height=4cm]{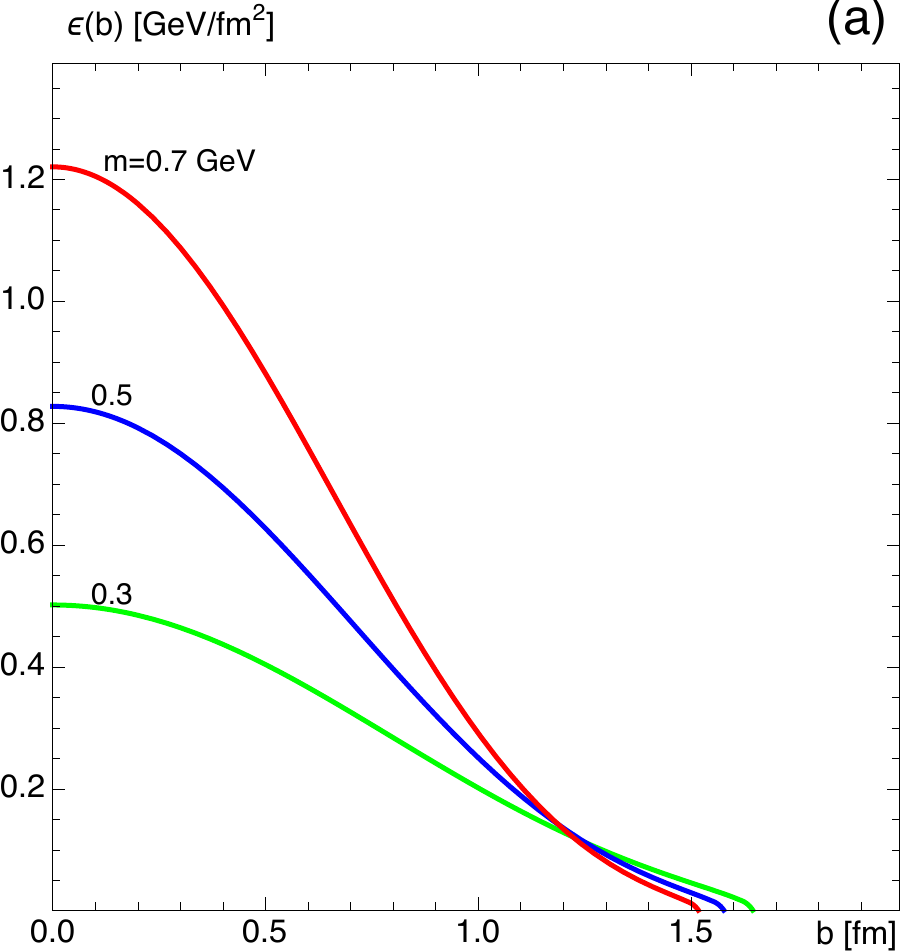} \hspace{4mm} \
\includegraphics[height=4cm]{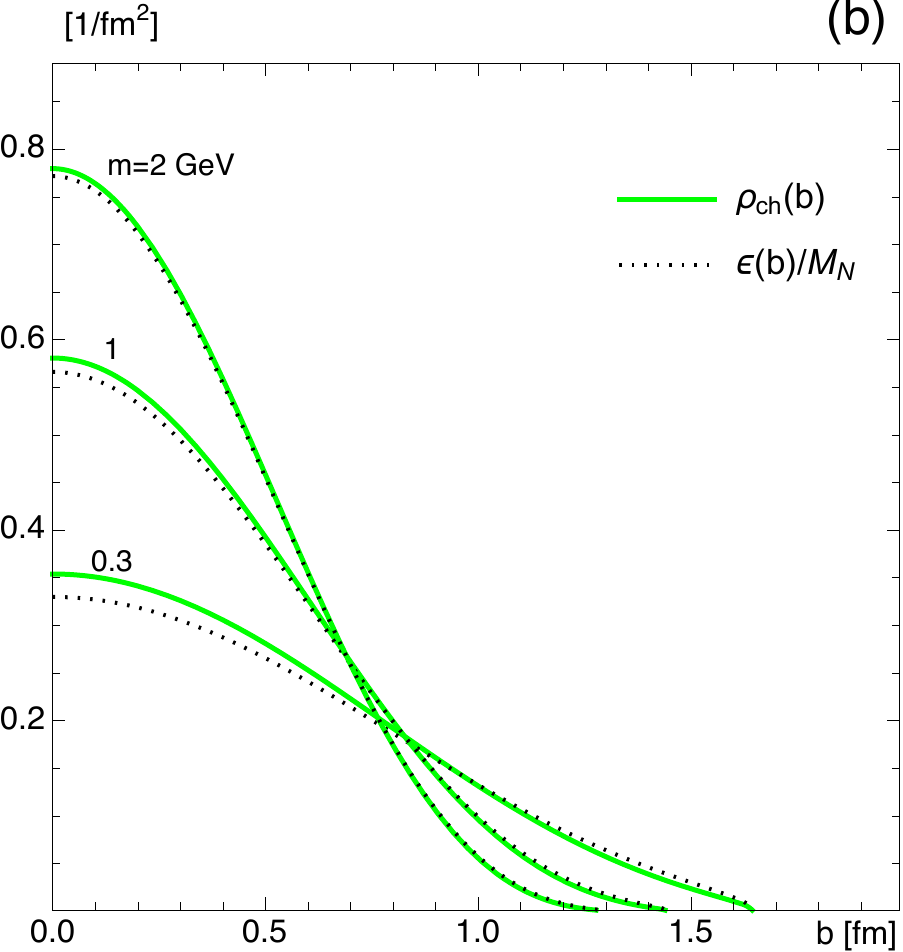} \hspace{4mm} \
\includegraphics[height=4cm]{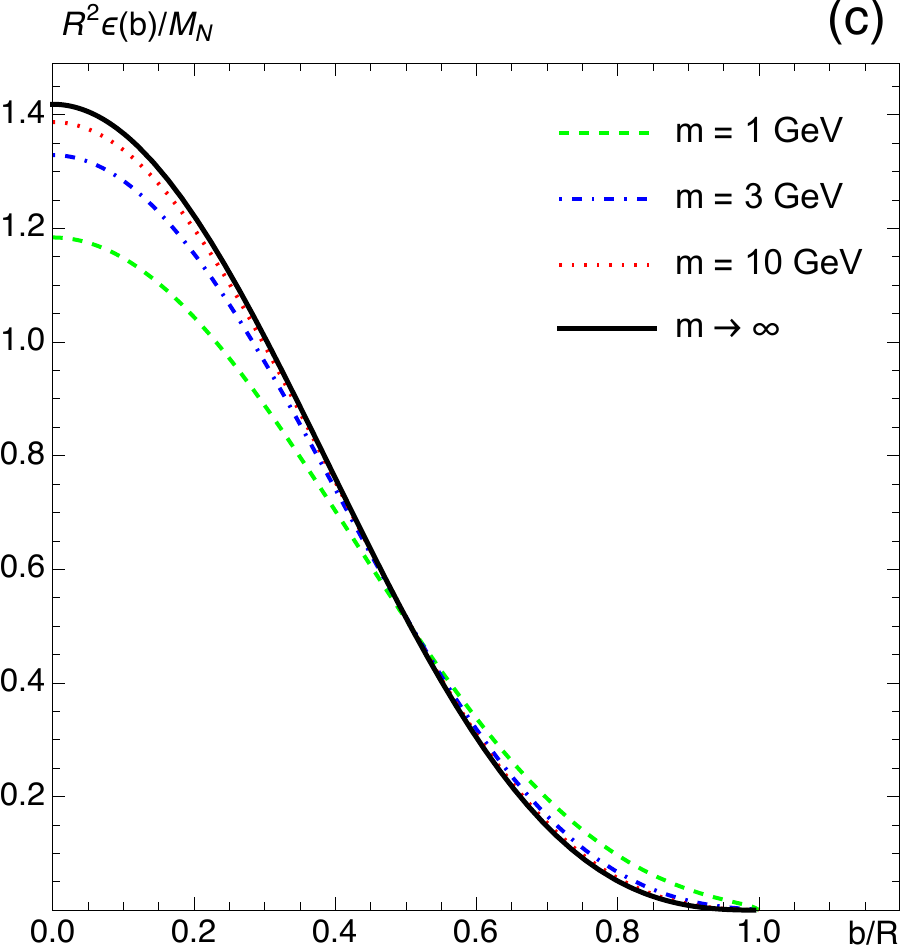} \hspace{4mm} \
\par\end{centering}
\caption{\label{Fig-2:energy-heavy-quark-T00} 
The 2D energy distribution in the bag model for 
fixed $B$ and increasing $m$.
(a) Energy distribution $\epsilon(b)$. 
(b) Normalized energy distribution $\epsilon(b)/M_N$
in comparison to the 2D electric charge distribution $\rho_{\rm ch}(b)$. 
(c) The scaling of $R^2\epsilon(b)/M_N$ for $m\to\infty$.}

\

\begin{centering}
\includegraphics[height=4cm]{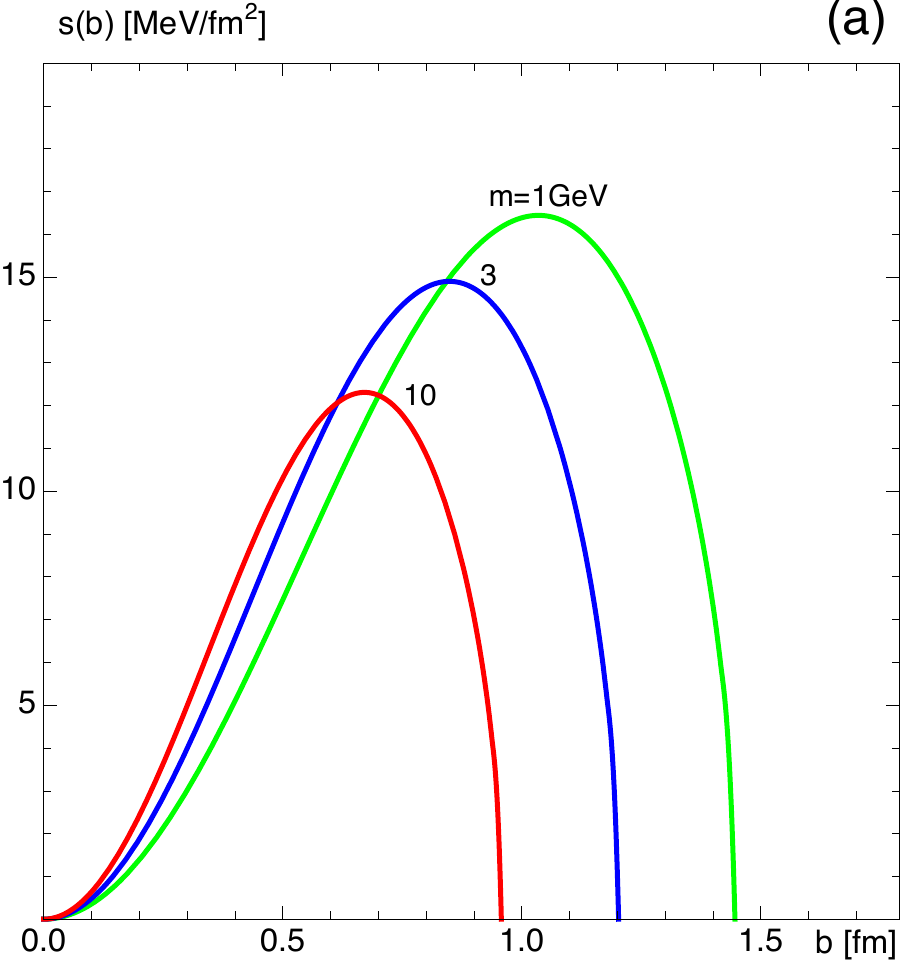} \hspace{4mm} \
\includegraphics[height=4cm]{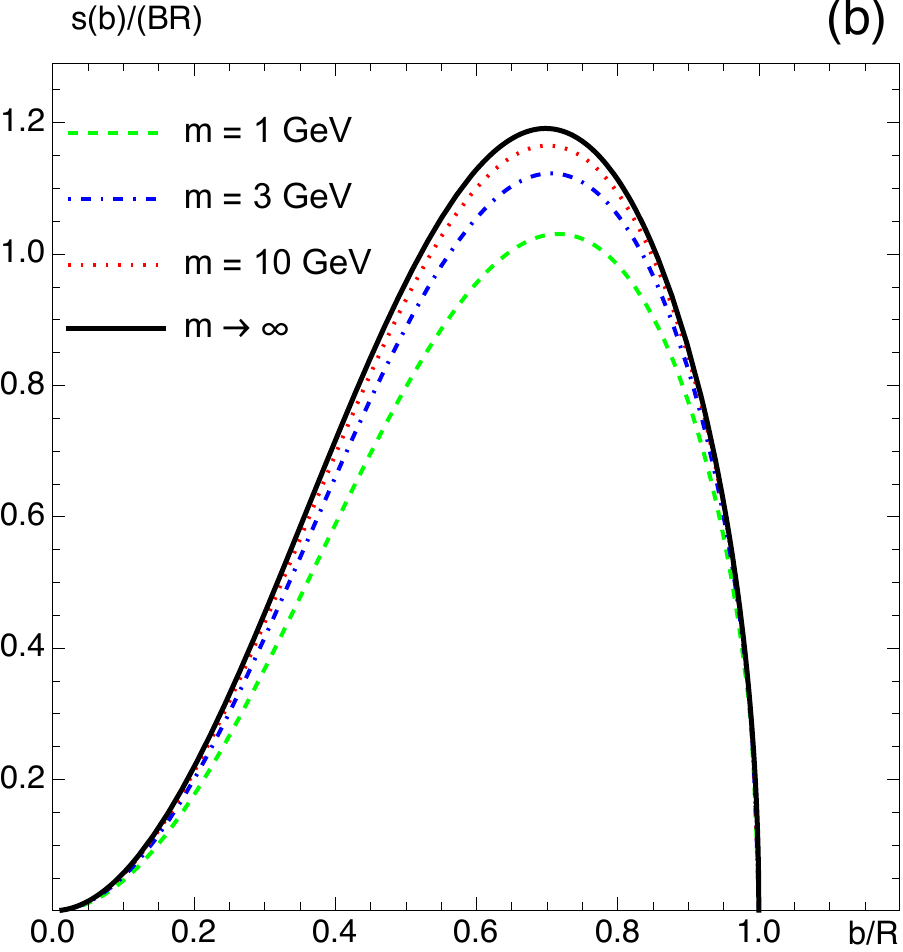} \hspace{4mm} \ 
\includegraphics[height=4cm]{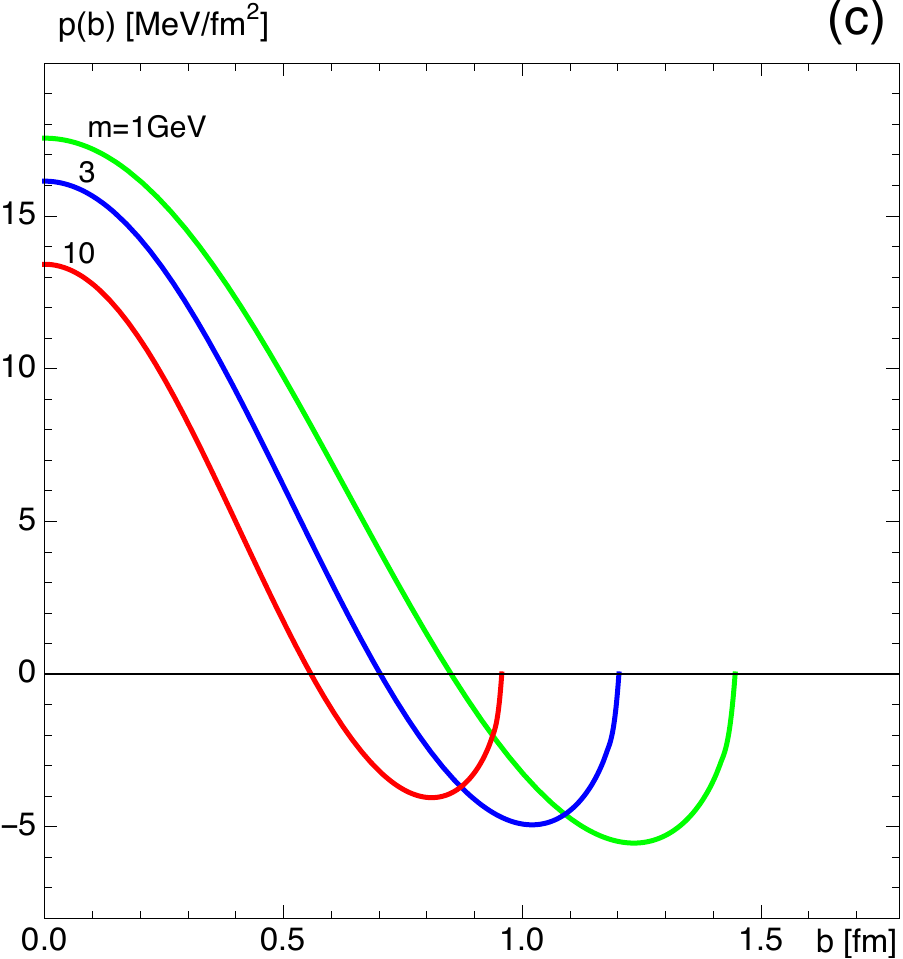} \hspace{4mm} \
\includegraphics[height=4cm]{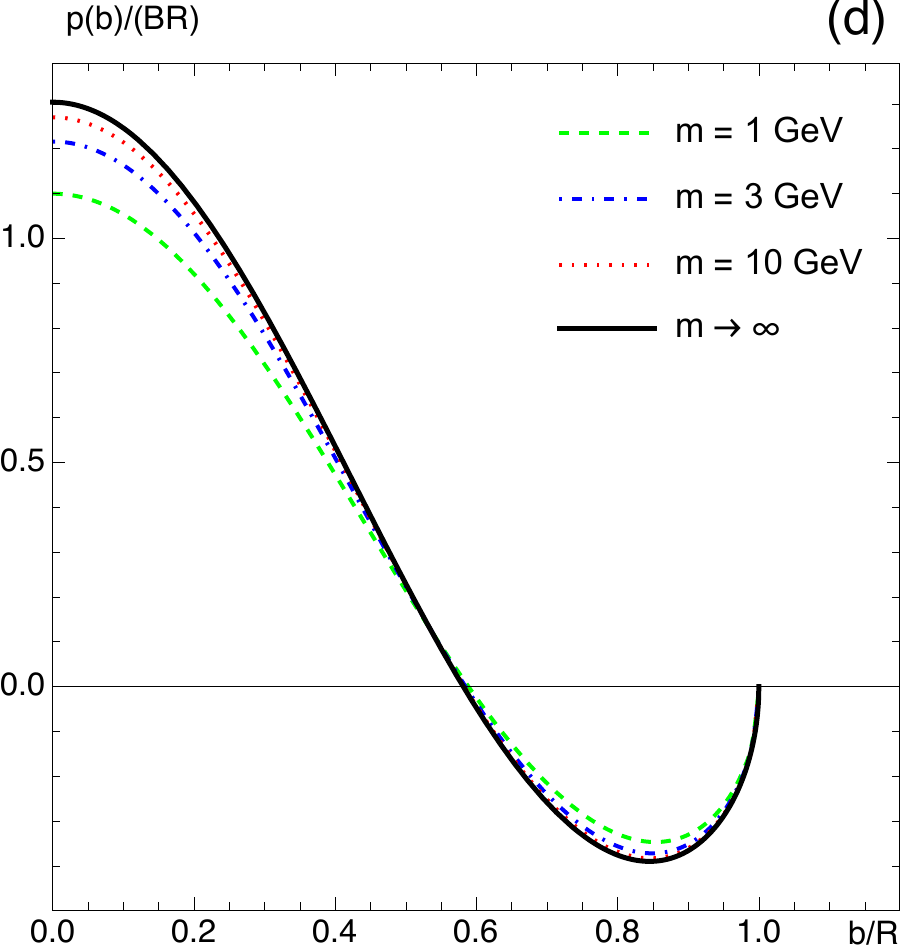} \hspace{4mm} \

\end{centering}
\caption{\label{Fig-3:heavy-quark-s-p} 
The 2D EMT shear force distribution $s(b)$ in the bag model for fixed $B$ and 
(a) selected increasing values of $m\ge 1\,{\rm GeV}$. 
(b) The rescaled dimensionless distribution $s(b)/(BR)$.
Similarly, the 2D pressure $p(b)$ for
(c) selected $m$, and  (d) the 
rescaled 2D pressure distribution $p(b)/(BR)$.}
\end{figure}

Having studied the 2D energy distribution in the 
physical situation for light quarks of $m=5\,{\rm MeV}$ 
in Sec.~\ref{Sec:EMT-physical-situation}, we now show 
$\epsilon(b)$ in Fig.~\ref{Fig-2:energy-heavy-quark-T00}a 
for selected heavier quark masses $m=0.3$, $0.5$, $0.7\,$GeV. 
While far from a heavy quark limit, these values clearly
show the trend: the energy distribution inside the nucleon
grows rapidly with increasing $m$ as one would intuitively
expect, because the mass of the nucleon grows while the
available ``2D-volume'' shrinks.

In Fig.~\ref{Fig-2:energy-heavy-quark-T00}b we compare the
rescaled energy distribution $\epsilon(b)/M_N$ to the 2D
electric charge distribution $\rho_{\rm ch}(b)$.
As Fig.~\ref{Fig-2:energy-heavy-quark-T00}b shows, 
$\epsilon(b)/M_N$ and $\rho_{\rm ch}(b)$ become more and 
more similar with increasing $m$: e.g., they become 
nearly indistinguishable for $m=2\,{\rm GeV}$ at the 
scale of Fig.~\ref{Fig-2:energy-heavy-quark-T00}b.
This is an interesting result. In general, viewing the 
nucleon structure through the distributions of electric 
charge or energy gives different pictures. But as the
constituents of the system become more massive, 
the difference between the two pictures becomes negligible. 
In the limit $m\to\infty$, the asymptotic expressions
for these two distributions become indeed equal. This can be seen 
by comparing the expression for $\epsilon(b)/M_N$ from
Eq.~\eqref{Eq:3D-energy-mR} and the expression for the 
electric charge distribution in Eq.~\eqref{Eq:3D-charge-mR}
of App.~\ref{Appendix:charge-distribution}.

The Figs.~\ref{Fig-2:energy-heavy-quark-T00}b also nicely
illustrates another intuitive feature. As the quark mass
increases, the 2D energy (and charge) distributions become
more strongly localized: for smaller $m$ the 2D energy and
charge distributions are small in the center and wide-spread
until the ``edge of the system'' (at $b=R$ where $R$ shrinks
as $m^{-1/5}$).
For larger $m$, the distributions grow in the center, and 
decrease in the region closer to the ``edge of the system.'' 
This result is intuitive because one naturally expects
fast-moving ultra-relativistic light quarks to have 
widely spread out distributions, while slowly-moving
non-relativistic heavy quarks are expected to have 
more localized distributions.

\begin{figure}[t!]
\begin{centering}
\includegraphics[height=4cm]{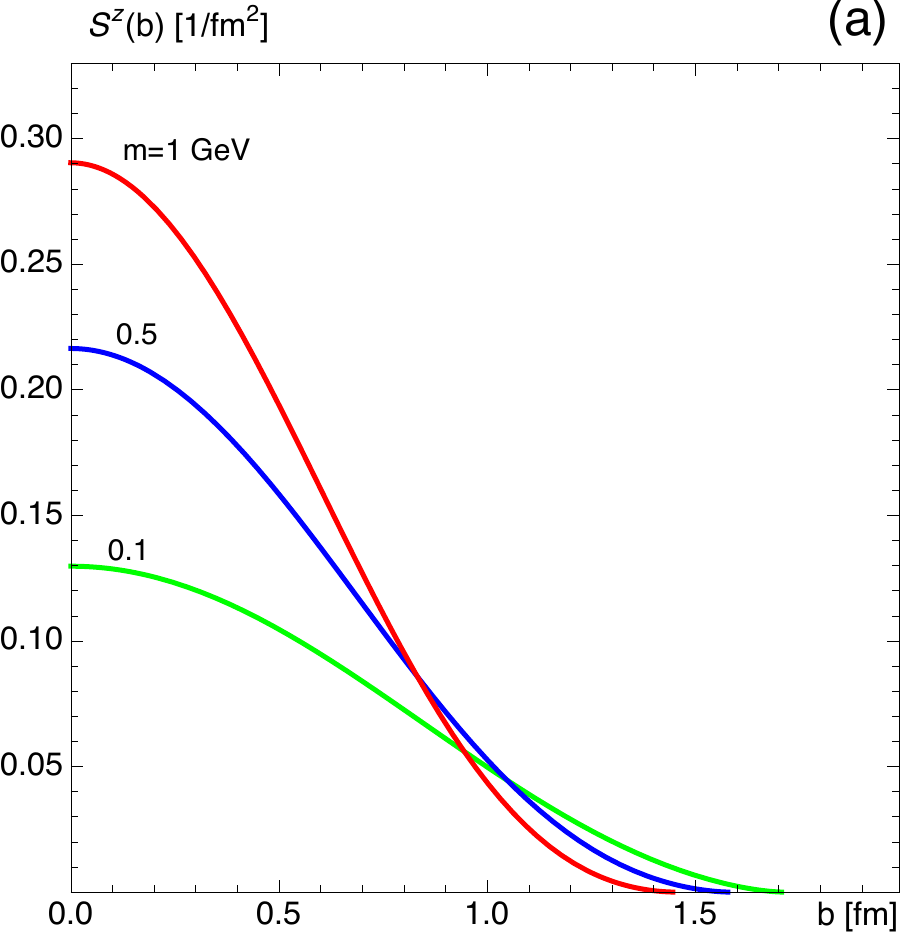} \hspace{4mm} \
\includegraphics[height=4cm]{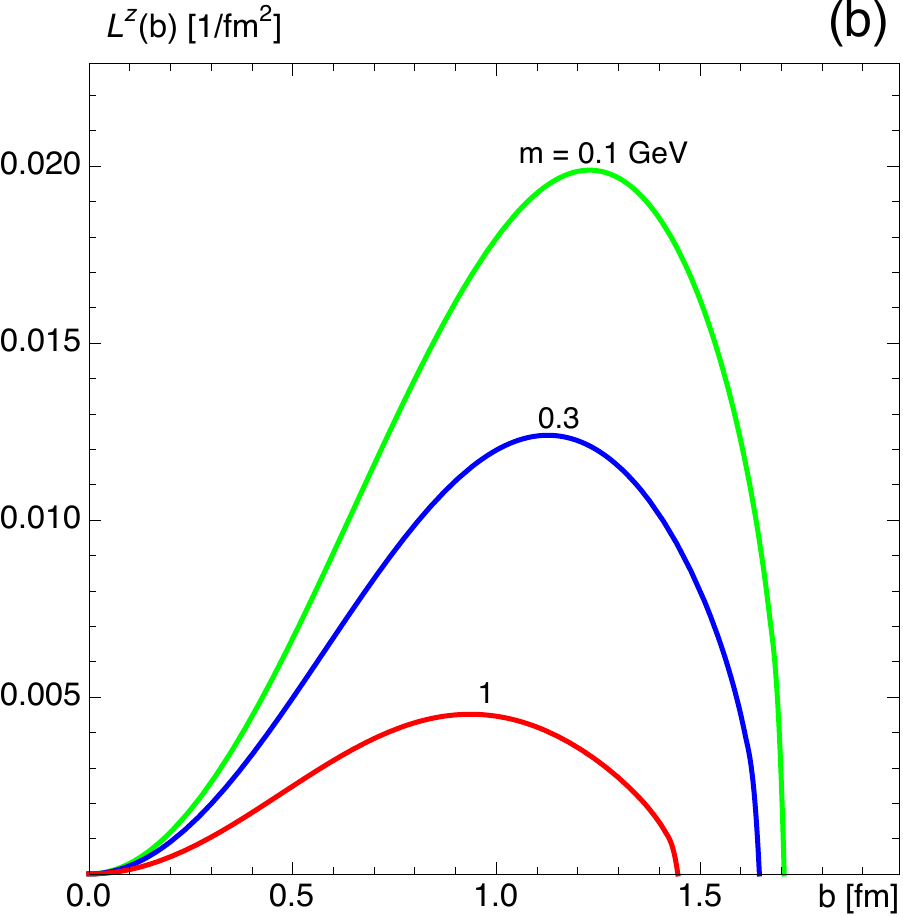} \hspace{4mm} \
\includegraphics[height=4cm]{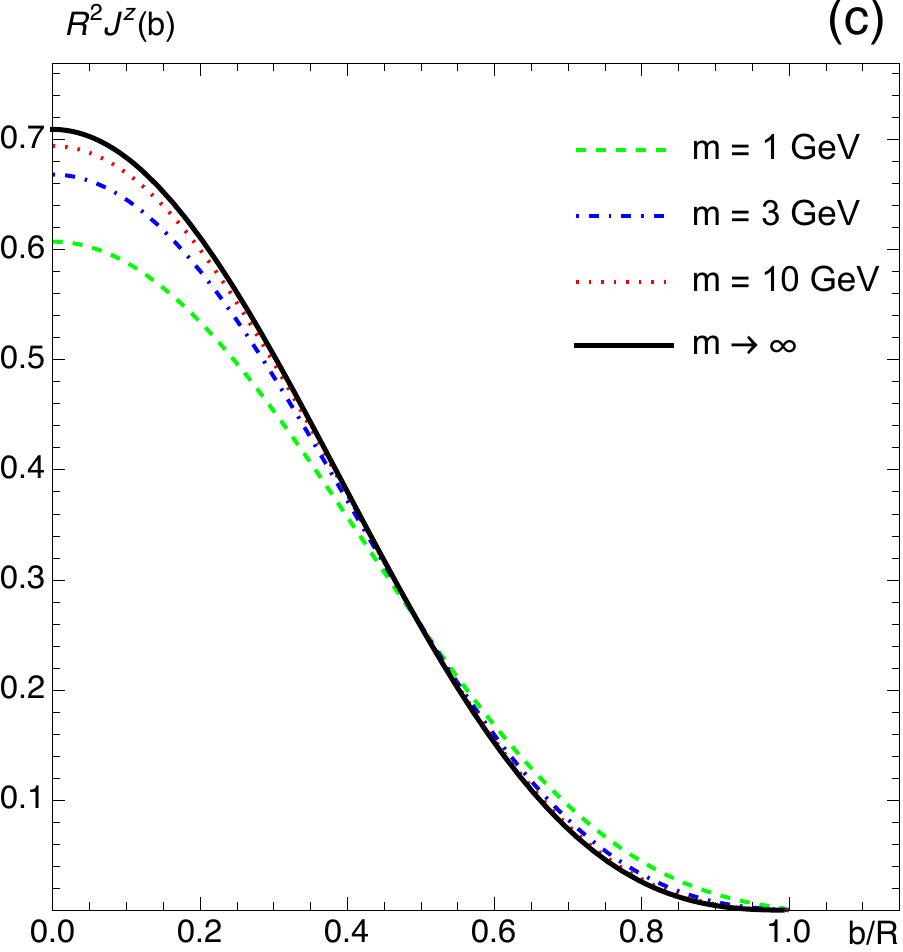} \hspace{4mm} \
\end{centering}
\caption{\label{Fig-4:AM-S-L-J-kinetic-heavy-quark} 
The 2D angular momentum distributions in the bag model for fixed $B$ and increasing $m$.
(a) Intrinsic spin $S^z(b)$ distributions. 
(b) Kinetic orbital angular momentum $L^z(b)$ distributions.
(c) The scaling of $R^2 J^z(b)$ for $m\to\infty$.}

\vspace{7mm}

\begin{centering}
\includegraphics[height=4cm]{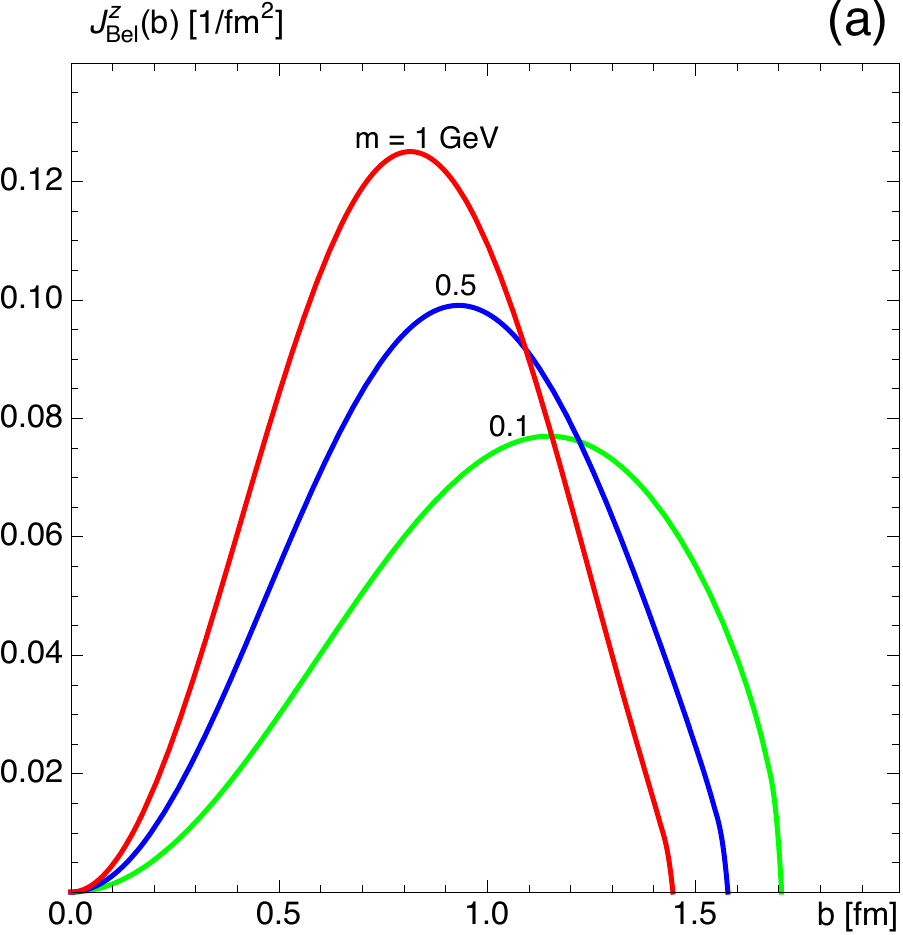} \hspace{4mm} \
\includegraphics[height=4cm]{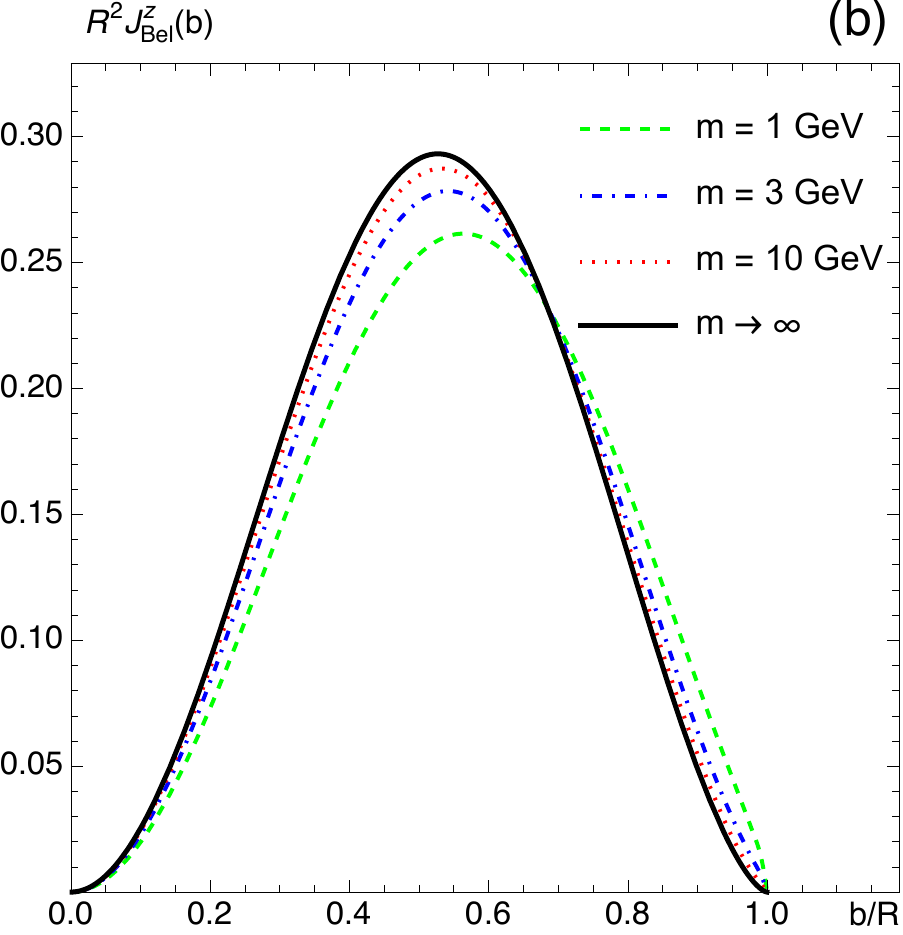} \hspace{4mm} \
\includegraphics[height=4cm]{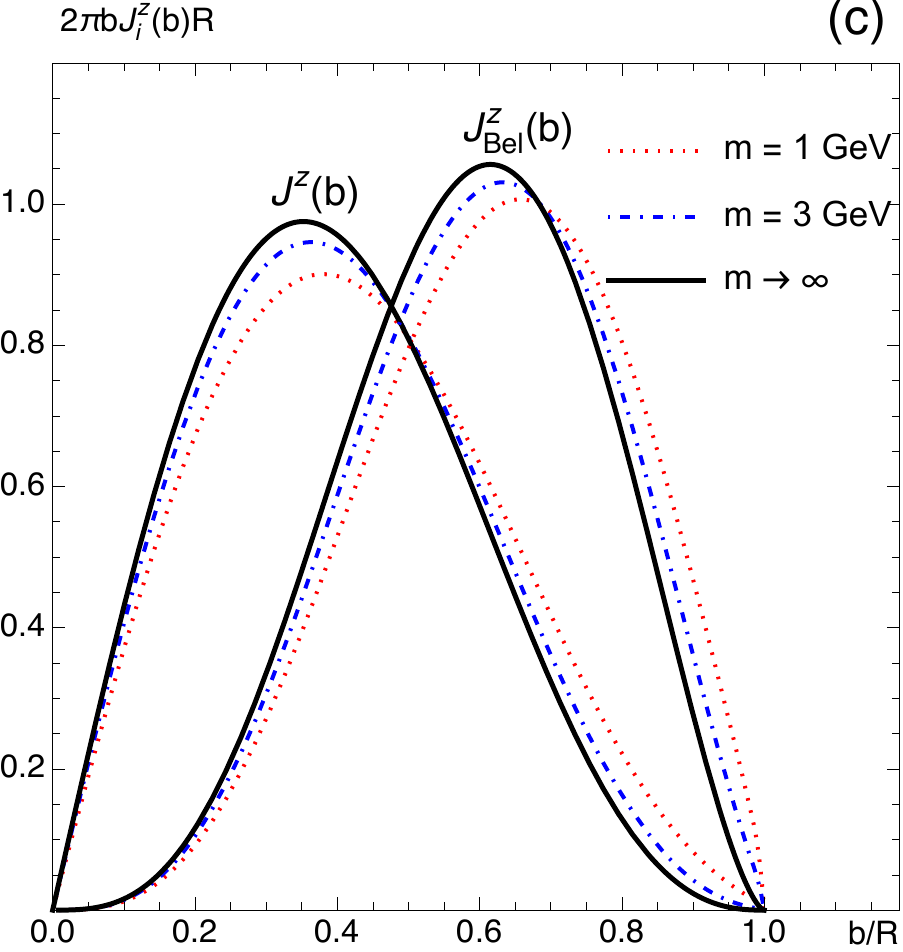} \hspace{4mm} \
\end{centering}
\caption{\label{Fig-5:AM-J-Belinfante-heavy-quark} 
The 2D angular momentum distribution in the bag model fixed $B$ and increasing $m$.
(a) Belinfante-form $J_\text{Bel}^z(b)$ angular momentum distributions. 
(b) The scaling of $R^2 J_\text{Bel}^z(b)$ for $m\to\infty$.
(c) The scaling of $2\pi b J_\text{i}^z(b) R$ for $m\to\infty$.}
\end{figure}

In the last plot related to $\epsilon(b)$ in 
Fig.~\ref{Fig-2:energy-heavy-quark-T00}c we show the
dimensionless rescaled distribution $R^2\epsilon(b)/M_N$ 
as function of $b/R$ for the values $m=1\,\rm GeV$,
$3\,\rm GeV$, $10\,\rm GeV$. This rescaled distribution 
has a well-defined finite limit $\lim_{m\to\infty}R^2\epsilon(b)/M_N$
which we include in the plot. Integrating this limiting curve 
over the rescaled 2D volume, $d^2b/R^2$, yields unity.
The  Fig.~\ref{Fig-2:energy-heavy-quark-T00}c shows
that the rescaled 2D energy distribution $R^2\epsilon(b)/M_N$
rapidly approaches its limiting shape. In fact, the curves
for $m=10\,\rm GeV$ and $m\to\infty$ agree within a few percent.
As the $m\to\infty$ limit is approached,  also
the rescaled distribution $R^2\epsilon(b)/M_N$ becomes more 
strongly localized towards the center.

Finally, we remark that the vertical slopes of the
2D distribution at $b=R$ observed for $m=5\,$MeV in 
Sec.~\ref{Sec:EMT-physical-situation} are in principle
present also for large $m$, but they become 
less and less pronounced. 

We discuss next the 2D force distributions in
Fig.~\ref{Fig-3:heavy-quark-s-p}. 
Initially, the 2D shear force distribution grows with increasing quark mass up to 
about $m \approx 0.8\,{\rm GeV}$. 
Being interested in the large-$m$ behavior, 
we do not show plots in this low-$m$ region. 
For $m>0.8\,{\rm GeV}$ the shear force
distribution starts to
decrease which is illustrated 
in Fig.~\ref{Fig-3:heavy-quark-s-p}a. In Fig.~\ref{Fig-3:heavy-quark-s-p}b we show the rescaled
dimensionless quantity $s(b)/(BR)$. Notice that
$\lim_{m\to\infty} s(b)/(BR)$ exists and assumes 
a well-defined value which is included in the plot
(it is convenient to include $B$ to have a dimensionless
quantity).
The 2D pressure distribution shows the same pattern:
the modulus of $p(b)$ increases with $m$ up to about
$0.9\,{\rm GeV}$, and starts to decrease for $m\gtrsim 0.9\,{\rm GeV}$ as 
shown in Fig.~\ref{Fig-3:heavy-quark-s-p}c.
Also the rescaled pressure $p(b)/(BR)$ has a well-defined
limit $\lim_{m\to\infty} p(b)/(BR)$ and Fig.~\ref{Fig-3:heavy-quark-s-p}d shows how this limit 
is approached. It is worth remarking that $p(b)$ at $b=0$
is proportional to the expression for the 3D surface 
tension defined as $\gamma = \int_0^\infty dr\,s(r)$.
The initial increase of the 2D pressure at $b=0$ and the 
subsequent decrease at $m\gtrsim0.8\,{\rm GeV}$ is therefore
tied to the $m$-dependence of the 3D surface tension
$\gamma$.
We stress that at any value of $m$ the distributions 
$s(b)$ and $p(b)$ satisfy the differential equation
(\ref{Eq:diff-eq-p-s-2D}), and $p(b)$ satisfies the 2D 
von Laue condition (\ref{Eq:von-Laue-2D}). This is true 
also for the limiting values of the rescaled quantities
$\lim_{m\to\infty} s(b)/(BR)$ and
$\lim_{m\to\infty} p(b)/(BR)$.

Next we proceed with the discussion of the 2D AM
distributions in bag model for selected quark masses $m$.
In Fig.~\ref{Fig-4:AM-S-L-J-kinetic-heavy-quark}a we
show the spin distribution $S^z(b)$ for $m=0.1$,
$0.5$, $1\,$GeV. We see that the spin distribution
continuously increases with increasing $m$ in the
inner region and decreases in the outer region,
i.e.\ it becomes more strongly localized. 
In contrast, the kinetic OAM
distribution continuously decreases as $m$ grows,
see Fig.~\ref{Fig-4:AM-S-L-J-kinetic-heavy-quark}b.
Already for the range of mass values selected in Figs.~\ref{Fig-4:AM-S-L-J-kinetic-heavy-quark}a 
and \ref{Fig-4:AM-S-L-J-kinetic-heavy-quark}b, the spin
distribution strongly dominates over the kinetic OAM distribution (notice that the scale
on the $y$-axis is 15 times larger in
Fig.~\ref{Fig-4:AM-S-L-J-kinetic-heavy-quark}a as
compared to Fig.~\ref{Fig-4:AM-S-L-J-kinetic-heavy-quark}b).
This is an interesting observation which can also be
intuitively understood. As $m$ increases, the inertia
of the quarks becomes larger and larger 
(i.e. quarks become more and more non-relativistic)
making orbital 
motion less and less important for the spin
budget of the nucleon. 
In Fig.~\ref{Fig-4:AM-S-L-J-kinetic-heavy-quark}c we
show the rescaled total kinetic AM
distribution $J^z(b)=L^z(b)+S^z(b)$ multiplied by $R^2$ which
for $m> 1\,$GeV practically coincides with $S^z(b)$.
Notice that this quantity has a well-defined limit
$\lim_{m\to\infty} R^2 J^z(b)$ which is included
in Fig.~\ref{Fig-4:AM-S-L-J-kinetic-heavy-quark}c.

In Fig.~\ref{Fig-5:AM-J-Belinfante-heavy-quark}a we 
show the 2D Belinfante AM distribution
$J^z_{\rm Bel}(b)$ for $m=0.1$, $0.5$, $1\,$GeV which
grows continuously with $m$. 
In Fig.~\ref{Fig-5:AM-J-Belinfante-heavy-quark}b 
we show the rescaled Belinfante AM 
distribution $R^2 J_{\rm Bel}^z(b)$ which has a well-defined 
limit $\lim_{m\to\infty} R^2 J_{\rm Bel}^z(b)$ included
in the figure. 
Also for the Belinfante AM distribution 
we observe that it becomes more strongly localized as 
$m$ grows. Note that by construction $J^z_\text{Bel}(0)=0$ whereas $J^z(0)=S^z(0)\neq 0$.

The kinetic and Belinfante AM distributions
are, however, much different even in the heavy quark limit.
In Fig.~\ref{Fig-5:AM-J-Belinfante-heavy-quark}c we show
the rescaled distributions $2\pi b\,R\,J^z(b)$ and 
$2\pi b\,R\,J_{\rm Bel}^z(b)$ as functions of $b/R$
which have both well-defined limits for $m\to\infty$.
Very clearly, as $m$ grows and the limit is reached,
the two distributions exhibit a much different 
behavior --- even though all curves in 
Fig.~\ref{Fig-5:AM-J-Belinfante-heavy-quark}c 
yield $\frac12$ upon integration over the rescaled
variable $b/R$. 

\newpage
\section{2D kinetic EMT distributions in the large system size limit}
\label{Sec:large-R-limit-m-fix}

In this section, we discuss 2D EMT distributions in 
the limit of large bag radius $R$ for fixed quark mass $m$
which will keep fixed at $5\,{\rm MeV}$, corresponding to the 
physical situation of Sec.~\ref{Sec:EMT-physical-situation}. 
The large-$R$ limit belongs to a class of limits,
in which the interaction in the bag model becomes small.
As in the heavy quark limit of
Sec.~\ref{Sec:heavy-mass-limit-B-fix}, also in this case 
the dynamics of the system becomes non-relativistic,
however for a different reason. 

In fact, even though both limits lead to non-relativistic situations, the physics is 
significantly different in the two cases. For instance,
the internal forces behave much differently in the two limits
which can be understood as follows.
In the limit $R\to\infty$ with $m$ fixed, the bag constant 
scales as $B \propto R^{-5}$ which follows from the virial theorem
(\ref{eq:virial-limit-mR}). 
We recall that the bag constant 
naturally sets the scaling for $p(r)$ and $s(r)$,
see Sec.~\ref{sec:limits-overview}.

The behaviour of the 3D energy distribution is different. As $R\to\infty$,
we have $N_c$ quarks bound by a ``mean field'' which is more and more
diluted as the size of the system grows and $B\sim {R^{-5}}$ decreases.
In this situation, the mass of the system approaches $M_N\sim N_c m$
which (is 15 MeV in our case, and) implies for the 3D energy distribution the scaling 
$\epsilon(r)\sim R^{-3}$. The total kinetic and Belinfante 
AM distributions scale as $R^{-3}$, and OAM as $R^{-5}$.
The 2D distributions are obtained by integrating out one spatial
dimension, and the associated 2D distributions scale as 
$\epsilon(b) \sim R^{-2}$, $J^z(b) \sim R^{-2}$, $J^z_{\rm Bel}(b) \sim R^{-2}$, $L^z(b) \sim R^{-4}$, $s(b) \sim R^{-4}$, $p(b) \sim R^{-4}$. 

\begin{figure}[b]
\begin{centering}
\includegraphics[height=4cm]{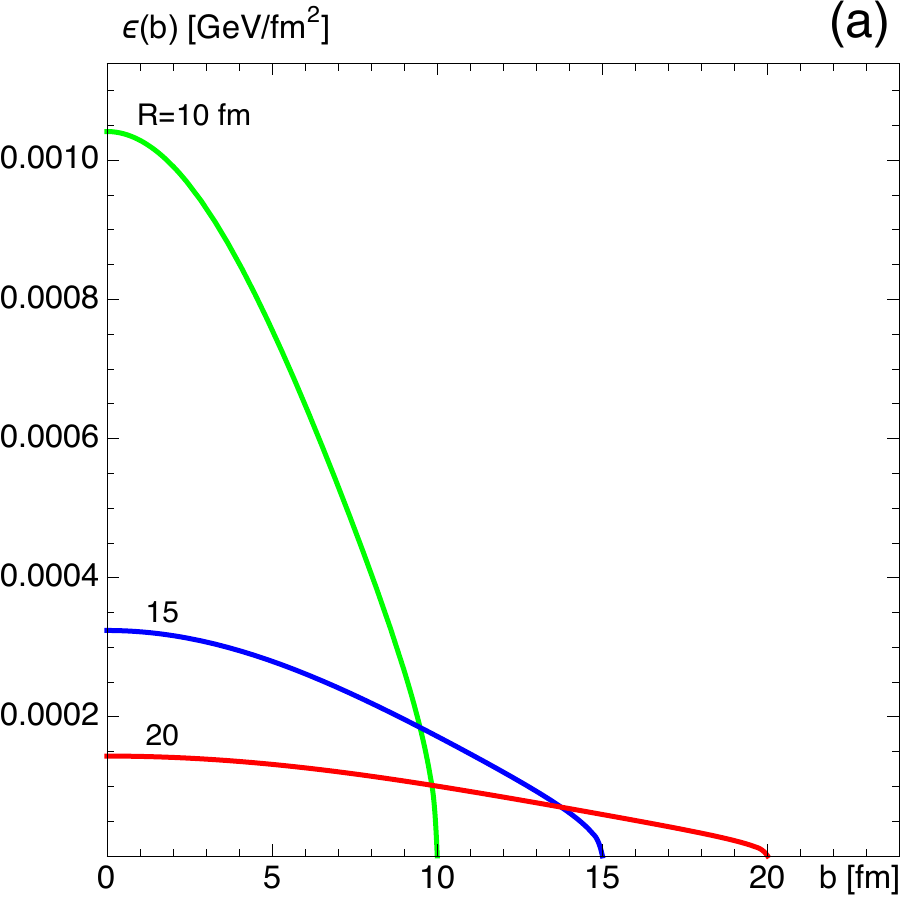} \hspace{4mm} \
\includegraphics[height=4cm]{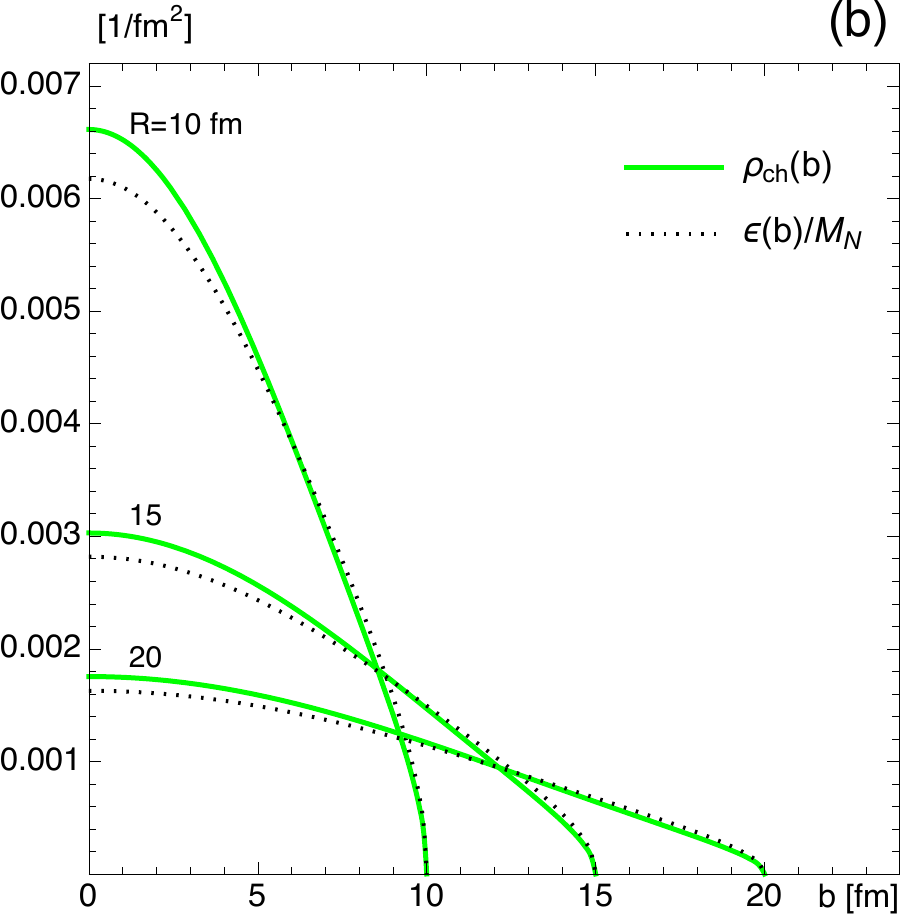} \hspace{4mm} \
\includegraphics[height=4cm]{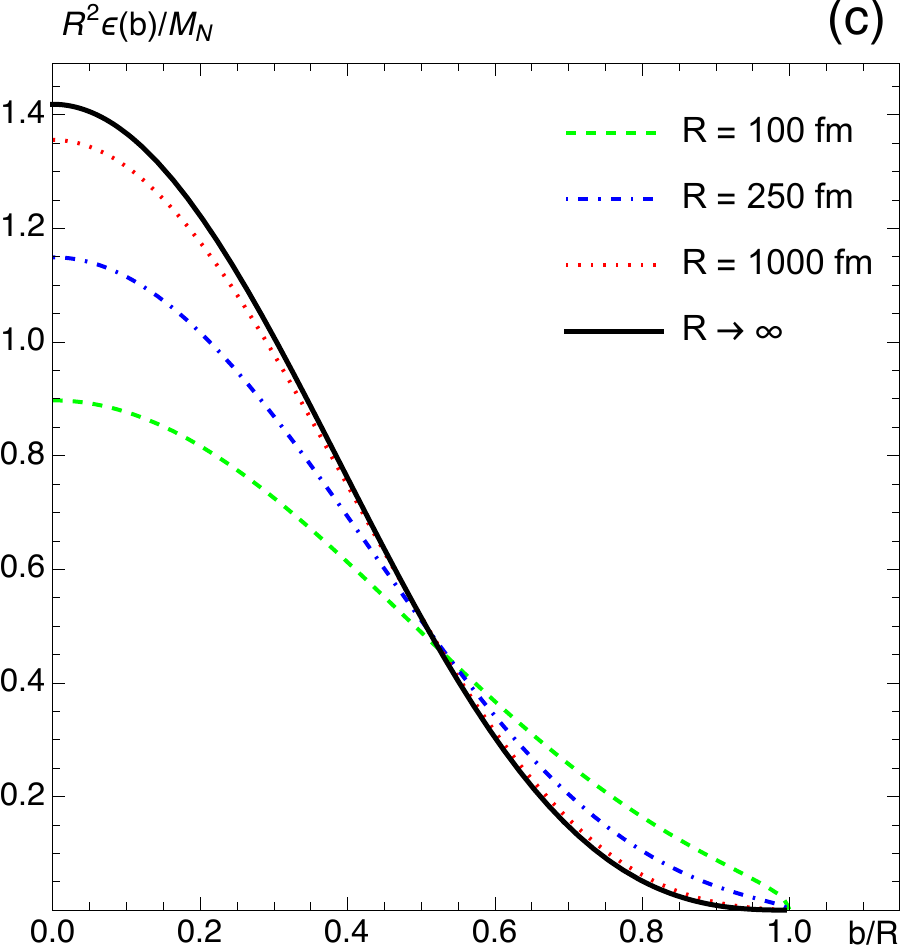} \hspace{4mm} \
\par\end{centering}
\caption{\label{Fig-6:energy-large-radius-T00} 
2D distributions in the bag model for 
fixed $m$ and increasing bag radii $R$.
(a) Energy distribution~$\epsilon(b)$. 
(b) Normalized energy distribution $\epsilon(b)/M_N$
in comparison to the 2D electric charge distribution $\rho_{\rm ch}(b)$.
(c) The scaling of $R^2\epsilon(b)/M_N$ for $R\to\infty$. }

\

\begin{center}
\includegraphics[height=4cm]{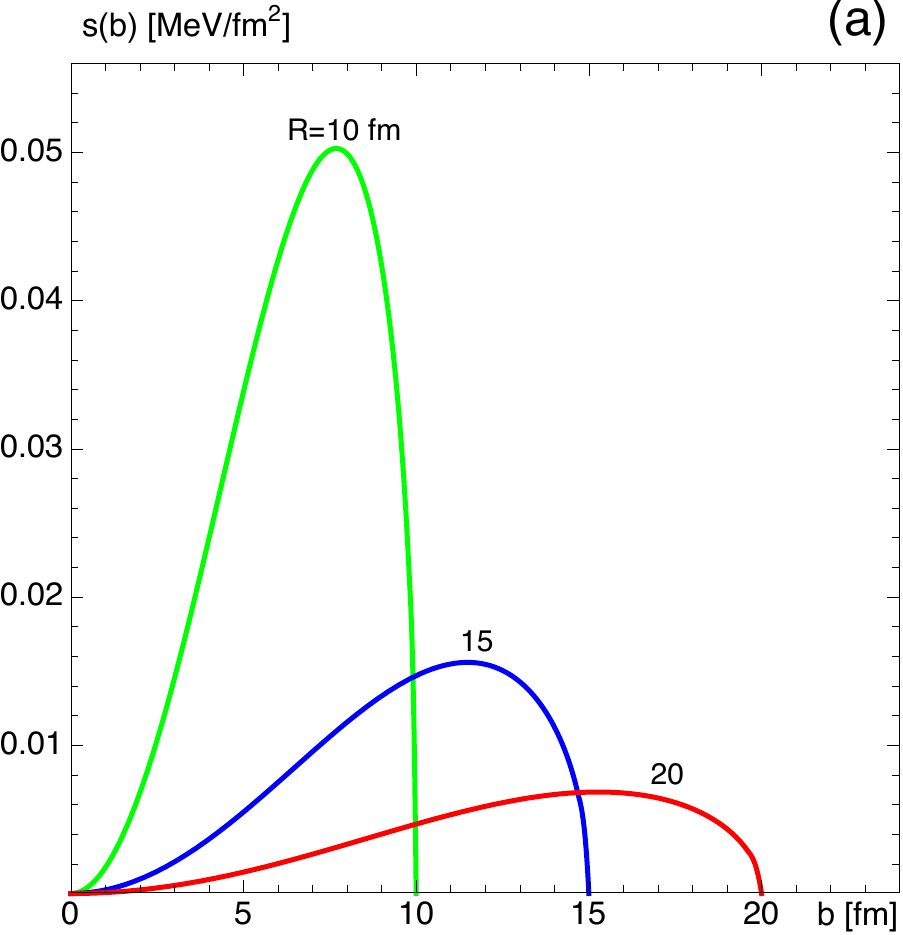} \hspace{4mm} \
\includegraphics[height=4cm]{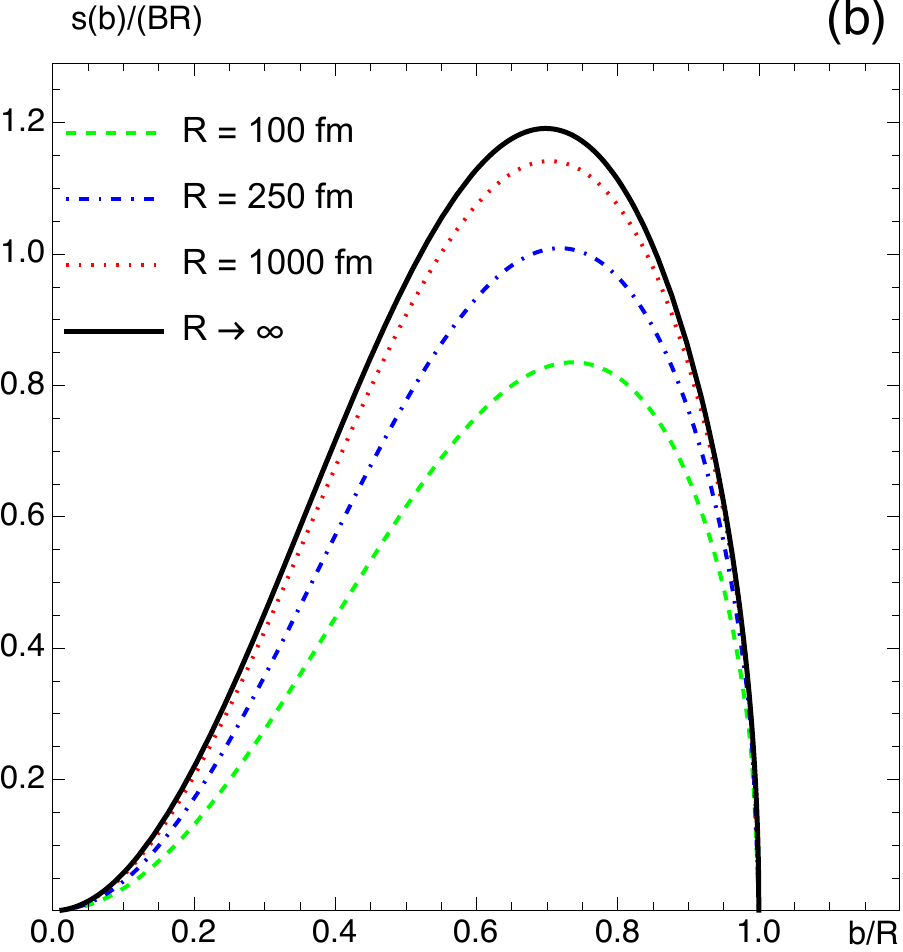} \hspace{4mm} \
\includegraphics[height=4cm]{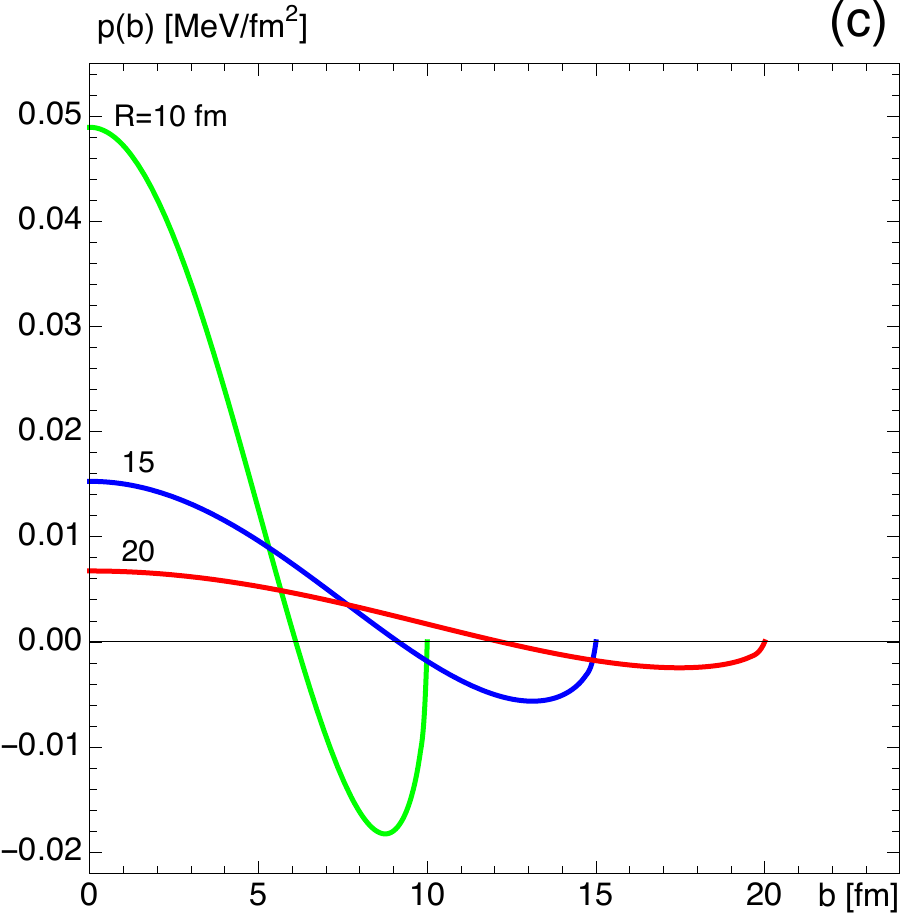} \hspace{4mm} \ 
\includegraphics[height=4cm]{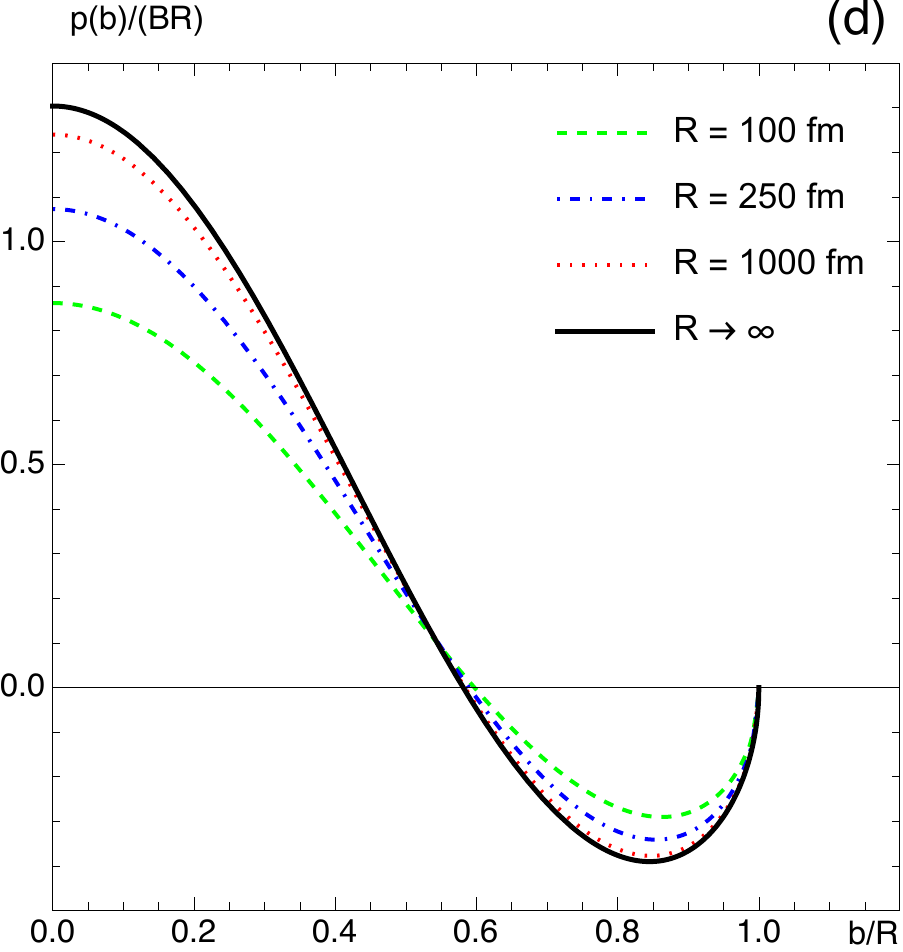}  
\end{center}
%
%
\caption{\label{Fig-7:large-radius-s-p} 
The 2D EMT shear force and pressure distributions in the bag model for fixed $m$ and increasing $R$.
(a) Shear force~$s(b)$. 
(b) The rescaled dimensionless distribution $s(b)/(BR)$.
(c) Pressure $p(b)$. 
(d) The rescaled dimensionless distribution $p(b)/(BR)$.}
\end{figure}

In Fig.~\ref{Fig-6:energy-large-radius-T00}a we depict
$\epsilon(b)$ as function of $b$ for increasing values 
of $R=10,\;15, \;20\,{\rm fm}$ which shows the trend of how the
system size grows and the energy distribution becomes more
and more diluted. 
The normalized energy distribution $\epsilon(b)/M_N$ is
shown in Fig.~\ref{Fig-6:energy-large-radius-T00}b in
comparison to the electric charge distribution $\rho_{\rm ch}(b)$
for selected values $R=10,\;15, \;20\,{\rm fm}$. Also here we 
see how the distribution becomes more and more diluted as the system size grows. In addition, we see that the difference
between $\epsilon(b)/M_N$ and $\rho_{\rm ch}(b)$ decreases 
as $R$ increases.
In Fig.~\ref{Fig-6:energy-large-radius-T00}c we display 
the scaling of the dimensionless quantity
$R^2\epsilon(b)/M_N$ for $R=100,\;250,\;1000\,{\rm fm}$. 
The limiting curve of
$\lim_{R\to\infty}R^2\epsilon(b)/M_N$ 
is included in the plot, and we see that 
it is approached very slowly. 
Even when $R$ is 3 orders of magnitude larger than in 
the physical situation, we can still distinguish
$R^2\epsilon(b)/M_N$ from its limiting curve. 
For $R = 1\,$\AA , when the size of the system 
corresponds to that of an atom, the model result would 
be indistinguishable from the limiting curve on the 
scale of Fig.~\ref{Fig-6:energy-large-radius-T00}c.

In Fig.~\ref{Fig-7:large-radius-s-p} we investigate 
the 2D force distributions. 
In Fig.~\ref{Fig-7:large-radius-s-p}a we depict the
2D shear force distribution $s(b)$ for increasing 
values of $R=10,\;15, \;20\,{\rm fm}$. The figure
shows that $s(b)$ strongly decreases for growing $R$. 
The ``scaling regime'' $s(b)\sim R^{-4}$ is, however,
approached only when $R$ is 2 orders of magnitude
above the physical value of $R=1.7\,{\rm fm}$  
as illustrated in Fig.~\ref{Fig-7:large-radius-s-p}b 
which shows the dimensionless quantity $s(b)/(BR)$ 
including its limit $\lim_{R\to\infty}s(b)/(BR)$. 
In the case of the pressure $p(b)$ shown in 
Fig.~\ref{Fig-7:large-radius-s-p}c and the
rescaled quantity $p(b)/(BR)$ displayed in
Fig.~\ref{Fig-7:large-radius-s-p}d we make
the same observations.

\begin{figure}[t]
\begin{center}
\includegraphics[height=4cm]{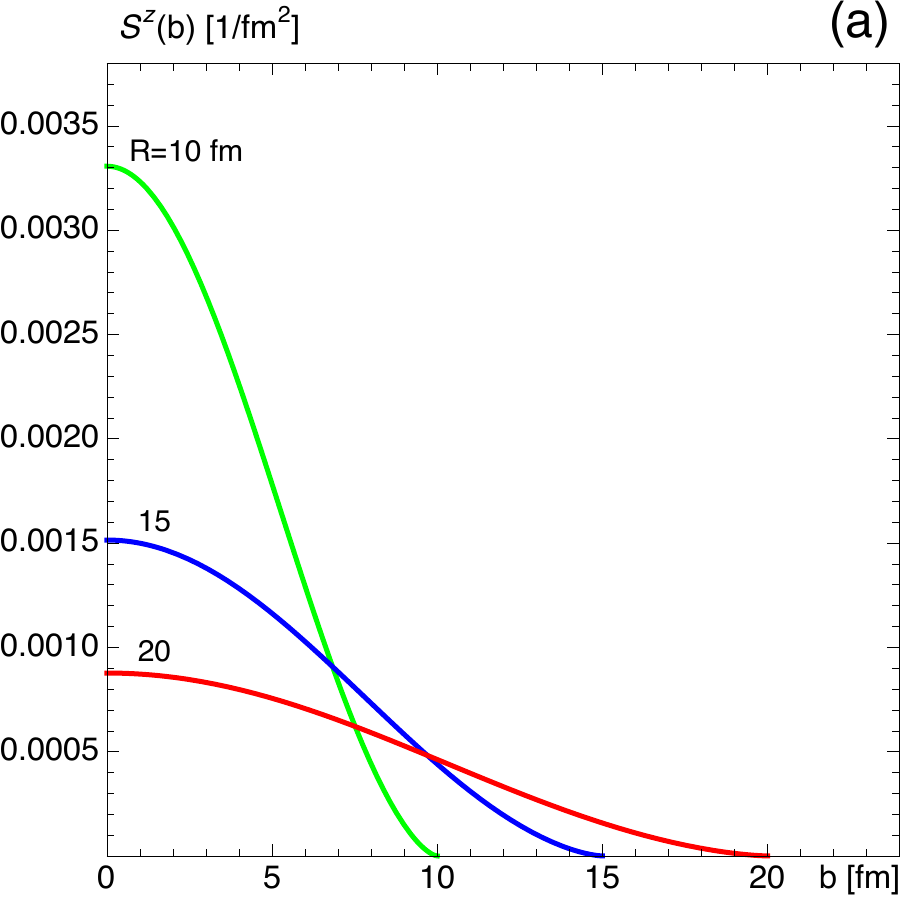} \hspace{4mm} \
\includegraphics[height=4cm]{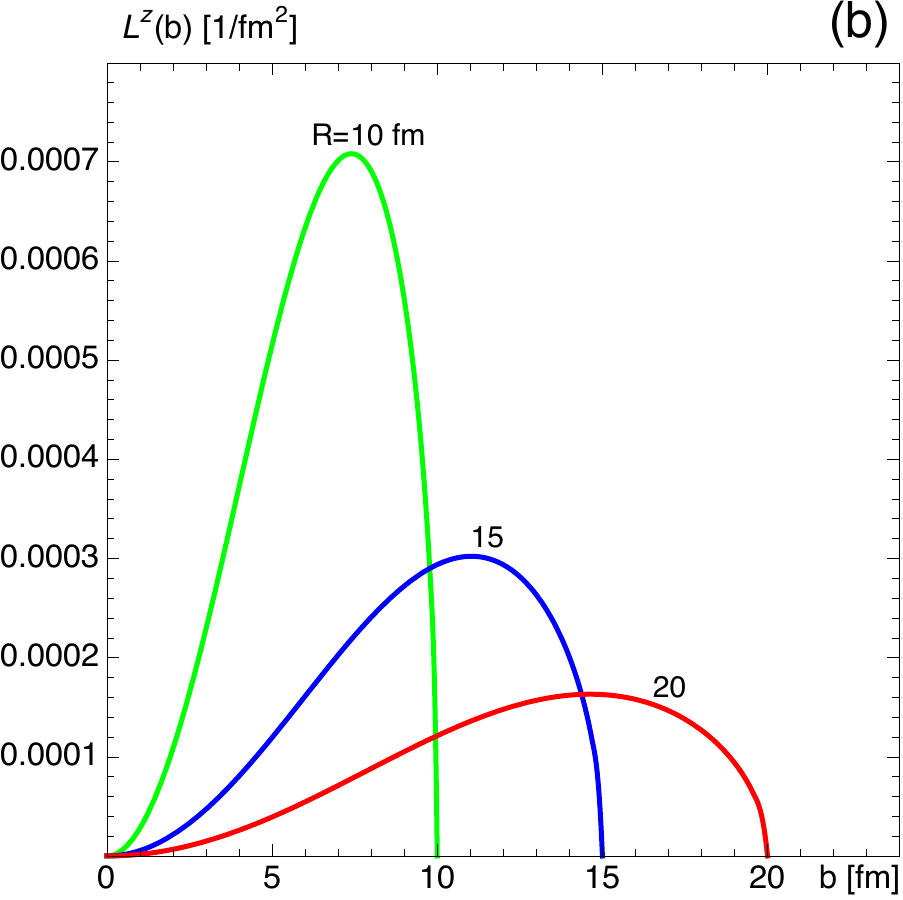} \hspace{4mm} \
\includegraphics[height=4cm]{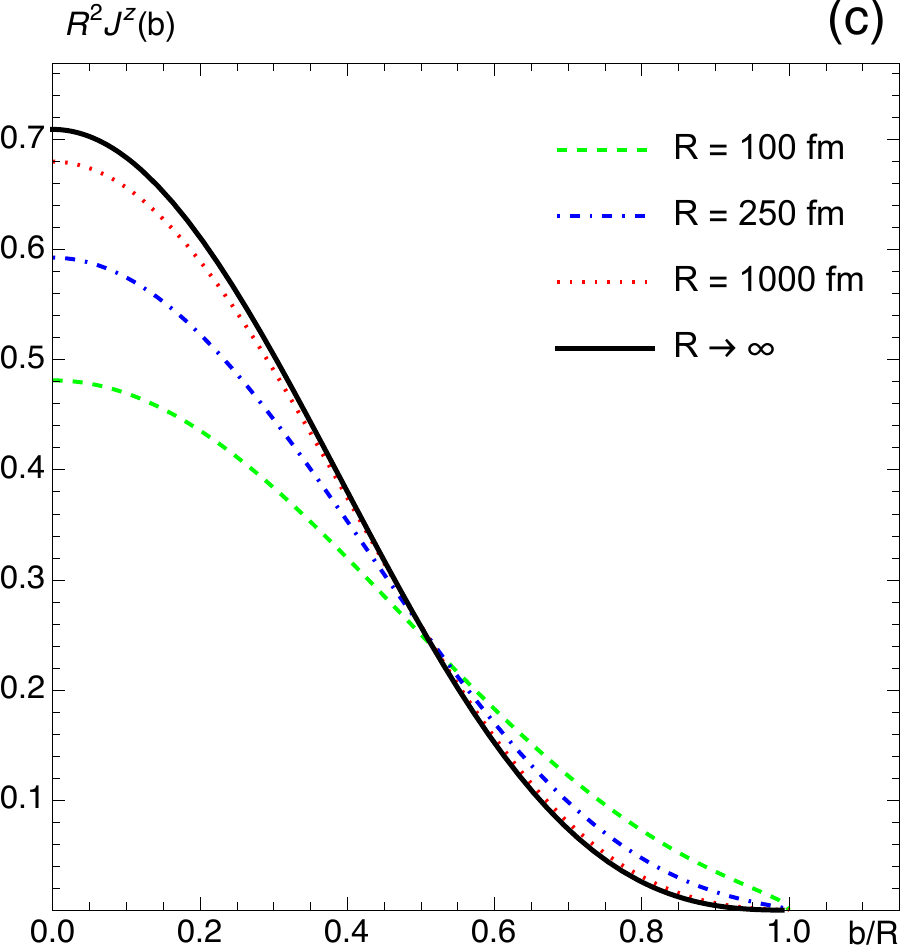}  
\end{center}
\caption{\label{Fig-8:AM-S-L-J-kinetic-large-radius} 
The 2D angular momentum distributions in the bag model for fixed $m$ and increasing bag radii $R$.
(a) Intrinsic spin $S^z(b)$ distributions. 
(b) Orbital angular momentum $L^z(b)$ distributions.
(c) The scaling of $R^2 J^z(b)$ for $R\to\infty$.}

\vspace{7mm}

\begin{center}
\includegraphics[height=4cm]{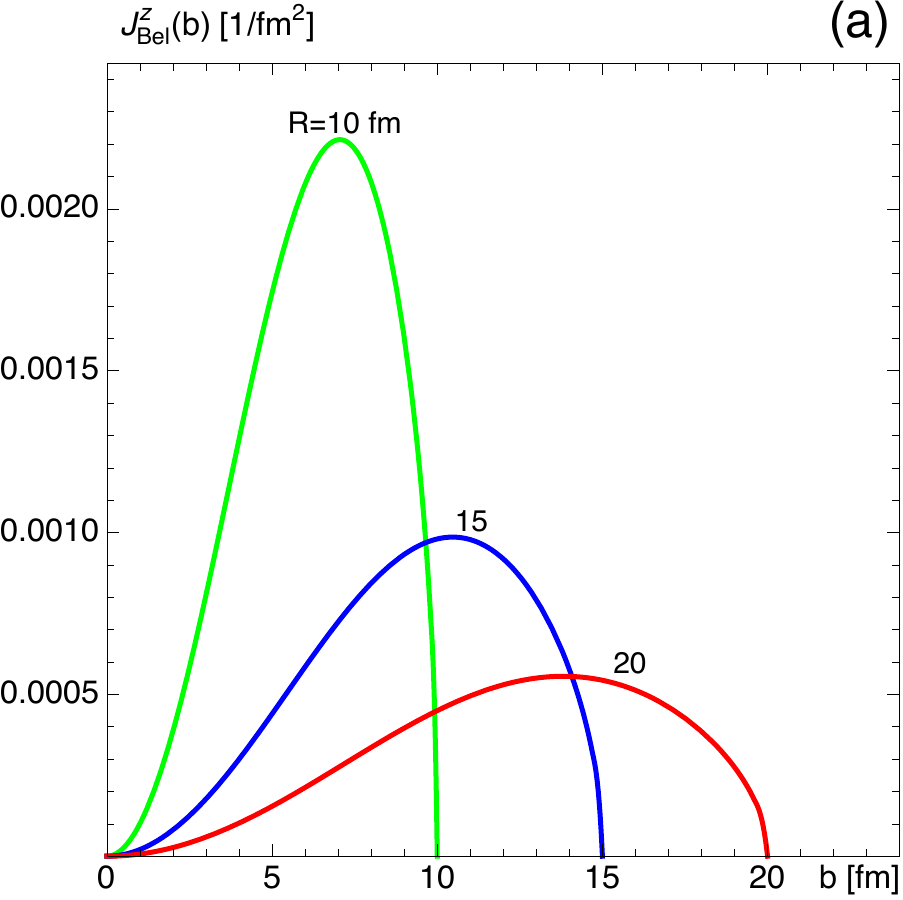} \hspace{4mm} \
\includegraphics[height=4cm]{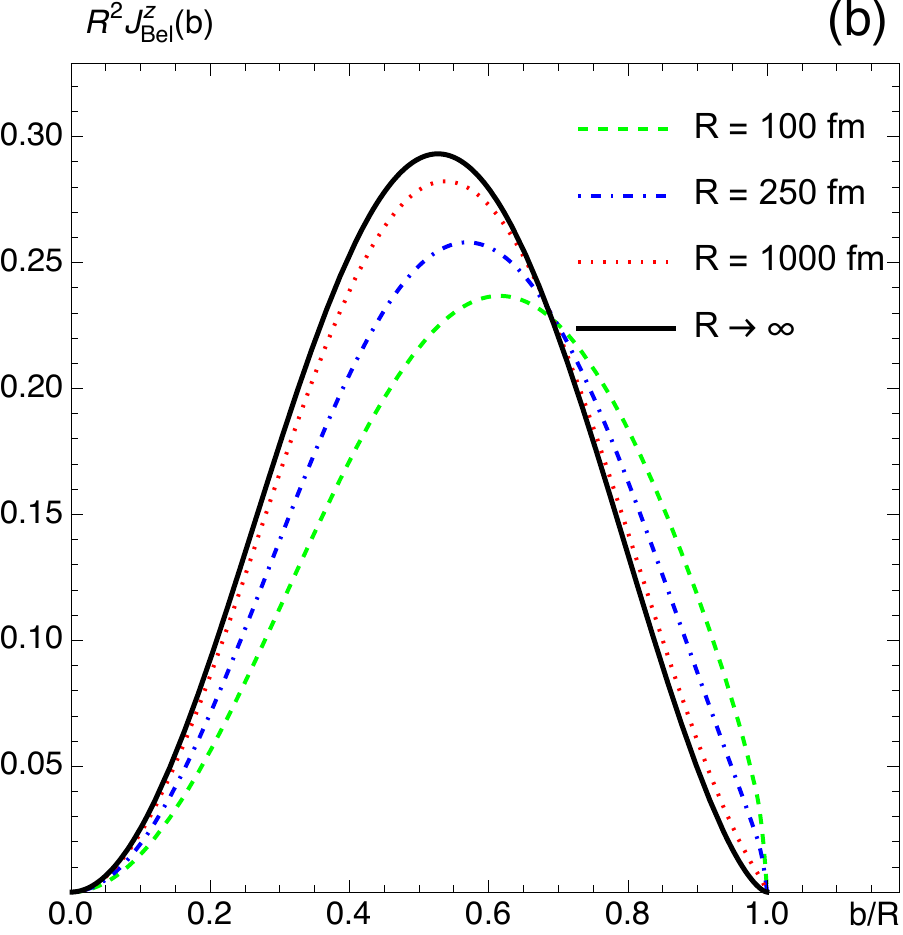} \hspace{4mm} \
\includegraphics[height=4cm]{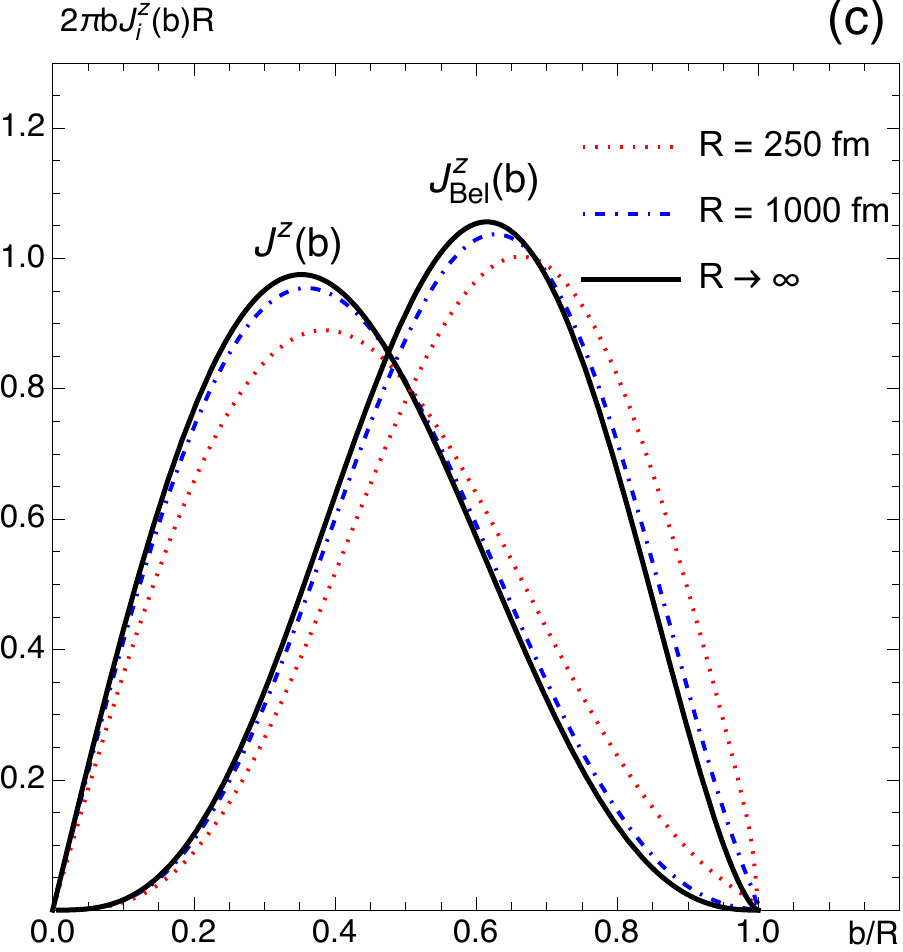}  
\end{center}
\caption{\label{Fig-9:AM-J-Belinfante-large-radius} 
The 2D angular momentum distribution in the bag model for
fixed $m$ and increasing bag radii $R$.
(a) Belinfante-form $J_\text{Bel}^z(b)$ angular momentum distributions. 
(b) The scaling of $R^2 J_\text{Bel}^z(b)$ for $R\to\infty$.
(c) The scaling of $2\pi b J_\text{i}^z(b) R$ for $R\to\infty$.} 

\end{figure}

The behavior of the 2D AM distributions 
in the large system size limit is shown in 
Fig.~\ref{Fig-8:AM-S-L-J-kinetic-large-radius}.
The intrinsic spin distribution $S^z(b)$ is shown 
in Fig.~\ref{Fig-8:AM-S-L-J-kinetic-large-radius}a
for $R=10,\;15, \;20\,{\rm fm}$, and that of the kinetic OAM distribution $L^z(b)$ is depicted 
in Fig.~\ref{Fig-8:AM-S-L-J-kinetic-large-radius}b
for the same values of $R$. Notice the different scales
in these two figures, showing that the OAM plays
a much smaller role in the spin budget as $R$
increases. In the limit $R\to\infty$, the OAM distribution becomes less and less 
important compared to the intrinsic spin distribution.
This is not apparent for the $R$ values chosen in 
Figs.~\ref{Fig-8:AM-S-L-J-kinetic-large-radius}a
and \ref{Fig-8:AM-S-L-J-kinetic-large-radius}b but
the intrinsic spin distribution decreases
as $S^z(b) \sim R^{-2}$, i.e.~much more slowly than the OAM
distribution which is suppressed as 
$L^z(b) \sim R^{-4}$. 

It is an interesting observation
that OAM becomes irrelevant as $R$
increases. It is important to keep in mind that the 
quarks can be light and one would expect that a relativistic description
is necessary for any $R$.
However, the increasing bag radius $R$ simulates a 
more and more weakly bound system amenable to a
non-relativistic 
description. This can be understood by invoking Heisenberg's uncertainty principle: with a larger volume provided to the quarks to ``fill out'', their motion becomes slower, and with that the role of OAM decreases.
The scaling of the kinetic angular momentum distribution
$R^2 J^z(b)$ for increasing $R$ is shown in
Fig.~\ref{Fig-8:AM-S-L-J-kinetic-large-radius}c
along with the limiting curve for 
$\lim_{R\to\infty} R^2J^z(b)$.
As in the case of the other EMT distributions, the
scaling behavior becomes apparent when $R$ is at
least 2 orders of magnitude larger than in the
physical situation.  

In Fig.~\ref{Fig-9:AM-J-Belinfante-large-radius}a 
we depict the 2D Belinfante AM 
distribution for selected values of 
$R=10,\,15,\,20\,{\rm fm}$ and 
Fig.~\ref{Fig-9:AM-J-Belinfante-large-radius}b
shows the dimensionless rescaled distibution
$R^2J^z_{\rm Bel}(b)$ as function of $b/R$ 
including its limiting curve
$\lim_{R\to\infty}R^2J^z_{\rm Bel}(b)$. 
Finally, in 
Fig.~\ref{Fig-9:AM-J-Belinfante-large-radius}c 
we compare respectively the dimensionless rescaled 
kinetic and Belinfante AM distributions
$2\pi b J^z(b) R$ and
$2\pi b J^z_{\rm Bel}(b) R$, 
including their $R\to\infty$ limits.
We see that the 2 different distributions
clearly differ also in the large system
size limit.

\newpage 
\section{2D kinetic EMT distributions in constituent quark limit}
\label{Sec:constituent-limit}

In this section, we discuss the behavior of 2D EMT 
distributions in the limit L3 where $m\to M_N/N_c$ 
with the nucleon mass kept fixed at its physical value.
For the following it is convenient to introduce 
the mass $\mx = M_N/N_c$, i.e.\ the maximal 
mass a quark can asymptotically take in the
limit L3. 
For massless quarks, 3/4 of the nucleon mass is due
to the kinetic energy of the ultra-relativistic quarks 
and 1/4 is due to the bag energy (we will say more about 
nucleon mass decomposition in Sec.~\ref{Sec:M-decompose}). 
As the limit $m\to \mx$ is approached, the quark mass
constitutes nearly all of the nucleon mass, while the 
contributions of quark kinetic energy and bag energy 
become negligible. The limit L3 can therefore be 
considered as a constituent quark limit. As a consequence
of the limit $m\to\mx$, the motion of the quarks becomes 
nonrelativistic.

\begin{figure}[b!]
\begin{centering}
\includegraphics[height=4cm]{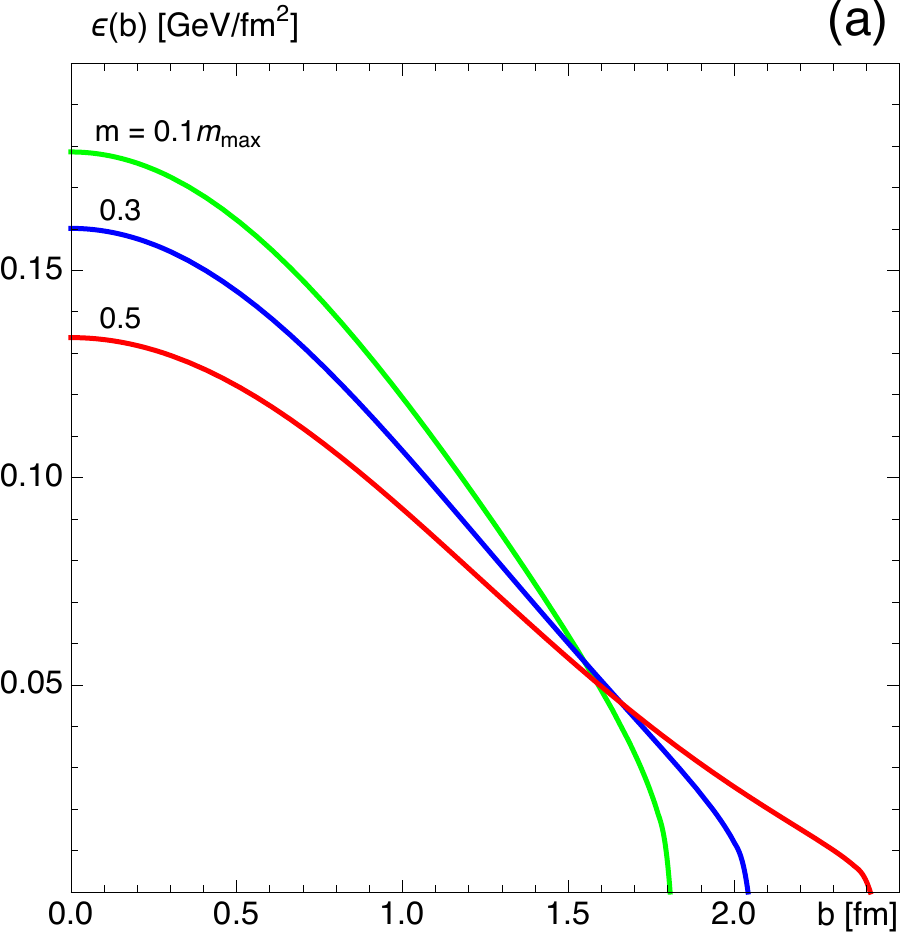} \hspace{4mm} \
\includegraphics[height=4cm]{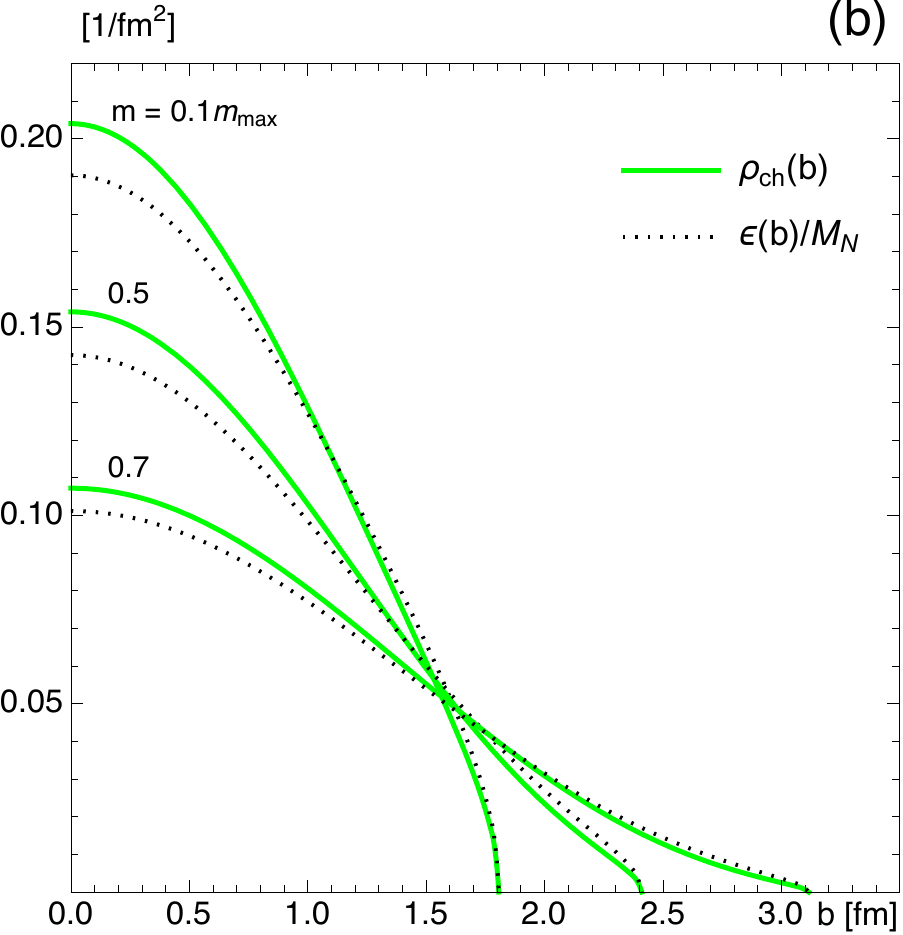} \hspace{4mm} \
\includegraphics[height=4cm]{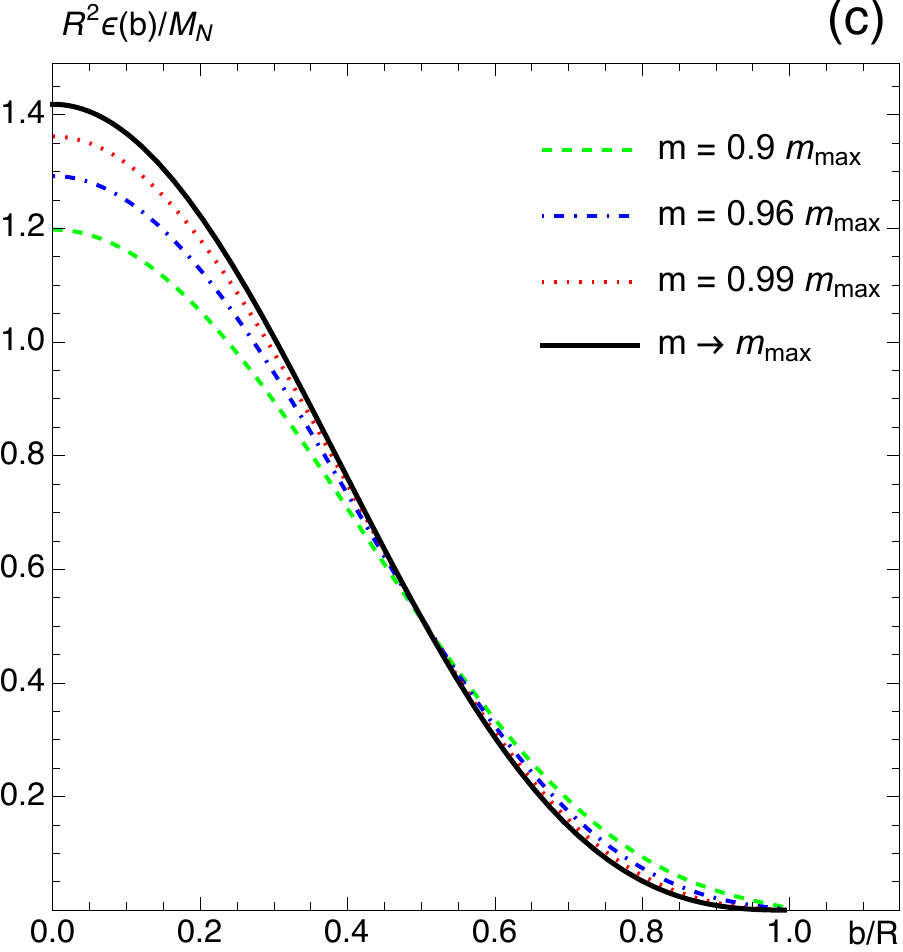} \hspace{4mm} \
\par\end{centering}
\caption{\label{Fig-10:energy-non-relativistic-T00} 
2D distributions in the bag model for 
fixed nucleon mass $M_N$ and increasing quark masses $m$.
(a) Energy distribution~$\epsilon(b)$. 
(b) Normalized energy distribution $\epsilon(b)/M_N$
in comparison to the 2D electric charge distribution $\rho_{\rm ch}(b)$. (c) The scaling of $R^2\epsilon(b)/M_N$ for $m\to m_{\text{max}}$. }


\

\begin{centering}
\includegraphics[height=4cm]{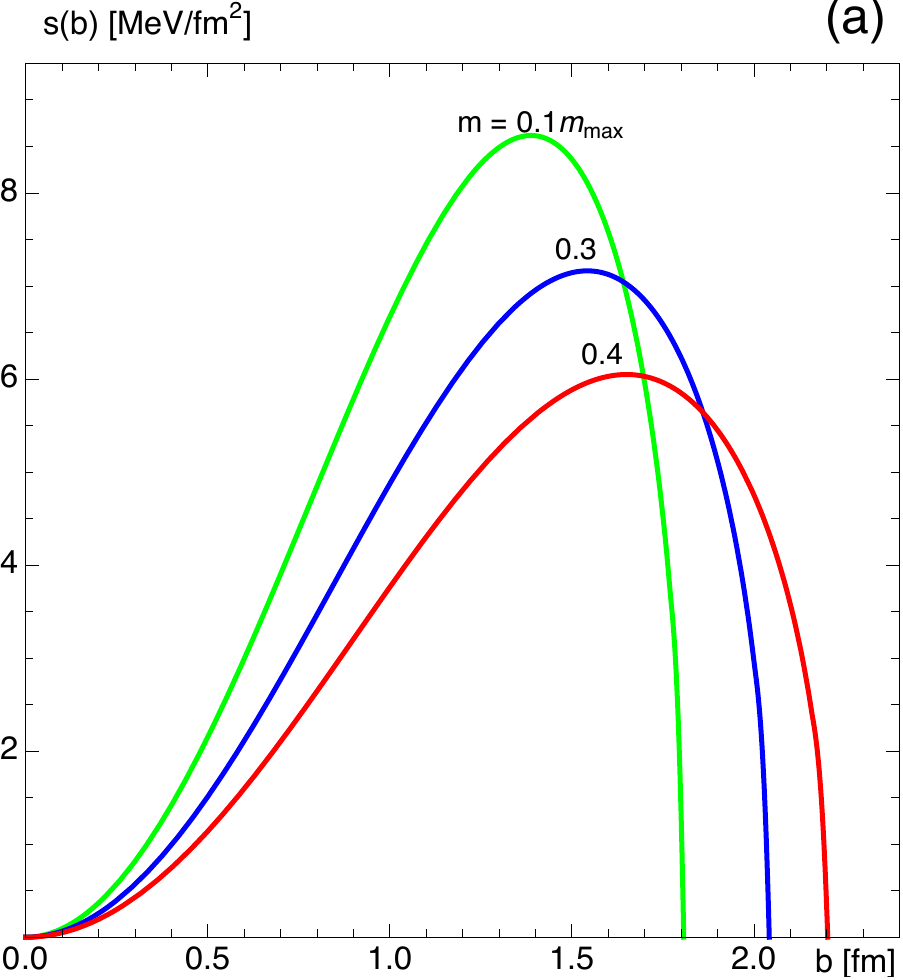} \hspace{4mm} \
\includegraphics[height=4cm]{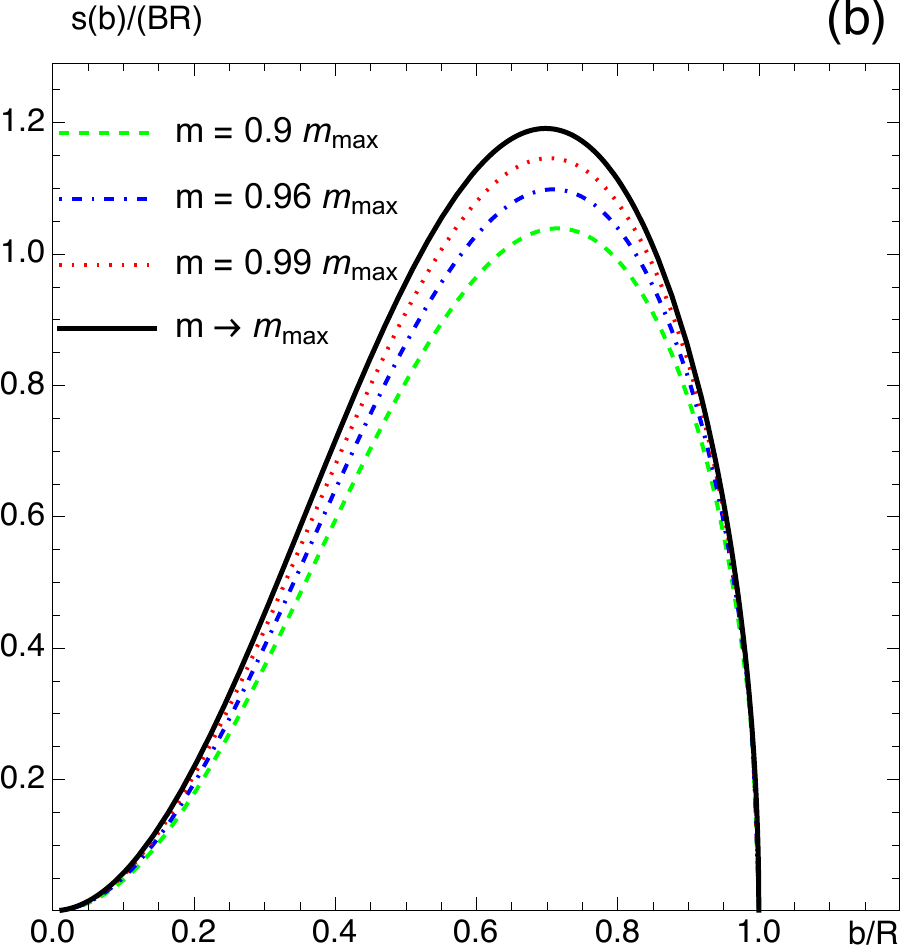} 
\hspace{4mm} \
\includegraphics[height=4cm]{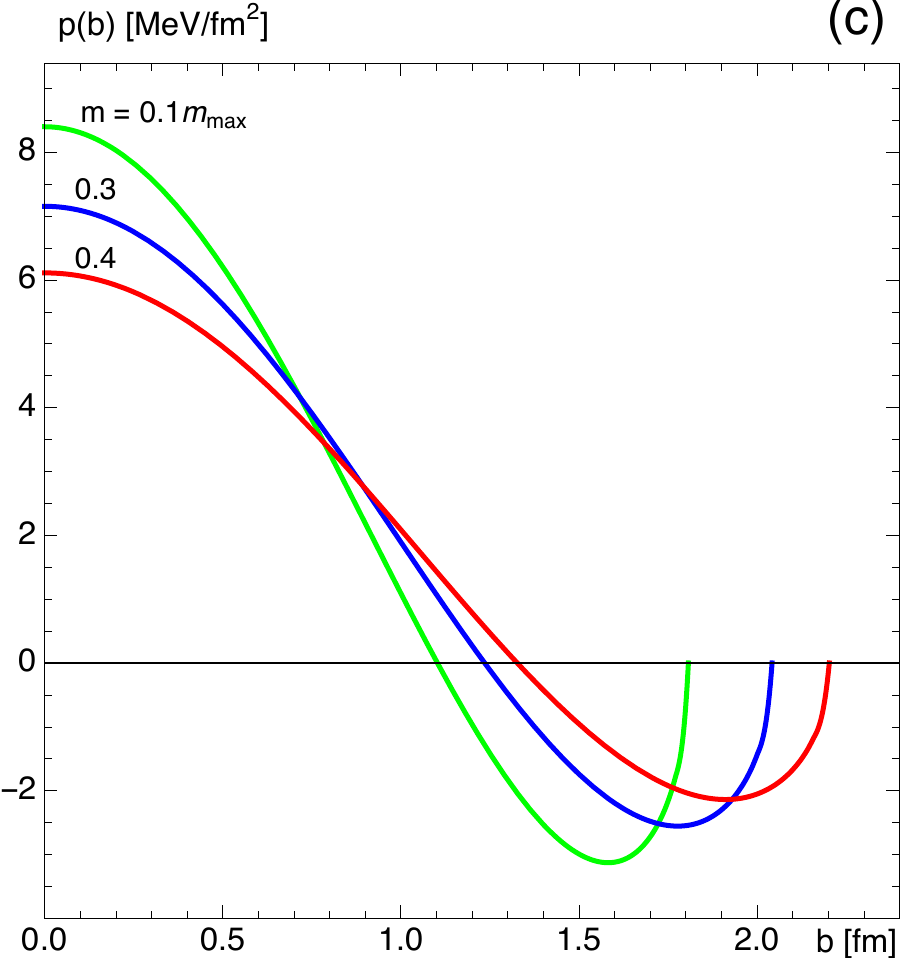} \hspace{4mm} \ 
\includegraphics[height=4cm]{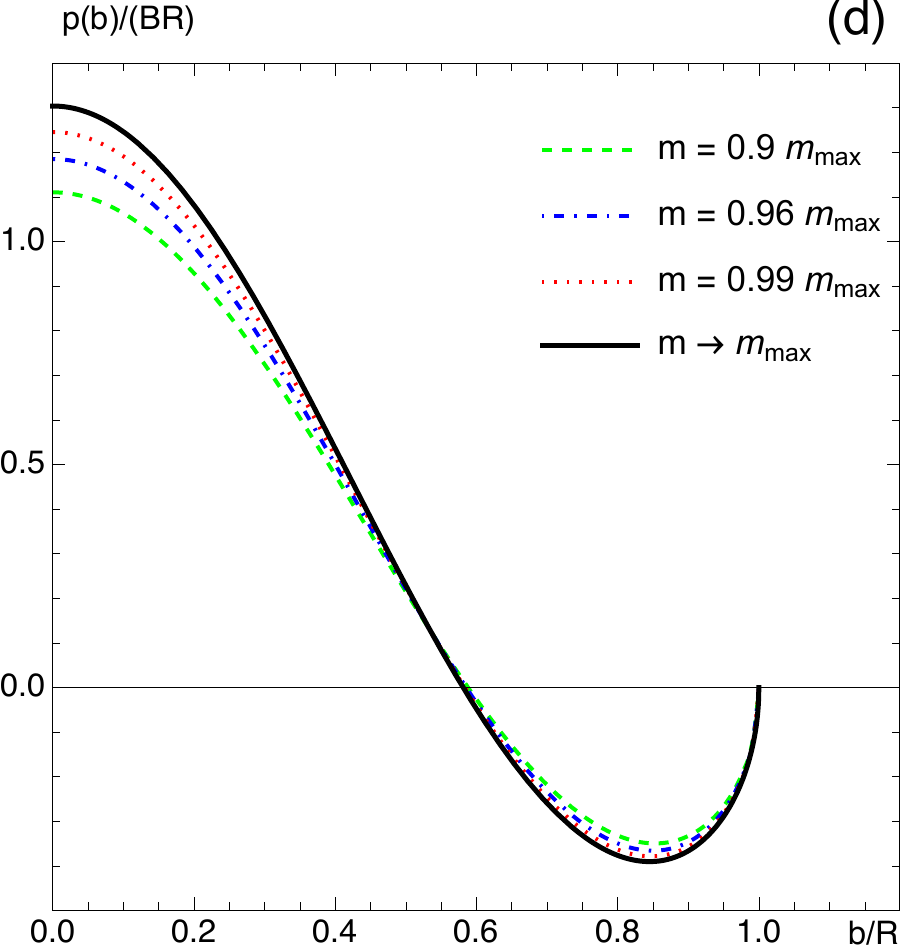}  
\end{centering}
\caption{\label{Fig-11:non-relativistic-s-p} 
2D distributions in the bag model for 
fixed nucleon mass $M_N$ and increasing quark masses $m$.
(a) Shear force~$s(b)$. 
(b) The rescaled dimensionless distribution $s(b)/(BR)$.
(c) Pressure $p(b)$. 
(d) The rescaled dimensionless distribution $p(b)/(BR)$.}
\end{figure}

In the limit L3, the 3D distributions scale as 
$\epsilon(r) \sim R^{-3}$, 
$S^{z}(\vec r) \sim R^{-3}$, 
$J^{z}(\vec r) \sim R^{-3}$, 
$J_{\rm Bel}^{z}(\vec r) \sim R^{-3}$,
$L^{z}(\vec r) \sim R^{-5}$,
$s(r) \sim R^{-5}$, $p(r) \sim R^{-5}$,
see Sec.~\ref{sec:limits-overview}.
Integrating the 3D distributions over the $z$-axis 
produces the scaling behaviour of the associated $2D$
distributions as 
$\epsilon(b) \sim R^{-2}$, 
$S^{z}(b) \sim R^{-2}$, 
$J^{z}(b) \sim R^{-2}$, 
$J_{\rm Bel}^{z}(b) \sim R^{-2}$,
$L^{z}(b) \sim R^{-4}$,
$s(b) \sim R^{-4}$, $p(b) \sim R^{-4}$.
We see that similarly to the large-system size
limit L2, also here the EMT distributions become
more and more diluted, although the underlying
physical situations are much different.
In fact, in L2 we start with a compact proton of
mass 938$\,$MeV made of 5$\,$MeV quarks and let the 
system size $R\to\infty$ which drives the total mass of 
the system asymptotically to $15\,\rm MeV$. In L3,
we start and end with a system mass of 938$\,$MeV and
vary $m$ from 5$\,$MeV to $\mx$ and as a response to
that the size of the system $R$ becomes large.

Fig.~\ref{Fig-10:energy-non-relativistic-T00}a illustrates
$\epsilon(b)$ as function of $b$ for increasing values of
$m=0.1,\;0.3,\;0.5\,\mx$.
We see how the size of the system increases.
As $M_N$ is kept constant and all contributions to the 
energy distribution and nucleon mass are positive, in the limit L3 
the kinetic energy of the quarks (as well as the bag energy 
$E_{\rm bag}=\frac43\,\pi R^3B$) must decrease.
By the Heisenberg uncertainty principle,
the kinetic energy of a bound quantum particle 
decreases if the particle is provided a larger
volume to fill out. Hence the bag radius grows
in this limit. With the mass of the nucleon being 
fixed at its physical value, the energy distribution 
inside the system becomes more dilute. The electric 
charge distribution follows a similar pattern which
we show in Fig.~\ref{Fig-10:energy-non-relativistic-T00}b 
where we compare the normalized energy distribution 
$\epsilon(b)/M_N$ with the electric charge distribution 
$\rho_{\rm ch}(b)$ for selected values of $m=0.1,\;0.5,\;0.7\,\mx$.
We see that the difference between two distributions becomes 
less and less apparent for larger quark masses. Finally in
Fig.~\ref{Fig-10:energy-non-relativistic-T00}c we depict the
scaling of the dimensionless quantity $R^2\epsilon(b)/M_N$ for
$m=0.9,\;0.96,\;0.99\,\mx,$ including the curve 
associated with $m \rightarrow \mx$. 
When $m=0.99\,\mx$ the size of the
system reaches $R= 17.76\,\rm fm$.

\begin{figure}[t]
\begin{centering}
\includegraphics[height=4cm]{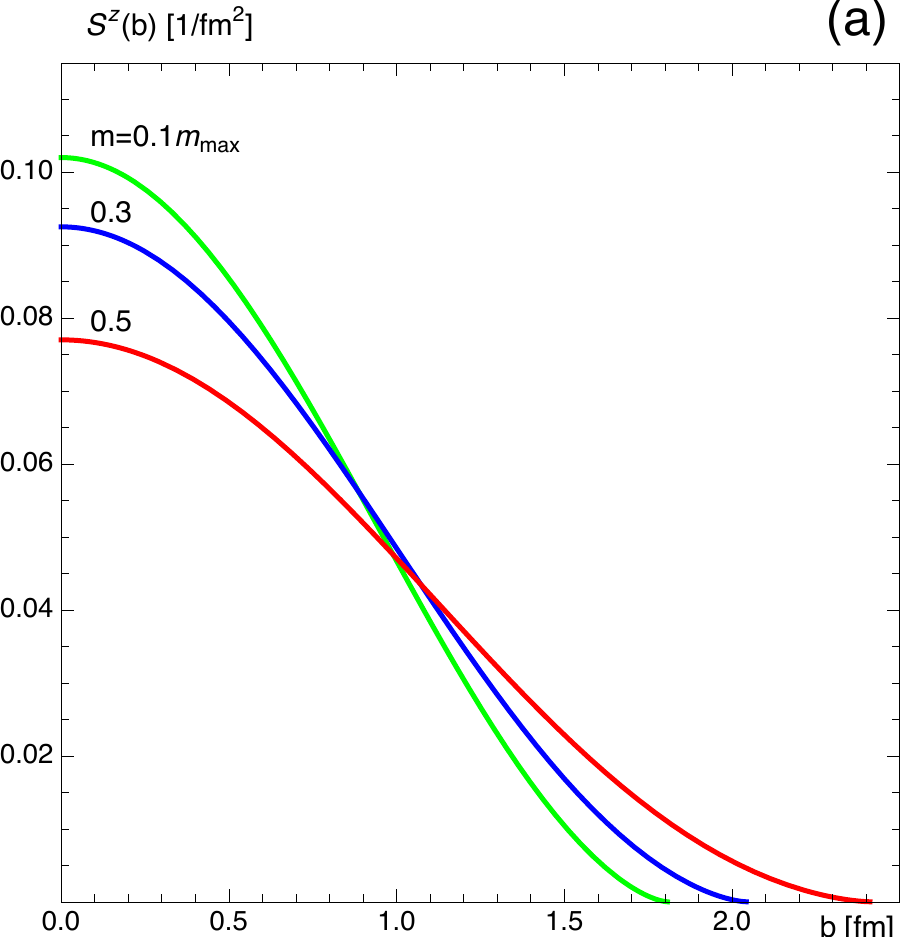} \hspace{4mm} \
\includegraphics[height=4cm]{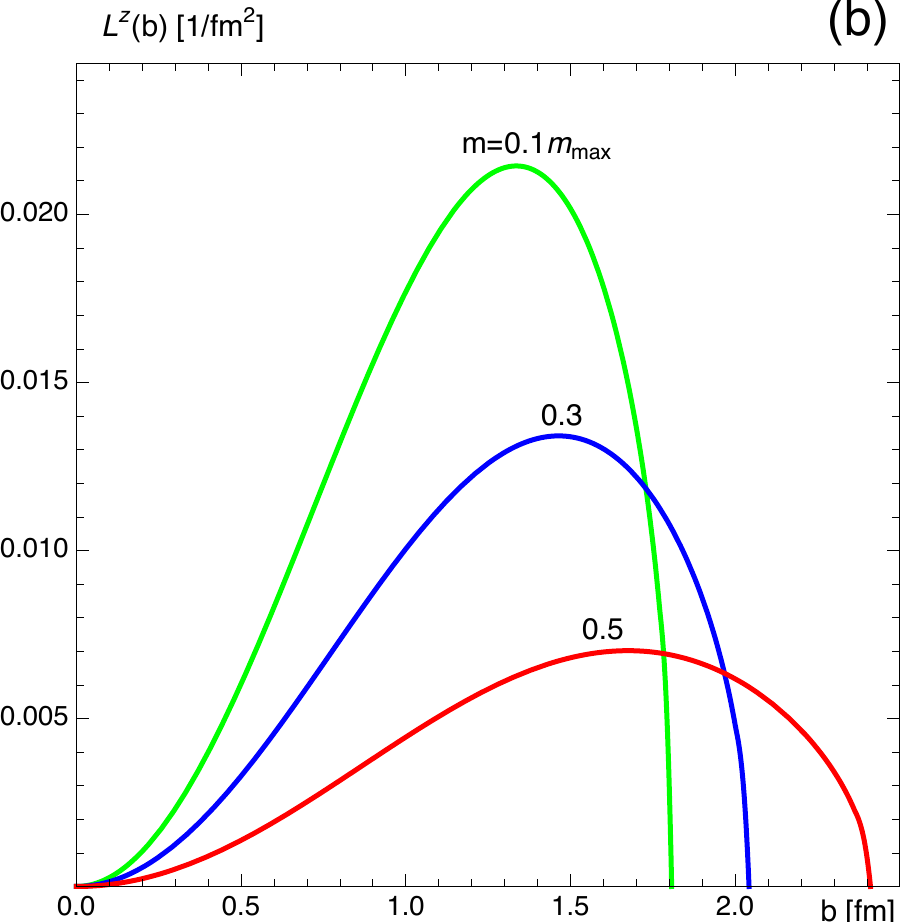} \hspace{4mm} \
\includegraphics[height=4cm]{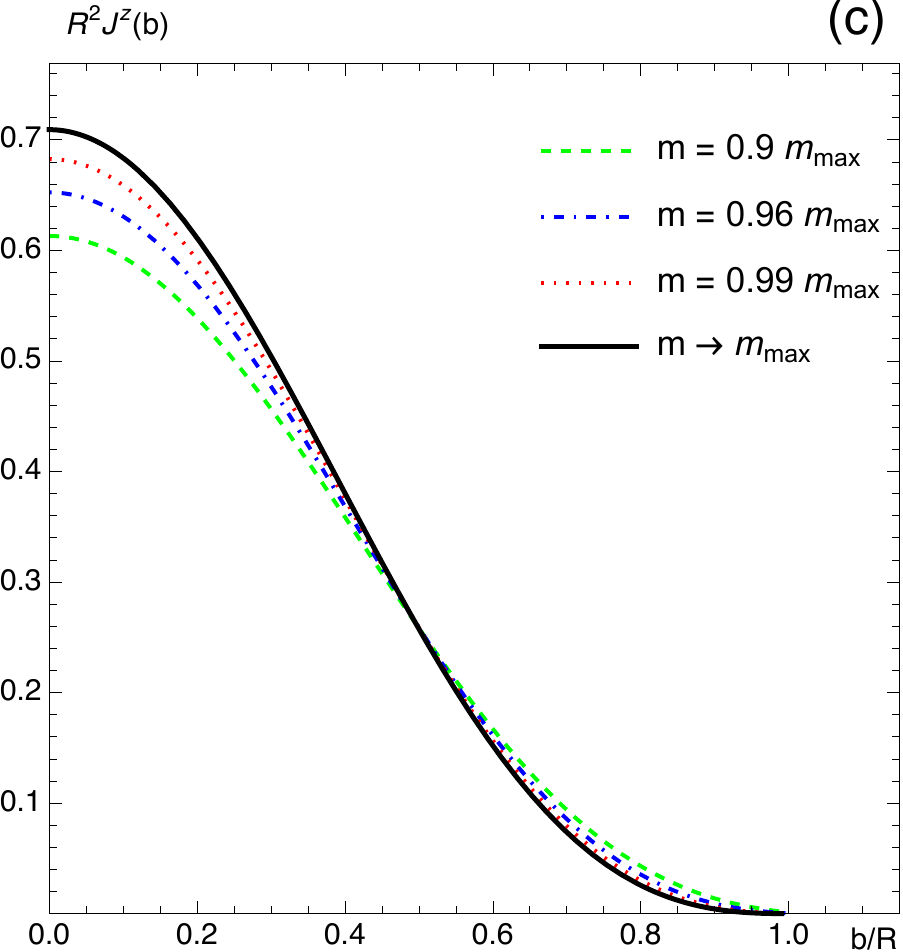} 
\end{centering}
\caption{\label{Fig-12:AM-S-L-J-kinetic-non-relativistic} 
The 2D angular momentum distributions in the bag model for fixed nucleon mass $M_N$ and increasing quark masses $m$
(a) Intrinsic spin $S^z(b)$ distributions. 
(b) Orbital angular momentum $L^z(b)$ distributions.
(c) The scaling of $R^2 J^z(b)$ for $R\to\infty$.}

\vspace{7mm}


\begin{centering}
\includegraphics[height=4cm]{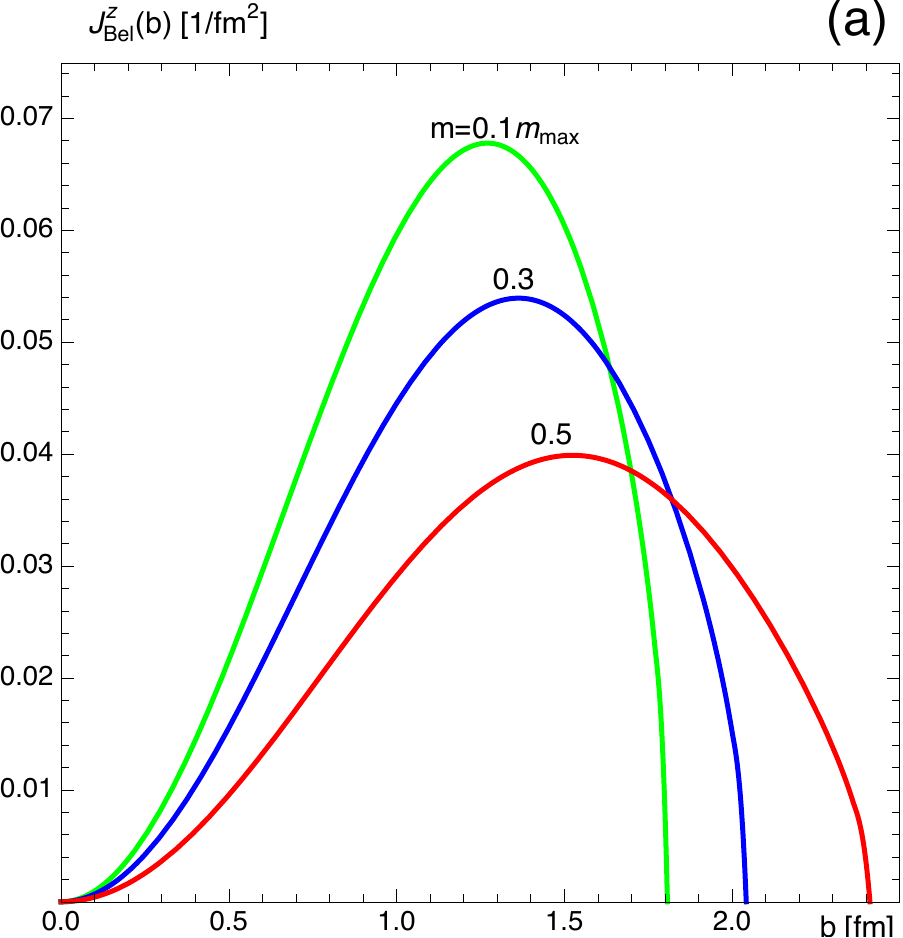} \hspace{4mm} \
\includegraphics[height=4cm]{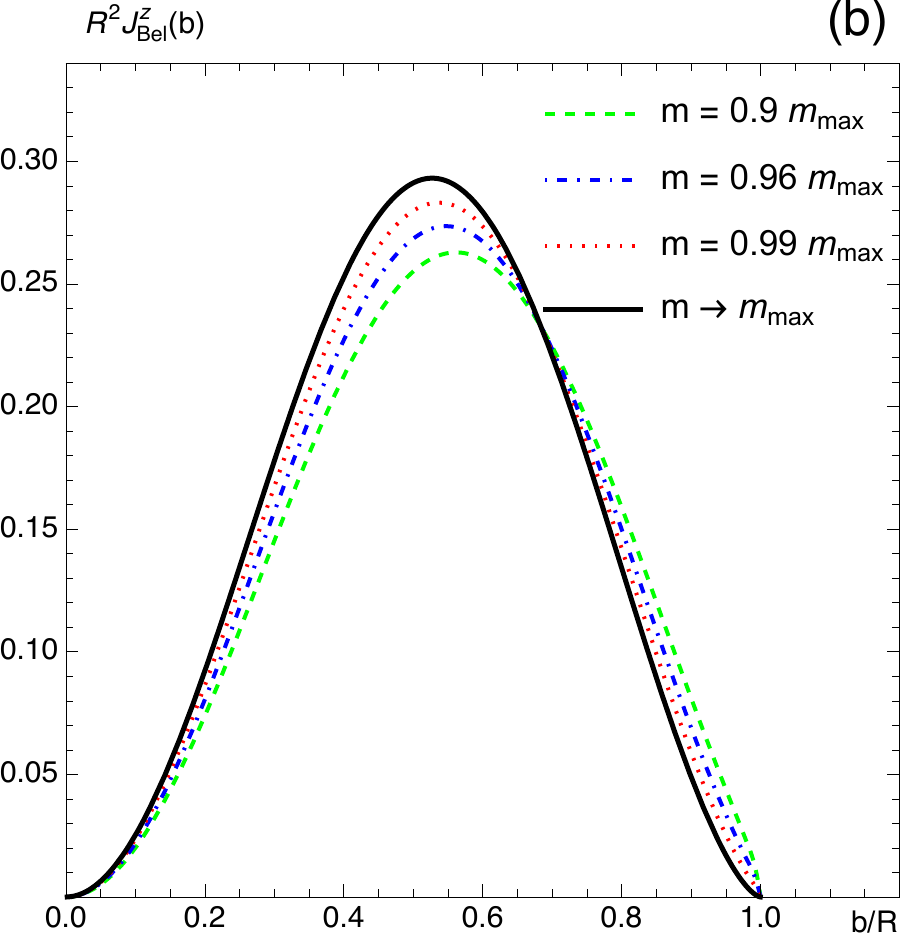} \hspace{4mm} \
\includegraphics[height=4cm]{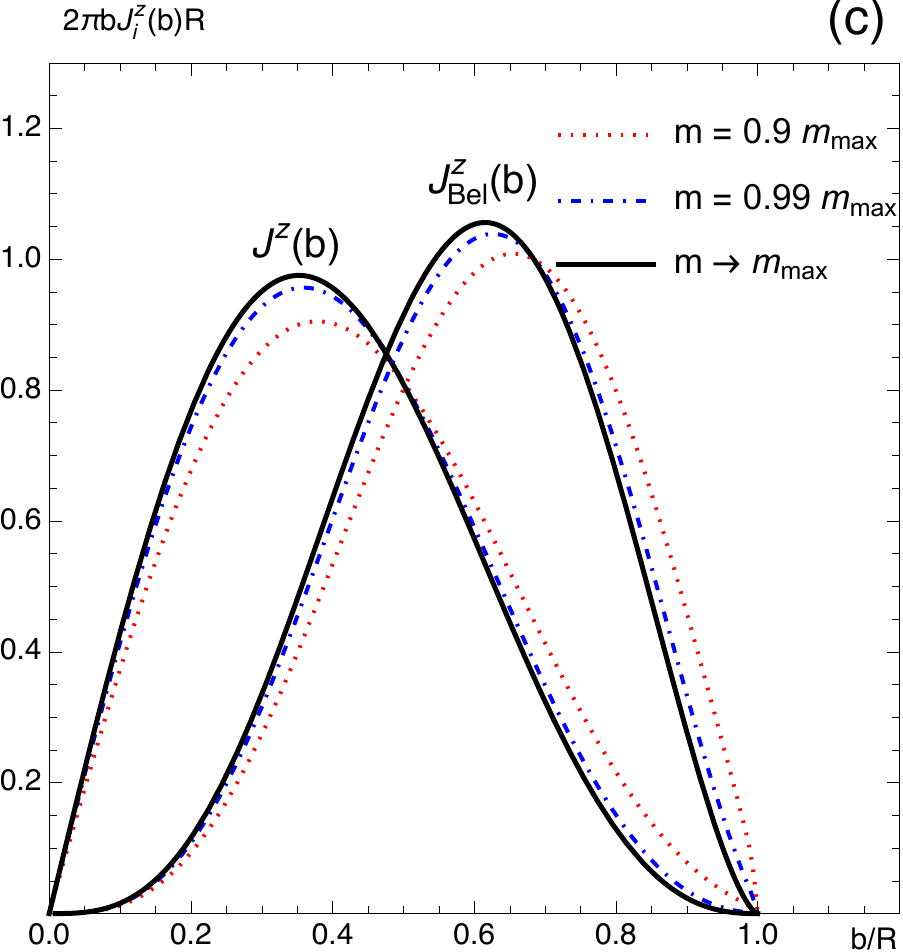} 
\end{centering}
\caption{\label{Fig-13:AM-J-Belinfante-non-relativistic} 
The 2D angular momentum distribution in the bag model for
fixed nucleon mass $M_N$ and increasing quark masses $m$
(a) Belinfante-form $J_\text{Bel}^z(b)$ angular momentum distributions. 
(b) The scaling of $R^2 J_\text{Bel}^z(b)$ for $R\to\infty$.
(c) The scaling of $2\pi b J_\text{i}^z(b) R$ for $R\to\infty$.}
\end{figure}

Next, we discuss our results regarding the 2D force distributions. In Figs.~\ref{Fig-11:non-relativistic-s-p}a and \ref{Fig-11:non-relativistic-s-p}c we depict the 2D shear force $s(b)$ and pressure $p(b)$ distributions for increasing values of $m=0.1,\;0.3,\;0.4\,\mx$. The figures illustrate how 2D force distributions decrease for $m\to\mx$. As the quark masses increase and constitute nearly the entire nucleon mass, the 2D force distributions scale as $R^{-4}$ which is to be contrasted with the $R^{-2}$ scaling of the energy distribution. Thus, the force distributions become much more dilute than the 2D energy distribution. This illustrates that the matter of the system is bound by weaker and weaker forces as the constituent quark limit is approached. This illustrates why the system size grows in this limit.   Figs.~\ref{Fig-11:non-relativistic-s-p}b and \ref{Fig-11:non-relativistic-s-p}d display the scaling behaviour of 2D force distributions in terms of the dimensionless quantities $s(b)/(BR)$ and $p(b)/(BR)$, respectively, for $m=0.9,\;0.96,\;0.99\,\mx,$ including the curves associated with $m \rightarrow \mx$.

Fig.~\ref{Fig-12:AM-S-L-J-kinetic-non-relativistic} shows how the 2D kinetic AM distributions behave in the constituent quark limit. The 2D intrinsic spin distributions $S^z(b)$ for quark masses $m=0.1,\;0.3,\;0.5\,\mx$ are shown in Fig.~\ref{Fig-12:AM-S-L-J-kinetic-non-relativistic}a, and the 2D OAM distributions $L^z(b)$ for the same quark masses are illustrated in Fig.~\ref{Fig-12:AM-S-L-J-kinetic-non-relativistic}b. The contribution of the two AM distributions to the total AM differs significantly by magnitude and the difference widens for growing $m$. As $m \to \mx$, the relative OAM contribution to the total AM approaches zero and the total AM is constituted solely by the intrinsic spin distribution. Finally, Fig.~\ref{Fig-12:AM-S-L-J-kinetic-non-relativistic}c displays the scaling of the dimensionless kinetic AM distribution $R^2 J^z(b)$ for increasing values of $m$ including the limiting curve associated with $m \rightarrow \mx$. 

In Fig.~\ref{Fig-13:AM-J-Belinfante-non-relativistic}a, the 2D Belinfante AM distribution is shown for selected values of $m=0.1,\;0.3,\;0.5\,\mx$, and in Fig.~\ref{Fig-13:AM-J-Belinfante-non-relativistic}b the dimensionless rescaled distribution $R^2J^z_{\rm Bel}(b)$ is shown as a function of $b/R$. Fig.~\ref{Fig-13:AM-J-Belinfante-non-relativistic}c compares the dimensionless rescaled kinetic and Belinfante AM distributions $2\pi b J^z(b) R$ and $2\pi b J^z_{\rm Bel}(b) R$, including the limiting curves associated with $m \rightarrow \mx$. Once again, the kinetic AM distribution is more skewed towards the bag center. In contrast, the Belinfante AM distribution shifts towards the bag boundary due to its orbital-like behavior.

\section{Mass decomposition in the bag model} \label{Sec:M-decompose}

The decomposition of the nucleon mass in QCD into 
contributions from quarks and gluons has attracted a 
lot of attention in the recent literature~\cite{Ji:1994av,Ji:1995sv,Lorce:2017xzd,Hatta:2018sqd,Metz:2020vxd,Ji:2021mtz,Lorce:2021xku}. It is interesting 
to address this question in a quark model framework
where technical difficulties due to quantum anomalies
do not occur.

Let us introduce the notation $\braket{O}=
\la N|\sum_q\int d^3r\,\psi_q^\dag{O}\psi_q|N\rangle$ for the
expectation value of a Dirac operator $O$ in the 
nucleon states in the rest frame, and consider
the quark Dirac Hamiltonian 
\be
    H_q=\vec{\alpha}\cdot\vec{p}+\gamma^0m
\ee
which we express in momentum space. In this notation and
considering the bag contribution (due to ``gluons''), 
the nucleon mass can be decomposed in the bag model
into three terms as 
\begin{equation}
\label{App:mass-decompose-1}
    M_N = \braket{\vec{\alpha}\cdot\vec{p}} 
      + \braket{\gamma^0 m} 
      + \frac{4}{3} \pi R^3\,B \, .
\end{equation}
The first term in (\ref{App:mass-decompose-1})
is the kinetic energy of the quarks inside the 
nucleon, and is given by 
\begin{equation}
\label{App:mass-decompose-2}
    E_{\rm kin} =
    \braket{\vec{\alpha}\cdot\vec{p}} = \frac{N_cA^2}{4\pi} \,\alpha_{+}\alpha_{-} \int d^3r \bigg[ j_0 j_1^\prime - j_0^\prime j_1 + \frac{2 j_0 j_1}{r}\bigg]\,
    \Theta_V\,.
\end{equation}
The second term is the quark mass contribution to the 
nucleon mass 
\ba
\label{App:mass-decompose-3}
    E_{\rm mass} &=& 
    \braket{\gamma^0 m} = m \; \frac{N_cA^2}{4\pi} 
    \int d^3r \bigg[ \alpha_{+}^2 j_0^2 - \alpha_{-}^2 j_1^2 \bigg]\, 
    \Theta_V\, ,
\ea
and the last term is the volume contribution 
from the bag vacuum energy
\begin{equation}
\label{App:mass-decompose-bag}
    E_{\rm bag} = \frac43\,\pi R^3\,B\,.
\end{equation}

It is worth noticing that the quark kinetic energy
in Eq.~(\ref{App:mass-decompose-2}) is exactly 3 times
the quark contribution to the volume integral over
the 3D pressure, see
Eq.~(\ref{Eq:3D-pressure}), where the factor
3 is the space dimension, i.e.\ we have 
\be
     \braket{\vec{\alpha}\cdot\vec{p}} =
     3\int d^3r \,
     p_q(r)\,
\ee
with $p_q(r)$ defined in Eq.~(\ref{Eq:3D-pressure}).
The term
$\int d^3r \,p_q(r)$
can be viewed as the pressure-volume work of quarks 
analogous to $PV$ in thermodynamics.
It is not accidental that the quark contribution
to the pressure makes an appearance in the mass 
decomposition. The deeper reason for that is the
connection between the von Laue
condition~\eqref{Eq:von-Laue-3} and virial
theorem~\eqref{virial-massive-MR}, which are equivalent
in the bag model~\cite{Neubelt:2019sou} and in other models 
like chiral quark-soliton model~\cite{Goeke:2007fp}, Skyrme 
model~\cite{Cebulla:2007ei} or $Q$-balls \cite{Mai:2012yc}.
Notice that $\la \gamma^0 m\ra$
in Eq.~\eqref{App:mass-decompose-3} is related to the
pion-nucleon sigma term and the sum rule
$\sum_q m_q\int dx\,e^q(x)$, where $e^q(x)$ is a twist-3 
parton distribution function~\cite{Lorce:2014hxa}
(recall that we use $m=m_u=m_d$ and neglect isospin 
violating effects in this work). 

We first focus on the case $m=0$ where 
$E_{\rm kin} = N_c\,\omega_0/R$ and obviously 
$E_{\rm mass}=0$.
Keeping the number of space dimensions $n$ general, the 
nucleon mass is $M_N(R) = N_c\,\omega_0/R + b_n\,R^nB$
where 
$b_n=2\pi^{n/2}/\Gamma(n/2)$.
The virial theorem~\eqref{virial-massive-MR} corresponds to
$M'_N(R)=0$ and yields $N_c\,\omega_0 = n\,b_n\,R^{n+1}B$ 
implying that for massless quarks $E_{\rm kin} = n\,E_{\rm bag}$.
Thus, in the physical situation in $n=3$ space dimensions,
3/4 of the nucleon mass is due to the quark kinetic energy
and 1/4 is due to the bag contribution which is a crude 
model for gluonic effects. In QCD such decompositions are
scale dependent, and the above decomposition of the nucleon
is valid at a low hadronic scale $\mu_0< 1\,\rm GeV$ associated
with the  bag model.
This relation is often used to eliminate the bag contribution
and express the nucleon mass in the bag model as 
$M_N = 4\omega_0/R$ for $N_c=3$ colors and $n=3$ 
space dimensions \cite{Yuan:2003wk}.

When $m\neq0$ the situation is different. 
Evaluating the integrals in 
Eqs.~(\ref{App:mass-decompose-2},~\ref{App:mass-decompose-3}) 
yields lengthy expressions for $E_{\rm kin}$ and 
$E_{\rm mass}$ which, making use of the transcendental
equation~\eqref{Eq:omega-transcendental-eq}, can be
rewritten as 
\sub{\label{App:mass-decompose}\ba
\label{App:mass-decompose-2a}
    E_{\rm kin} &=& 
    \frac{2(\Omega_0-1)\omega_0^2}
    {2\Omega_0(\Omega_0-1)+m R}\;
    \frac{N_c}{R}\;,\\
\label{App:mass-decompose-3a}
    E_{\rm mass} 
    &=&
    \frac{2(\Omega_0-1)mR+ \Omega_0}
    {2\Omega_0(\Omega_0-1)+mR}\;
    N_c\,m\,.
\ea}
The kinetic and mass
contributions to the nucleon mass add up to
\be\label{Eq:Equark}
    E_q
    = E_{\rm kin} + E_{\rm mass}
    = \braket{\vec{\alpha}\cdot\vec{p}} + \braket{\gamma^0 m} 
    = \frac{N_c\,\Omega_0}{R}
    = \braket{H_q}\,,
\ee
i.e.\ to the total quark contribution to the nucleon mass,
$E_q$, which corresponds to the expectation value of the
quark Hamiltonian operator $H_q$.

\begin{table}[t!]
\begin{tabular}{clrrrrcc}
\hline
\hline
    \hspace{5mm} situation \hspace{5mm}
    & parameters 
    & \ \ $M_N$/MeV \
    & \ $E_{\rm kin}$/MeV \
    & \ $E_{\rm bag}$/MeV \
    & \ $E_{\rm mass}$/MeV \ 
    & \ $E_{\rm kin}:E_{\rm bag}:E_{\rm mass}$ & \spacer\\
    \hline
      physical    
    & $R=1.72\,{\rm fm}$, $m=5\,{\rm MeV}$
    & 938.272  
    & 698.233
    & 233.744
    &   7.295
    & $3:1:0.031$ & \spacer\\
      limit L1
    & $R=1.29 \,\rm fm$, $m=2\,\rm GeV$
    & 6257.562
    & 299.615
    &  99.872
    & 5858.075
    & $3:1:58.656\!\!$ & \spacer \\
    limit L2
    & $R = 1000\,\rm fm$,  
      $m=5\,\rm MeV$
    & 15.182
    & 0.216
    & 0.0719
    & 14.895
    & $3:1:207.225\!\!\!\!\!$
    & \spacer \\
      limit L3 
    & $R = {2.70}\,\rm fm$, $m = 0.6\, \mx$
    & 938.272
    & 375.342
    & 125.114
    & 437.817
    & $3:1:3.499$ & \spacer \\
\hline
\hline
\end{tabular}

\caption{\label{Tab2} 
The mass decomposition in the bag model in 
the physical situation and for selected values
as encountered in the limits L1, L2, L3.
The respective parameters $m$, $R$, $M_N$ 
and individual contributions
$E_{\rm kin},\;E_{\rm bag},\;E_{\rm mass}$ are
listed along with the relative partitioning 
$E_{\rm kin}:E_{\rm bag}:E_{\rm mass}$ with
the bag energy as reference point.
The ratio $E_{\rm kin}:E_{\rm bag}$
is equal to $3:1$ exactly in all cases.
Recall that $\mx$ is defined as one third 
of the physical proton mass 938.272$\,$MeV.}
\end{table}


In Table~\ref{Tab2} we show the nucleon mass decomposition 
in the physical situation, and for selected examples from 
the limits L1, L2, L3. Interestingly, the relative ratio 
$E_{\rm kin}:E_{\rm bag} = 3:1$ remains valid 
(in 3 space dimensions) not only in the massless case as
discussed above, but for any $m$ which is non-trivial.
When $m\neq 0$ it is important to keep in mind that the 
quark energy $E_q = N_c\Omega_0/R$ in Eq.~(\ref{Eq:Equark})
depends for $m\neq 0$ on $R$ also through 
$\Omega_0=\sqrt{\omega_0^2+(mR)^2}$, where $\omega_0=\omega_0(mR)$
is an implicit function of $R$ due to 
Eq.~(\ref{Eq:omega-transcendental-eq}). Noting that the
variation of $\omega_0$ with respect to $R$ can be expressed 
as \cite{Neubelt:2019sou}
\be
    \frac{\partial\omega_0}{\partial R}
    = \frac{m\omega_0}{2\Omega_0(\Omega_0-1)+mR}\,,
\ee
we obtain the remarkable identity 
\be\label{Eq:identity-Eq-Ekin}
      \frac{\partial E_q}{\partial R}
    = \frac{\partial}{\partial R}
      \biggl(\frac{N_c\Omega_0}{R}\biggr) 
    = - \frac{E_{\rm kin}}{R}\,,
\ee    
i.e.\ in the bag model the variation of the total quark 
energy $E_q=E_{\rm kin}+E_{\rm mass}$ with respect to $R$
is simply related to the quark kinetic energy. Equipped
with the identity (\ref{Eq:identity-Eq-Ekin}) we can 
express the virial theorem (\ref{virial-massive-MR}) as 
\be
     E_{\rm kin} = 3\,E_{\rm bag}\,
\ee
which holds for any $m$. However, as illustrated 
Table~\ref{Tab2} this is only the relative partition of 
the quark kinetic and bag energy. For $m\neq0$
in addition the mass term $E_{\rm mass}$ enters
whose contribution is not given by a simple ratio. 

\begin{figure}[t]
  \begin{center}
  \includegraphics[height=6.0cm]{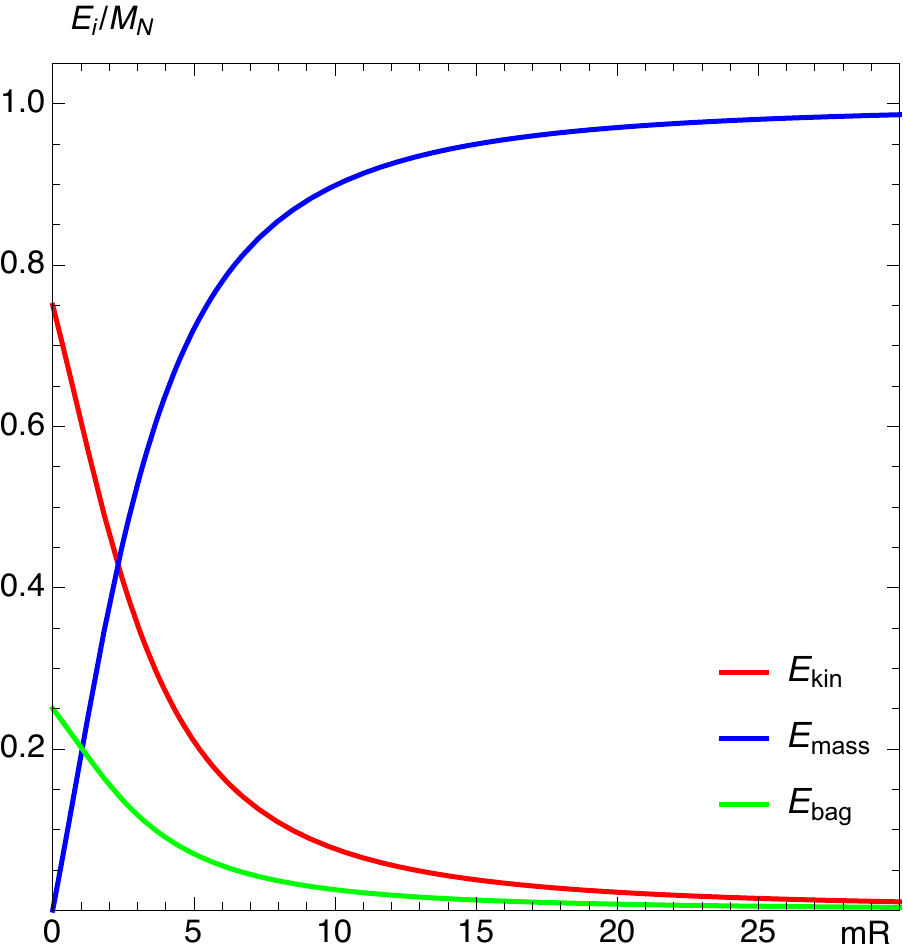} 
  \end{center}
\caption{\label{Fig-14:mass-decomposition} 
Nucleon mass decomposition in the bag model. Shown
are the relative contributions of $E_{\rm kin}/M_N$, 
$E_{\rm bag}/M_N$, $E_{\rm mass}/M_N$ as 
functions of $mR$.}
\end{figure}

The Table~\ref{Tab2} illustrates that one deals
with much different nucleon mass decompositions in the 
different limits. This is not surprizing because,
as explained in Sec.~\ref{sec:limits-overview}, the three
limits correspond to different physical situations. The three 
limits have in common that $E_{\rm mass}$ contributes
for $mR\to\infty$
asymptotically $100\,\%$ of the nucleon mass, while the
contributions of $E_{\rm kin}$ and $E_{\rm bag}$ vanish. 
But the underlying physics is much different. In fact,
in each case we ``start'' with the physical nucleon mass,
but we end up asymptotically at very different values 
for $M_N$, namely (cf.\ Table~\ref{Tab1})
\begin{itemize}
\item in L1 ($m\to\infty$, $B$ fixed): \ 
$M_N\to N_c m\to\infty$,

\item
in L2 ($R\to\infty$, $m=5\,{\rm MeV}$ fixed): \ 
$M_N\to N_c  m = 15\,{\rm MeV}$,

\item
in L3 ($m\to \mx = M_N/3=\;$fixed): \
$M_N=938\,$MeV is fixed at its physical value. 
\end{itemize}
Considering the different physical situations, 
it is remarkable that the {\it relative} contributions 
to the nucleon mass defined as 
$E_{\rm kin}/M_N$, $E_{\rm bag}/M_N$, $E_{\rm mass}/M_N$
and plotted as function of $mR$ (which in all limits goes 
to infinity, albeit for different reasons), 
all coincide and are described by
universal curves in Fig.~\ref{Fig-14:mass-decomposition}.

$E_{\rm kin}/M_N$ and $E_{\rm bag}/M_N$ assume respectively
the values 3/4 and 1/4 at $mR=0$, and are monotonically 
decreasing. They go to zero for $mR\to\infty$ satisfying 
$E_{\rm kin}:E_{\rm bag} =3:1$ at each value of $mR$.
The mass contribution is zero at $mR=0$, and
$E_{\rm mass}/M_N$ is monotonically increasing for 
finite $mR$ approaching $100\,\%$ as $mR\to\infty$. 
When $mR\approx 1$ the relative contributions of the
bag energy and the mass term become equal.
When $mR\approx 2.3$ the relative contribution of the
mass term catches up to that of the quark kinetic
energy, and becomes the dominant contribution beyond
that.

This was the nucleon mass decomposition in the bag model 
as based on the bag energy and two quark contributions 
in the Hamiltonian, namely quark kinetic energy
$\braket{\vec{\alpha}\cdot\vec{p}}$ and quark 
mass term $\la\gamma^0m\ra$.
In literature, it was proposed~\cite{Ji:1994av,Ji:1995sv,Ji:2021mtz} 
that the nucleon mass should be decomposed 
in terms of the trace part (rank-0 scalar operator, contributing~1/4) 
and the traceless part of the EMT (rank-2 tensor, 
contributing~3/4 to the nucleon mass), i.e.\
\be
      T^{\mu\nu} 
      = \frac14\,g^{\mu\nu} {T^\alpha}_\alpha +
      T^{\mu\nu}_{\rm traceless} \,,    
\ee
with the latter simply defined as
$T^{\mu\nu}_{\rm traceless}=T^{\mu\nu}-\frac14\,g^{\mu\nu} 
{T^\alpha}_\alpha$. In QCD, such a decomposition is natural. 
For instance, the trace part receives a contribution from 
the trace anomaly and is twist-4, while the traceless part 
of the EMT is related to matrix elements of twist-2 operators whose 
quark and gluon contributions are constrained by information 
on parton distribution functions from deep-inelastic scattering 
experiments. One obtains a nucleon mass decomposition based on 
contributions from the trace part $\frac14\,g^{00} {T^\alpha}_\alpha$ 
and the traceless part 
$T^{00}_{\rm traceless}$~\cite{Ji:1994av,Ji:1995sv,Ji:2021mtz}.

In the bag model, the situation is simpler as there is no 
trace anomaly, and all matrix elements of the EMT are explicitly 
known, see Eq.~\eqref{Eq:EMT-kin}. The trace contributes to
the nucleon mass the portion
\be
    M_{N,\, \rm trace} 
    = \frac14\,g^{00} \int d^3r\, {T^\alpha}_\alpha(\vec{r}) 
    = \frac14\,\int d^3r\, 
      \biggl(T^{00}_q+T^{00}_{\rm bag}-3\,p(r)\biggr) 
    = \frac14\,\biggl(\frac{N_c\Omega_0}{R}+\frac43\,\pi R^3B\biggr)
    = \frac14\,M_N,
\ee
where we used the von Laue condition, Eq.~(\ref{Eq:von-Laue-3}).
The contribution from the traceless part is 
\be
    M_{N,\, \rm traceless} 
    = \int d^3r\, {T^{00}_{\rm traceless}(\vec{r})}
    = \frac34\,\int d^3r\, 
      \biggl(T^{00}_q+T^{00}_{\rm bag}+p(r)\biggr) 
    = \frac34\,\biggl(\frac{N_c\Omega_0}{R}+\frac43\,\pi R^3B\biggr)
    = \frac34\,M_N
\ee
using again the von Laue condition. 
While it is correct, one does not gain much insight from considering the trace and traceless parts separately. This is consistent with the general discussion of Refs.~\cite{Lorce:2017xzd,Lorce:2021xku}.


\newpage
\section{Conclusions}
\label{Sec:conclusions}

This work was dedicated to the study of 2D energy-momentum tensor (EMT) distributions of the nucleon.
We have obtained several general results, and presented 
results from the quark model calculations in the bag model.
Among the model-independent results are explicit proofs of 
several conditions for 2D EMT distributions based on 
mechanical stability criteria. 
Another important model-independent result is the demonstration that
the different definitions of 2D EMT distributions in the Breit,
elastic and infinite-momentum frames coincide in the large-$N_c$ limit for a longitudinally polarized nucleon. 
(For AM distributions in a transversely polarized nucleon this
is not the case, due to a trivial contribution from the
center-of-mass motion.)

We then employed the bag model formulated in the large-$N_c$
limit to study these 2D EMT distributions. 
The large-$N_c$ limit is important for the 3D interpretation
EMT distributions~\cite{Goeke:2007fp} and to make calculations of EMT 
form factors in the bag model justified~\cite{Neubelt:2019sou}.
We have presented numerical results for the 2D EMT distributions,
and demonstrated the consistency of the model description.
In the physical situation, for which we chose to use a current
quark mass of 5$\,$MeV and bag radius of 1.7$\,$fm,
the distributions of mass and electric charge in the proton
resemble each other. The 2D pressure distribution obeys 
the pertinent von Laue condition, and the kinetic AM 
is dominated by the intrinsic spin contribution which
contributes 66$\,\%$ of the nucleon spin, with the
remaining 34$\,\%$ being due to orbital angular momentum (OAM). 

We then studied the EMT distributions in three different
limits, which helps deepen our understanding of the 2D
structure of the nucleon. In the ``heavy-quark 
limit'' limit L1, we 
increased the quark mass $m\to\infty$ while
keeping the strength of the strong forces (mimicked by the bag
constant $B$) fixed. In this limit the nucleon mass grows like 
$M_N\to N_cm\to\infty$ while the nucleon size shrinks, which 
implies, for instance, an increase of the 2D energy distribution.
In the large system size limit L2, we kept the mass of the 
quarks fixed at $5\,\rm MeV$ and gave them a larger and larger 
volume to fill out by taking the bag radius $R\to\infty$.
All EMT distributions become diluted in this limit which
is supported by numerical results. As $R\to\infty$
with $m=5\,\rm MeV$ fixed, the nucleon mass goes to
$N_cm = 15\,\rm MeV$. The forces encoded in the bag
constant decrease like $B\sim R^{-5}$, which implies
for the 2D distributions $s(b)$ and $p(b)$ a scaling
of the type $R^{-4}$. 
In the constituent-quark limit L3, we let the quark mass
approach $M_N/N_c$ while the nucleon mass $M_N$ was kept 
at its physical value. Thus, this limit creates a situation
where the nucleon mass is nearly entirely due to the masses of the
quarks. By taking $m\to M_N/N_c$ drives the bag radius to
become larger and $B$ to decrease. Both limits L2 and L3 
belong to a class of ``weak-binding limits''. Even
though the binding forces decrease, the quarks remain
always confined in the bag model.

In all three limits, one effectively deals with 
non-relativistic dynamics. Also the distinction between the 
energy and the electric charge distributions becomes less and less apparent.
Asymptotically we have $\epsilon(b)/M_N=\rho_{\rm ch}(b)$
in the three limits, i.e.\ the mass and electric charge 
in the proton are distributed in exactly the same way.
Another interesting observation is that in all three limits 
the quark OAM becomes negligible compared to the intrinsic 
spin distribution. 
The kinetic AM (defined in terms of the asymmetric EMT) and the Belinfante AM (associated with the symmetric part
of the EMT) have significantly different shapes, even though
both consistently integrate to the value 1/2 for the nucleon
spin. The difference has two different origins, namely
(i) a quadrupole contribution which is present in 3D as well
as in 2D Belinfante AM but not in the kinetic AM, and (ii)
a total derivative term. The characteristic difference of 
these two AM distributions is not only present in the 
physical situation, but persists in all considered limits.

We have also studied the mass decomposition. In the bag model,
one can unambiguously define three contributions to the nucleon
mass, namely due to (i) quark kinetic energy 
$E_{\rm kin}=\la\vec{\alpha}\cdot\vec{p}\ra$, (ii) quark mass
$E_{\rm mass}=\la\gamma^0m\ra$, and (iii) bag energy 
$E_{\rm bag}=\frac43\,\pi R^3B$ which simulates the
confining effects of gluons within the bag model. 
We showed that the ratio of quark
kinetic energy to bag energy is $3:1$ independently of the
quark mass. This is the case in the physical situation, and
in the limits. Another interesting insight is that the 
relative mass decompositions
$E_{\rm kin}/M_N$, $E_{\rm mass}/M_N$, $E_{\rm bag}/M_N$
as functions of the product $mR$ are described by the same universal 
curves in all three limits. This is remarkable considering
the different physical situations in the three limits.
Finally we note that starting from the EMT distributions, the contributions to the mass do not separate naturally in the bag model into quark mass and kinetic terms.
Rather one directly encounters a decomposition into two terms,
the bag energy and total quark energy. The latter can of course
be further decomposed into the kinetic energy and mass term of 
quarks, but this requires the evaluations of the expectation
values of the separate operators $\vec{\alpha}\cdot\vec{p}$ and $\gamma^0m$
in the Dirac Hamiltonian.

We hope our study will stimulate further model
investigations of 2D EMT distributions. One interesting and
natural extension of this work could be the consideration 
of effects due to chiral symmetry as modelled e.g.\ in the
cloudy bag model~\cite{Owa:2021hnj} similarly to what has
been done in the chiral quark-soliton model~\cite{Kim:2021jjf}.
As illustrated by the present work, the studies in models play an important role for the 
understanding and interpretation of the nucleon structure.

\ \\
{\bf Acknowledgments.}
C.L.~and P.S.~acknowledge the uncountable discussions with Maxim Polyakov from the very beginning of their careers, and mourn the loss of a great colleague, mentor, and friend. The work of P.S.~was supported by the National Science Foundation under the Awards No.~1812423
and 2111490.
The work of K.T. was supported by the U.S. Department of Energy under Contract No. de-sc0012704.


\appendix
\section{Stability requirements for 2D BF distributions}\label{App-2D-constraints}

\noindent 

In this Appendix we provide the detailed proofs of the
stability requirements for the 2D EMT distributions in 
the BF discussed in Sec.~\ref{Sec-IID}. In this 
Appendix we do not work in any specific limit, 
e.g.\ the number of colors $N_c$ is finite, 
and the proofs are general and model-independent.

The 3D EMT distributions satisfy certain criteria which are
necessary (but not sufficient) requirements for mechanical
stability. In particular, in a 3D stable system, the following
conditions are expected~\cite{Lorce:2018egm} 
\begin{enumerate}
\item $\epsilon(r)|_{r=0} < \infty$, $p(r)|_{r=0} < \infty$ and $s(r)|_{r=0} = 0$ , 
\item $\frac{d\epsilon(r)}{dr} < 0$ and $\frac{dp_r(r)}{dr} < 0$ , 
\item $\epsilon(r) > 0$ and $p_r(r) > 0$ ,
\item (Null Energy Condition) $\epsilon(r) + p_i(r) \geq 0$ ,
\item (Weak Energy Condition) $\epsilon(r) + p_i(r) \geq 0$ and $\epsilon(r) \geq 0$ ,
\item (Strong Energy Condition) $\epsilon(r) + p_i(r) \geq 0$ and $\epsilon(r) + 3\,p(r)\geq 0$ ,
\item (Dominant Energy Condition) $\epsilon(r) \geq |p_i(r)|$ where $i=r,t$.
\end{enumerate} 

Owing to Eq.~\eqref{Eq:EMT-2D-Breit-frame}, 
analogous conditions exist for the 2D EMT distributions
in the BF. Some of these conditions were
mentioned in the main text in Sec.~\ref{Sec-IID}.
Below we will state all conditions and provide
explicit proofs that if the corresponding 3D condition 
is true, then also its 2D counterpart is true.
To the best of our knowledge, these 2D  
conditions and their proofs 
have not been discussed explicitly in literature before 
and will be presented and proven below for the first time.

The above-stated 3D stability conditions 
can be translated into 2D stability conditions as follows: \\
\begin{enumerate}
\item $\epsilon(b)|_{b=0} < \infty$, $p(b)|_{b=0} < \infty$ and $s(b)|_{b=0} = 0$. 
\\ \\
Proof: Let us write $\epsilon(b)$ as 
\be
  \epsilon(b) = \int_{-\infty}^{\infty} dz \, \epsilon(r) = 2\int_b^{\infty} dr \, \frac{r}{\sqrt{r^2-b^2}}\,\epsilon(r)\, .
\ee
Then, at $b=0$ we get $\epsilon(b)|_{b=0} = 2\int_0^{\infty} dr\, \epsilon(r) < \infty$ 
where it is clear that the integral is finite because
$M_N=\int d^3r\, \epsilon(r)$ is finite.
Similarly,
\ba
  p(b) & = & \int_{-\infty}^{\infty} dz \, \bigg[p(r) + \frac{b^2 - 2 z^2}{6 r^2}\,s(r) \bigg] \nonumber \\
  & = & 2\int_b^{\infty} dr \, \frac{r}{\sqrt{r^2-b^2}}\, \bigg[p(r) + \frac{3b^2 - 2 r^2}{6 r^2}\,s(r) \bigg].
\ea 
At $b=0$, the expression yields $p(b)|_{b=0} = 2\int_0^{\infty} dr \big[ p(r) - \frac13 \,s(r) \big]$. Therefore, by using the 1D von Laue stability condition Eq.~\eqref{Eq:von-Laue-1}, we get
\be
  p(b)|_{b=0} = 2\int_0^\infty dr\, s(r) = 2\gamma < \infty.
\ee 
Finally, $s(b)|_{b=0} = 0$ is satisfied by the definition of $s(b) = \int_{-\infty}^\infty dz \, \frac{b^2}{r^2}\,s(r)$.


\item $\frac{d\epsilon(b)}{db} \leq 0$ and $\frac{dp_r(b)}{db} \leq 0$. \\\\
Proof: First, let us suppose $\frac{d\epsilon(r)}{dr} < 0$. Then
\be
  \frac{d\epsilon(b)}{db} = 2\int_b^\infty dr \, \frac{b}{\sqrt{r^2-b^2}} \, \frac{d\epsilon(r)}{dr} \leq 0.
\ee
Similarly by using the equation $\frac{dp_r(b)}{db} = -\frac{s(b)}{b}$, as given in~\cite{Lorce:2018egm}, 
we get
\be
  \frac{dp_r(b)}{db} = -\frac{2}{b}\int_b^\infty dr\, \frac{b^2}{r\sqrt{r^2-b^2}}\underbrace{s(r)}_{>0} \leq 0 \, ,
\ee
where we used the equation $\frac{dp_r(r)}{dr} = -\frac{2s(r)}{r}$ and the 3D stability condition $\frac{dp_r(r)}{dr} < 0$ to determine the sign of $s(r)$. 
\item\label{2D-E-RP-positivity} $\epsilon(b) \geq 0$ and $p_r(b) \geq 0$. 

\ \\
Proof: Suppose $\epsilon(r) > 0$, then
\be
  \epsilon(b) = \int_{-\infty}^{\infty} dz \, \epsilon(r) = 2\int_b^{\infty} dr \,\frac{r}{\sqrt{r^2-b^2}} \, \epsilon(r) \geq 0.  
\ee
Next, writing $p_r(b)$ in terms of 
\be
  p_r(b) = 2\int_b^\infty dr \,\frac{r}{\sqrt{r^2-b^2}} \big[ p_r(r) - \frac{r^2-b^2}{r^2}s(r) \big] \, ,
\ee
yields at $b=0$
\be
  p_r(b)|_{b=0} = 2\int_0^\infty dr \,\big[ p(r) - \frac{1}{3}s(r) \big]\, .
\ee
Then by using the 1D von Laue relation Eq.(\ref{Eq:von-Laue-1})
we conclude that 
\be
  p_r(b)|_{b=0} = 2\int_0^\infty dr \,s(r) = 2\gamma > 0.
\ee
On the other hand, 
$p_r(b)|_{b\rightarrow \infty} = 0$. Moreover, from the condition 2 above, we know that $\frac{dp_r(b)}{db} \leq 0$. As a result, we conclude that the radial pressure $p_r(b)$ decreases monotonically from $b=0$ to $b\to\infty$ and can only take non-negative values, i.e., $p_r(b)\geq 0$. 

\item\label{NEC-2D} (Null Energy Condition) $\epsilon(b) + p_i(b) \geq 0$. \\\\
Proof: 
First, by using the 2D condition \ref{2D-E-RP-positivity}, we conclude that $\epsilon(b) + p_r(b) \geq 0$. Next, 
let us suppose $\epsilon(r) + p_t(r) \geq 0$. Then  
\be
  \epsilon(b) + p_t(b) = \int_{-\infty}^{\infty} dz\, \epsilon(r) + \int_{-\infty}^{\infty} dz\, p_t(r) \geq 0.
\ee
\item (Weak Energy Condition) $\epsilon(b) + p_i(b) \geq 0$ and $\epsilon(b) \geq 0$. \\\\
Proof: This condition is satisfied as a result of the 2D conditions \ref{2D-E-RP-positivity} and \ref{NEC-2D}.
\item (Strong Energy Condition) $\epsilon(b) + p_i(b) \geq 0$ and $\epsilon(b) + 2\,p(b)\geq 0$. \\\\
Proof: Suppose $\epsilon(r) + p_t(r)\geq 0$. Then 
\ba
  \epsilon(b) + 2\,p(b) & = & \int_{-\infty}^{\infty} dz \,\bigg[ \underbrace{\epsilon(r) + 2\,p(r)}_{\geq\,p(r) + \frac13\,s(r)} + \frac{b^2-2z^2}{3r^2}\,s(r)\bigg] \nonumber \\
  & \geq & \int_{-\infty}^{\infty} dz\, \bigg[ p(r) + \frac13\,s(r) + \frac{b^2-2z^2}{3r^2}\,s(r) \bigg] = p_r(b).
\ea  
Since $p_r(b) \geq 0$, we get $\epsilon(b) + 2\,p(b)\geq 0$.
\item (Dominant Energy Condition) $\epsilon(b) \geq |p_i(b)|$. \\\\
Proof: First, let us suppose that $\epsilon(r) \geq |p_r(r)|$. Since $\epsilon(r) > 0$ and $p_r(r) > 0$, we get 
\be
  \int dz\, \epsilon(r) \geq \int dz\, |p_r(r)|.
\ee
On the other hand, by taking into account $p_r(b) \geq 0$ as well as $s(r) > 0$, we obtain 
\be
  \int dz\, p_r(r) \geq \int dz\, \big[ p_r(r) - \frac{z^2}{r^2} \,s(r) \big] = p_r(b).
\ee
Therefore
\be
  \epsilon(b) \geq |p_r(b)|.
\ee
The proof that $\epsilon(b) \geq |p_t(b)|$ follows directly from the definitions. Suppose 
$\epsilon(r) \geq |p_t(r)|$. Then 
\be
  \int dz\, \epsilon(r) \geq \int dz\, |p_t(r)| \geq \bigg| \int dz \,p_t(r) \bigg|.
\ee
Hence
\be
  \epsilon(b) \geq |p_t(b)|.
\ee
\end{enumerate}

\section{Relation of kinetic and Belinfante AM distributions}
\label{App:Jkin-vs-JBel}

In this section of the appendix, we explicitly show that the difference between the kinetic and Belinfante AM distributions is a total derivative which yields zero under the volume integral. From Eq.~\eqref{Eq:OAM-z-distribution} and Eq.~\eqref{Eq:Spin-z-distribution}, the total kinetic AM distribution reads
\be
    J^z(\vec{r}) = \frac{A^2}{8\pi}\,\bigg[ \alpha_{+}^2\,j_0^2 + \alpha_{-}^2\,j_1^2 \bigg]\,\Theta_V,
\ee
whereas the total Belinfante AM can be expressed as 
\be\label{Eq:App-JzBel}
    J^z_{\rm Bel}(\vec{r}) = \frac{A^2}{8\pi}\,\biggl[\frac{2\omega_0}{R}\,
    r\,j_0\,j_1+\alpha_-^2\,j_1^2\biggr]\,(1-\cos^2\theta)\,\Theta_V.
\ee
One can decompose the Belinfante AM distribution in terms of its monopole and quadrupole contributions by using the relation $(1-\cos^2\theta) = \frac{2}{3}\,P_0(\cos\theta) - \frac{2}{3}\,P_2(\cos\theta)$ as follows
\ba
    J^z_{\rm mono}(\vec{r}) &=& \frac{A^2}{12\pi}\,\biggl[\frac{2\omega_0}{R}\,
    r\,j_0\,j_1+\alpha_-^2\,j_1^2\biggr]\,\Theta_V\, , \\
    J^z_{\rm quad}(\vec{r}) &=& -\frac{A^2}{12\pi}\,\biggl[\frac{2\omega_0}{R}\,
    r\,j_0\,j_1+\alpha_-^2\,j_1^2\biggr]\,P_2(\cos\theta)\,\Theta_V\, .
\ea
The difference between the kinetic and Belinfante AM distributions can therefore be written as 
\be
    r^2\,(J^z-J^z_{\rm Bel})(\vec{r}) = \frac{A^2\,R^2}{24\,\omega_0^{2}\,\pi}\,\biggl[3\,\alpha_{+}^2\,x^2\,j_0^2(x) + \alpha_{-}^2\,x^2\,j_1^2(x) - 4\,x^3\,j_0(x)\,j_1(x)\biggr]\,\Theta_V - r^2\,J^z_{\rm quad}(\vec{r}),
\ee
where we defined a new variable $x = \omega_0\,r / R$. By using the spherical Bessel function relations $j^{\prime}_0(x) = -j_1(x)$ and $j^{\prime}_1(x) = j_0(x) - \frac{2}{x}\,j_1(x)$ one can express the difference in terms of a total derivative and a quadrupole term 
\be\label{Eq:App-II-6}
    r^2\,(J^z-J^z_{\rm Bel})(\vec{r}) = \frac{A^2\,R^2}{24\,\pi\,\omega_0^2}\,\frac{d}{dx}\,\biggl(x^3\left[\alpha_{+}^2\,j_0^2(x) - \alpha_{-}^2\,j_1^2(x)\right]\biggr)\,\Theta_V - r^2\,J^z_{\rm quad}(\vec{r}).
\ee
Under volume integration, the quadrupole term drops out, while
the contributions from the monopole terms in~\eqref{Eq:App-II-6} 
correspond to a total derivative with respect to $r$. The latter
evidently vanishes at the lower integration limit, and is proportional
to $\alpha_{+}^2\,j_0^2(\omega_0) - \alpha_{-}^2\,j_1^2(\omega_0)$ at the upper integration limit
which is zero due to the transcendental equation~\eqref{Eq:omega-transcendental-eq}.

\section{Axial form factors, intrinsic spin distribution, and proof of Eq.~(\ref{Eq:relation-AM-distributions})} 
\label{Appendix:Sz}

In this section of the appendix, let us first include for 
completeness the definition of the nucleon axial form factors 
and their relation to the 3D quark spin density 
(\ref{Eq:spin-density-def}) in the BF which are given by
\be\label{Eq:axial-FF-and-spin-distribution-NEW}
    \la p^\prime,\vec s^{\,\prime}| 
    \bar\psi_q(0)\gamma^i\gamma_5\psi_q(0)|p,\vec s\rangle 
    = \bar u(p^\prime,\vec s^{\,\prime})
    \biggl[ \gamma^i\gamma_5 \, G^q_A(t) + \, \frac{\Delta^i\gamma_5}{2M_N}\, G^q_P(t) 
    \biggr]u(p,\vec s)
    = 2P^0\int d^3r\;e^{i\vec{\Delta}\cdot\vec{r}}\,2S^i_q(\vec{r})\, .
\ee
Since it is defined in terms of two independent 
form factors, the monopole and quadrupole contributions to 
$S^i_q(\vec{r})$ are independent of each other as mentioned 
in Sec.~\ref{Subsec:3D-distributions-in-BF}.
This is in contrast to the other "orbital-like" angular
distributions related to a single form factor 
like, e.g., $J^i_{\text{Bel},q}(\vec{r})$ which is
defined solely in terms of $J_q(t)$.

Evaluating the bag model expression for the contribution of the quark flavor $q$ to the axial form factor in Eq.~(\ref{Eq:axial-FF-and-spin-distribution-NEW}) in the large-$N_c$ limit yields the result
\be\label{Eq:axial-form-factor} 
    G_A^q(t) = 4\pi A^2 R^6 \int \frac{d^3k}{(2\pi)^3} \Big[ \alpha^{2}_{+} \, t_0(k) \, t_0(k^\prime) - \alpha^{2}_{-} \, e^3_{k} \, e^3_{k^\prime}\, t_1(k) \, t_1(k^\prime)\Big] \, ,
\ee
where $\vec{k}^\prime = \vec{k} + \vec{\Delta}$ and $k=|\vec{k}|$, $k^\prime=|\vec{k}^\prime|$.
The $t_i(k)$ are defined in terms of Fourier transforms of the spherical Bessel functions in the bag \cite{Neubelt:2019sou}. 
The model expression for the form factor 
$S_q(t)$ was derived in the Appendix of 
Ref.~\cite{Neubelt:2019sou}. 
It is important to remark that in the bag model these two form factors  
satisfy the general relation\footnote{Notice the notations $2S_q(t)|_{\mbox{\tiny this work}} = -\,F^q_{\rm can}(t)|_{\mbox{\tiny Ref.\cite{Neubelt:2019sou}}} = -\,D_q(t)|_{\mbox{\tiny Ref.\cite{Lorce:2017wkb}}}$ for the form factor associated with the antisymmetric part of the kinetic EMT, while the $D$-term form factor is denoted as $D_q(t)|_{\mbox{\tiny this work, Ref.\cite{Neubelt:2019sou}}}= 4\,C_q(t)|_{\mbox{\tiny Ref.\cite{Lorce:2017wkb}}}$ and analogous for gluons.} \cite{Bakker:2004ib,Leader:2013jra,Lorce:2017wkb}
\be
    S_q(t) = \tfrac{1}{2}\,G_A^q(t)\,.
\ee
This is another consistency test of the model \cite{Neubelt:2019sou}.

To show that in the bag model the difference between the
kinetic and Belinfante AM can be expressed as the total derivative of the intrinsic spin distribution, let us 
first rewrite the right-hand side of Eq.~(\ref{Eq:relation-AM-distributions}) as
\begin{equation}
    \tfrac{1}{2}\nabla^j\!
    \left(r^j\,
    [S_q^i(\vec r)]_{s^\prime  s}-\delta^{ji}\,\vec r\cdot
    [\vec S_q(\vec r)]_{s^\prime  s}\right)= 
    [S^i_q(\vec r)]_{s^\prime  s} + \tfrac{1}{2} r^j\left(\nabla^j 
    [S^i_q(\vec r)]_{s^\prime  s} - \nabla^i 
    [S^j_q(\vec r)]_{s^\prime  s}\right)\, ,
\end{equation}
where we use the spin density notation for a nucleon polarized
along a general direction. In the main text, the AM distributions
are defined for a nucleon in a spin-up state with 
respect to a chosen polarization axis.
Then, Eq.~(\ref{Eq:relation-AM-distributions}) is equivalent to
\begin{equation}
\label{Eq:equivalent-relation-AM-distributions}
    [L^i_q(\vec r)]_{s^\prime  s} -
    [J^i_{{\rm Bel},q}(\vec r)]_{s^\prime  s} 
    =\tfrac{1}{2} r^j\biggl(
    \nabla^j [S^i_q(\vec r)]_{s^\prime  s} - 
    \nabla^i [S^j_q(\vec r)]_{s^\prime  s} 
    \biggr)\,.
\end{equation}
The evaluation of the  spin (\ref{Eq:spin-density-def},~\ref{Eq:axial-FF-and-spin-distribution-NEW}), OAM (\ref{OAMdef}a), and Belinfante AM (\ref{OAMdef}c) quark distributions for a nucleon polarized along an arbitrary $i$-direction yields in the large-$N_c$ limit the bag model expressions
\begin{align}
    [S_q^i(\vec r)]_{s^\prime  s}
    &= \phantom{-}\frac{P_q\,A^2}{8\pi}\,
    \biggl[\alpha_{+}^2\,j_0^2\,\sigma_{s^\prime s}^i+\,\alpha_{-}^2\,j_1^2\,
    \bigl(2\,e_r^i\,  \hat{e}_r\cdot\vec{\sigma}_{s^\prime s}\,-\,\sigma_{s^\prime s}^i\,\bigr)\biggr]\,\Theta_V,
    \label{Eq:bag-3D-S-density-arbitrary-spin}\\
    [L_q^i(\vec r)]_{s^\prime  s}
    &= -\frac{P_q\,A^2}{4\pi}\,
    \biggl[\alpha_{-}^2\,j_1^2\,
    \bigl(e_r^i\,\hat{e}_r\cdot\vec{\sigma}_{s^\prime s}\,-\,\sigma_{s^\prime s}^i\,\bigr)\biggr]\,\Theta_V,\\ 
    [J^i_{{\rm Bel},q}(\vec r)]_{s^\prime  s} &= 
    -\frac{P_q\,A^2}{8\pi}\,\biggl[\frac{2\Omega_0}{R}\,
    \alpha_+\alpha_-\,r\,j_0\,j_1+\alpha_-^2\,j_1^2\biggr]
    \bigl(e_r^i\,\hat{e}_r\cdot\vec{\sigma}_{s^\prime s}\,-\,\sigma_{s^\prime s}^i\,\bigr)\,\Theta_V\,,
\end{align}
where $\sigma_{s^\prime s}^i = \chi^\dag_{s^\prime}\sigma^i\chi_s$. The left hand side of Eq.~(\ref{Eq:equivalent-relation-AM-distributions}) then can be written as 
\begin{equation}\label{generic-AM-difference}
    [L^i_{q}(\vec r)]_{s^\prime  s}-
    [J^i_{{\rm Bel},q}(\vec r)]_{s^\prime  s} 
    = \frac{P_q\,A^2}{8\pi}\,\bigl(e_r^i\,\hat{e}_r\cdot\vec{\sigma}_{s^\prime s}\,-\,\sigma_{s^\prime s}^i\,\bigr)\,
    \biggl[\frac{2\omega r}{R}\,j_0\,j_1\, -\,\alpha_{-}^2\,j_1^2\,\biggr]\,\Theta_V\,.
\end{equation}
To evaluate the right-hand side of Eq.~(\ref{Eq:equivalent-relation-AM-distributions}), we first compute
\begin{equation}
\begin{aligned}\label{spin-derivative}
    \nabla^j [S^i_{q}(\vec r)]_{s^\prime  s} = \frac{P_q\,A^2}{8\pi}\,
    \biggl[& e_r^j\,\frac{\omega}{R} \Big(2\, \alpha_{+}^2\,j_0\,j_0^\prime\,\sigma_{s^\prime s}^i+\,2\, \alpha_{-}^2\,j_1\,j_1^\prime\,
    \bigl(2\,e_r^i\,\hat{e}_r\cdot\vec{\sigma}_{s^\prime s}\,-\,\sigma_{s^\prime s}^i\,\bigr) \Big) \\ 
    & + \frac{2}{r}\,\alpha_{-}^2\,j_1^2\,\Bigl(\bigl(\delta^{ij} - 2\,e_r^i\,e_r^j\bigl)\,\hat{e}_r\cdot\vec{\sigma}_{s^\prime s}\,+\,e_r^i\,\sigma_{s^\prime s}^{j}\,\Bigr) \biggr]\,\Theta_V\,,
\end{aligned}
\end{equation}
and obtain a similar expression for 
$\nabla^i [S^j_{q}(\vec r)]_{s^\prime  s}$ 
by exchanging $i\leftrightarrow j$ in Eq.~(\ref{spin-derivative}). Then, by using the Bessel function identities $j_0^\prime(x) = -j_1(x)$ and $j_1^\prime(x) = j_0(x) - \frac{2}{x}\, j_1(x)$, one obtains
\begin{equation}
    \nabla^j [S^i_{q}(\vec r)]_{s^\prime  s} - 
    \nabla^i [S^j_{q}(\vec r)]_{s^\prime  s} = \frac{P_q\,A^2}{4\pi}\,\bigl(e_r^i\,\sigma_{s^\prime s}^{j}\,- e_r^j\,\sigma_{s^\prime s}^{i}\bigl)\,\biggl[\frac{2\omega}{R}\,j_0\,j_1\, - \, \frac{1}{r}\,\alpha_{-}^2\,j_1^2\, \biggr]\,\Theta_V\,.
\end{equation}
Therefore, 
\begin{equation}
    \tfrac{1}{2} r^j\biggl(
    \nabla^j [S^i_{q}(\vec r)]_{s^\prime  s} - 
    \nabla^i [S^j_{q}(\vec r)]_{s^\prime  s}\biggr) = \frac{P_q\,A^2}{8\pi}\,
    \bigl(e_r^i\,\hat{e}_r\cdot\vec{\sigma}_{s^\prime s}\,-\,\sigma_{s^\prime s}^i\,\bigr)\,\biggl[\frac{2\omega r}{R}\,j_0\,j_1\, -\,\alpha_{-}^2\,j_1^2\,\biggr]\,\Theta_V
\end{equation}
yields the same result as in Eq.~(\ref{generic-AM-difference}).


\section{Electric charge distribution of the proton} \label{Appendix:charge-distribution}

In this appendix, we derive the bag model expression for the electric charge distribution of the proton which is used in the main text for a comparison to the energy distribution. 
The matrix elements of the electromagnetic current operator $j^\mu$ can be parametrized in terms of electric and magnetic Sachs form factors, $G_E$ and $G_M$, as follows \cite{Lorce:2020onh}
\begin{equation}\label{Eq:Sachs-form-factors}
    \la p^\prime,\vec s^{\,\prime}|  j^{\mu}(0) |p,\vec s\rangle
    = \bar u(p^\prime,\vec s^{\,\prime})\biggl[ \frac{M_N P^\mu}{P^2} \, G_E(t) + \, \frac{i\epsilon^{\mu\alpha\beta\lambda} \Delta_\alpha P_\beta \gamma_\lambda \gamma_5}{2P^2} \, G_M(t) \biggr]u(p,\vec s)\, .
\end{equation}
The electric Sachs form factor $G_E(t)$ encodes the charge distribution which can be obtained by the Fourier transform 
\begin{equation}\label{Eq:charge-distribution}
    \rho_\text{ch}(\vec{r}) = \int \frac{d^3\Delta}{(2\pi)^3} \, e^{-i\vec{\Delta}\cdot\vec{r}} \, \frac{M_N}{P^0} \, G_E(t)\, .
\end{equation}
To obtain $G_E(t)$ from Eq.(\ref{Eq:Sachs-form-factors}), one can choose $\mu = 0$ in the Breit frame, i.e. $\vec{P} = \vec 0$, and set $\vec s=\vec s^{\,\prime}$.
This yields   
\begin{equation}\label{Eq:Sachs-Electric}
    \la p^\prime,\vec s|  \overline{\psi}\gamma^{0}\psi |p,\vec s\rangle =2\,M_N\, G_E(t)
    \,.
\end{equation}
We evaluate the electric Sachs form factor $G_E(t)$ in the bag model in the large-$N_c$ limit,
by choosing the nucleon polarization along the $z$-axis and momentum transfer $\vec{\Delta} = (0, 0, \Delta^3)$. The result then reads
\begin{equation}
    G_E(t) = 4\pi A^2 R^6 \int \frac{d^3k}{(2\pi)^3} \Big[ \alpha^{2}_{+} t_0(k) t_0(k^\prime) + \alpha^{2}_{-} \vec{e}_{k} \cdot\vec{e}_{k^\prime} t_1(k) t_1(k^\prime)\Big] \, ,
\end{equation}
with $\vec{k}^\prime$ and $\vec{k}$ as
defined in Eq.~(\ref{Eq:axial-form-factor}).
Carrying out the Fourier transform in Eq.~\eqref{Eq:charge-distribution} yields the
charge distribution
\begin{equation}\label{Eq:bag-model-charge-distribution}
    \rho_\text{ch}(r) = \frac{A^2}{4\pi} \Big[ \alpha^{2}_{+} j_0^2 + \alpha^{2}_{-} j_1^2\Big] \, \Theta_V \, .
\end{equation}
In the limit $mR\to\infty$ which may be 
realized in various physical situations,
see Sec.~\ref{sec:limits-overview}, the electric charge 
distribution of the proton becomes
\be\label{Eq:3D-charge-mR}
     \rho_\text{ch}(r) = c_0\,j_0(\kappa r)^2\,\Theta_V 
      + \dots \;,
\ee   
where the dots indicate terms which are suppressed by
powers of $1/(mR)$. The constants $\kappa$ and $c_0$ are defined in 
sequel of Eq.~(\ref{Eqs:limits-3D}). The normalization is such that 
$\int\di^3r\,c_0\,j_0(\kappa r)^2\,\Theta_V=1$,
see Sec.~\ref{sec:limits-overview}.

\end{document}